%% file: main.tex
\documentclass[prb,twocolumn,superscriptaddress,showpacs,amsmath,amssymb]{revtex4-2}

\usepackage{natbib}
\usepackage{graphicx}
\usepackage{latexsym}
\usepackage{amsmath}
\usepackage{amssymb}
\usepackage{amsfonts}
\usepackage{placeins}
\usepackage{units}
\usepackage{siunitx}
\usepackage{color}
\usepackage{bm}
\usepackage{bbm}
\usepackage{hyperref}
\usepackage{ulem}

\usepackage{verbatim}

\newcommand\commacirc{%
  \mathrel{\ooalign{\hss$,$\hss\cr%
  \kern0.0ex\raise0.6ex\hbox{\scalebox{0.7}{$\circ$}}}}}
\DeclareMathOperator{\Tr}{Tr}
\DeclareMathOperator{\tr}{tr}
\DeclareMathOperator{\sgn}{sgn}

\DeclareMathAlphabet\mathbfcal{OMS}{cmsy}{b}{n}

\renewcommand{\Re}{\mathop{\mathrm{Re}}}
\renewcommand{\Im}{\mathop{\mathrm{Im}}}
\renewcommand{\vec}[1]{\bm{#1}}

\usepackage{xcolor}
\definecolor{TTH-color}{rgb}{0.0,0.0,1}

\definecolor{MS-color}{rgb}{1,0.0,0.0}

\newcommand{\MSedit}[1]{{\color{MS-color}#1}}

\definecolor{PV-color}{rgb}{0.97,0.57,0.11}

\definecolor{RO-color}{rgb}{0.6,0,0.6}

\definecolor{remove-color}{rgb}{0.0,0.5,0.0}

\begin{document}

\title{Superconductivity provides a giant enhancement to the spin battery effect}

\author{Risto Ojaj\"arvi}
\affiliation{Department of Physics and Nanoscience Center, University of Jyvaskyla, P.O. Box 35 (YFL), FI-40014 University of Jyvaskyla, Finland}

\author{Tero T. Heikkil\"a}
\affiliation{Department of Physics and Nanoscience Center, University of Jyvaskyla, P.O. Box 35 (YFL), FI-40014 University of Jyvaskyla, Finland}

\author{P. Virtanen}
\affiliation{Department of Physics and Nanoscience Center, University of Jyvaskyla, P.O. Box 35 (YFL), FI-40014 University of Jyvaskyla, Finland}

 \author{M.A.~Silaev}
\affiliation{Department of Physics and Nanoscience Center, University of Jyvaskyla, P.O. Box 35 (YFL), FI-40014 University of Jyvaskyla, Finland}
\affiliation{Moscow Institute of Physics and Technology, Dolgoprudny, 141700 Russia}
\affiliation{Institute for Physics of Microstructures, Russian Academy of Sciences, 603950 Nizhny Novgorod, GSP-105, Russia}

 \begin{abstract} 
  We develop a theory of the spin battery effect in superconductor/ferromagnetic insulator (SC/FI) systems taking into account the magnetic proximity effect. We demonstrate that the spin-energy mixing enabled by the superconductivity leads to the enhancement of spin accumulation by several orders of magnitude relative to the normal state.  %
 This  finding can explain the recently observed  giant inverse spin Hall effect generated by thermal magnons in the SC/FI system.
 We suggest a non-local electrical detection scheme which can directly probe the spin accumulation driven by the magnetization dynamics.
 We predict a giant Seebeck effect converting the magnon temperature bias into the non-local voltage signal. We also show how this can be used to enhance the sensitivity of magnon detection even up to the single-magnon level.
    \end{abstract}

 \maketitle

Generation and detection of pure spin signals is one of the main paradigms in spintronics\cite{wolf2001spintronics,
vzutic2004spintronics} and spin caloritronics\cite{Bauer2012a}. 
Spin pumping \cite{tserkovnyak2002enhanced,brataas2002spin,RevModPhys.77.1375}
in ferromagnet/metal multilayers
affects ferromagnetic resonance (FMR) and spin Hall magneto-resistance measurements\cite{nakayama2013spin, weiler2013experimental}. 
Spin Seebeck effect\cite{Uchida2010,weiler2013experimental} converts thermal nonequilibrium states into pure spin currents and can be used for the detection of magnons propagating through FI materials without electrical losses 
\cite{chumak2015magnon,cornelissen2015long}. Pure spin current flowing from the ferromagnet into the adjacent metal leads to the build up of spin accumulation known as the spin battery effect\cite{brataas2002spin}.

Recently it has been discovered\cite{yang2010extremely,quay2013spin,Wolf2013, hubler2012long,Kolenda2017,heidrich2019nonlocal} that 
superconductivity strongly increases spin relaxation times and lengths, which makes superconducting materials promising for spintronics\cite{linder2015superconducting, Han2019,RevModPhys.90.041001}. 
Long-range non-equilibrium spin states created in superconductors by  electrical and thermal  
injection of Bogoliubov quasiparticles have been intensively studied \cite{ozaeta2014predicted,Silaev2015,RevModPhys.90.041001,heikkila2019thermal,
krishtop2015nonequilibrium,
bobkova2015long,
bobkova2017thermospin,virtanen2016stimulated,kuzmanovic2020evidence}. The question of how the weak spin relaxation in superconductors shows up in spin pumping properties have remained unexplored and is addressed in the present Letter. 

Most of the  experimental 
\cite{bell2008spin, wakamura2015quasiparticle, quay2013spin, Kolenda2017,hubler2012long,Jeon2018,jeon2020tunable,PhysRevApplied.14.024086,Jeon2020giant}
and theoretical works studying magnetization dynamics in superconductor/ferromagnet systems focus on the FMR properties\cite{brataas2004spin,morten2008proximity,inoue2017spin,kato2019microscopic, silaev2020large, silaev2020finite,tanhayi2020superconductivity} and spin torques\cite{trif2013dynamic,ojajarvi2020nonlinear}.  

Here we consider the spin battery effect\cite{brataas2002spin}, that is the static spin accumulation of  Bogoliubov quasiparticles in a superconductor (SC) generated either by the coherent FMR drive or by the thermal magnons in the adjacent FI.  
Our study is motivated by the recent experiment demonstrating that magnons induce giant inverse spin-Hall signal in the transition state of Nb/YIG superconductor/ferromagnetic insulator system\cite{Jeon2020giant}.  Due to the close relation between the spin Hall signal and spin density, this observation hints that the spin accumulation induced by thermal magnons is modified in a highly non-trivial way by the superconducting correlations. 

 \begin{figure}
 \centering
 \includegraphics[width=\linewidth]{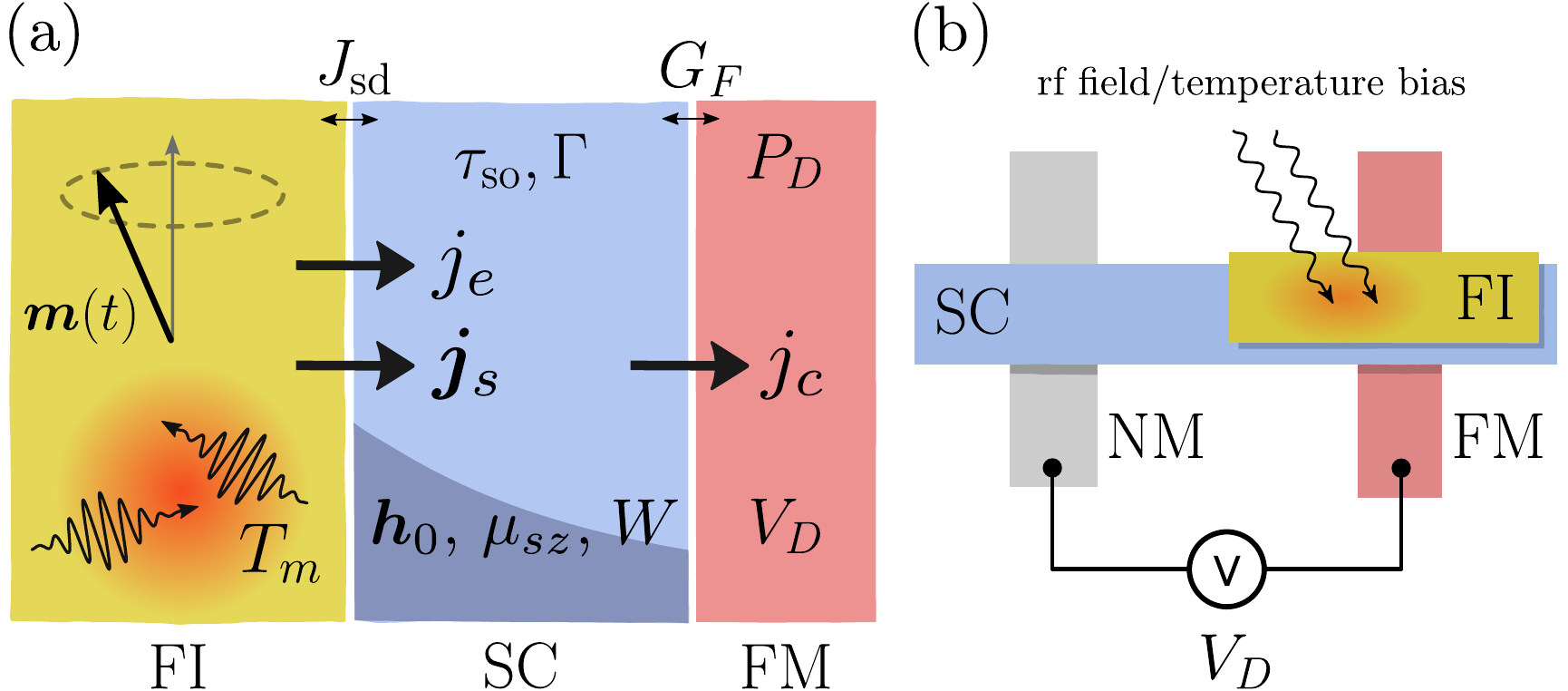}
  \caption{\label{Fig:SchematicNonLocal}
           Schematic FI/SC setup to measure spin accumulation induced by magnons. (a) 
           Nonequilibrium magnon distribution in FI, generated either by a coherent FMR drive or a temperature bias, induces spin and energy currents $\bm j_s$ and $j_e$ to the SC, which create spin and energy accumulations $\bm\mu_s$ and $W$ in SC. Proximity to FI also induces a static exchange field $\vec h_0$ in the SC. 
           The spin accumulation is converted to electrical voltage $V_D$ in the ferromagnetic electrode (FM) with the polarization $P_D$.
           (b) Non-local circuit to measure magnon-induced voltage $V_D$. 
  }
 \end{figure}

The considered setup is detailed in  Fig.~\ref{Fig:SchematicNonLocal}a which shows the time-averaged quasiparticle spin accumulation 
$\langle {\bm \mu}_s\rangle$
generated in SC. It can be
can be measured\cite{yang2010extremely,quay2013spin, hubler2012long,Kolenda2017} in the non-local circuit Fig.~\ref{Fig:SchematicNonLocal}b 
 consisting of the spin-polarized tunnel contact with a
 metallic ferromagnet (FM) near FI and the distant normal metal electrode (NM).
   The dc voltage $V_D$ induced into this tunnel contact in the absence of a charge current through it is \cite{heikkila2019thermal}  
       \begin{align} \label{Eq:VD}
           V_D = \frac{G_{Fn}}{G_{F}}  \bm P_D \cdot \langle {\bm \mu}_s\rangle.
       \end{align}
       Here $G_F =G_{Fn} \int_0^\infty d\varepsilon N (\varepsilon)\partial_\varepsilon n_0 $ is the linear local tunneling conductance and $\bm P_D$ the spin polarization of the SC/FM junction,  $N(\varepsilon)$ is the density of states in the superconductor and $n_0= \tanh (\varepsilon/2T)$ is the equilibrium distribution function.

 In the superconducting case the information carried by the strength of the spin pumping which determines the FMR linewidth
  is different from that in $\langle {\bm \mu}_s\rangle$. 
It is generally proportional to the amplitude of magnetization dynamics $\langle {\bm \mu}_s\rangle\propto\langle \bm m \times \partial_t \bm m \rangle $, where  $\bm m(t)$ is the unit vector of magnetization direction in FI.
In superconductors, however, the proportionality constant of $\langle \bm\mu_s\rangle$ is sensitive to the magnitude of energy relaxation
time $\Gamma^{-1}$. In the typical case 
$\Gamma^{-1}\gg \tau_{s}$  the resulting non-local voltage $V_D$ can be parametrically larger in the superconducting state than in the normal state by the factor $\sim(\Gamma \tau_{s} )^{-1}$.
In  superconductors 
Nb and Al these times are of the order\cite{
hubler2012long, jeon2018spin}   $ \tau_{s} \approx 0.1T_c^{-1}$ in Al and
and $ \tau_{s} \approx T_c^{-1}$ in Nb, while 
 $ \Gamma^{-1} (T_c) \approx 10^3 T_c^{-1} $ in both materials\cite{gershenzon1990electron, klapwijk1986electron}.
Therefore in these superconductors one can expect  an  enhancement of spin accumulation induced by spin pumping by the factor of $ (\Gamma \tau_{s} )^{-1} \sim 10^2-10^3$ as compared to the normal state.  


 The origin of the very large spin accumulation  in FI/SC contacts is twofold. 
 First, 
 magnetization dynamics  
 results in the  energy current   \cite{tserkovnyak2002enhanced,brataas2008scattering} 
 $j_e
=  \alpha \langle |\partial_t \bm m|^2 \rangle
$, where $\alpha$ is the contribution to the Gilbert damping coefficient due to the contact.  
%
  \begin{figure}
  \centering
   \includegraphics[width=0.95\linewidth]{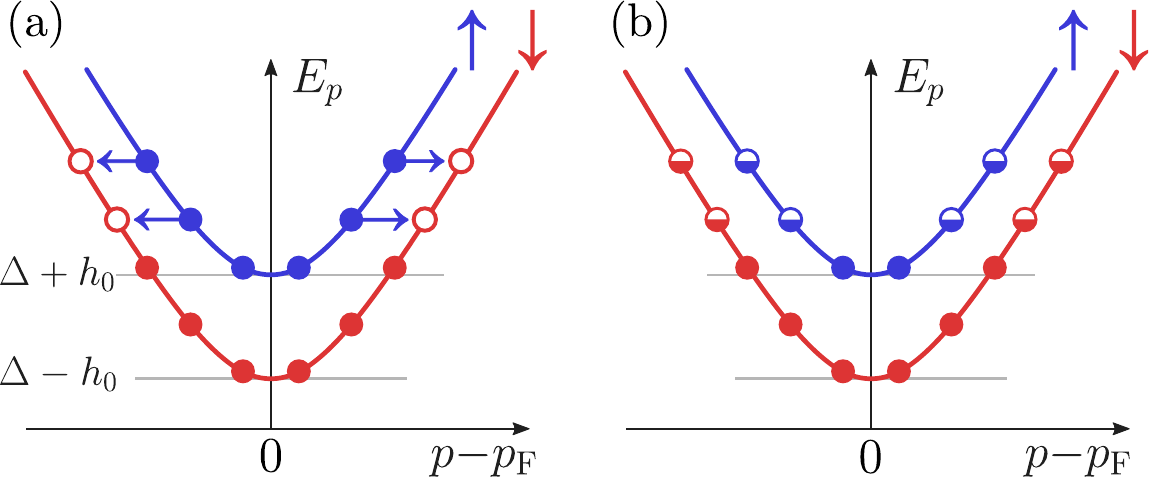}
   \caption{
   \label{Fig:SchematicWSzRelaxation}  
   Spin-split Bogoliubov  spectrum in SC and its occupation driven by magnons. (a) State with pure energy accumulation $W$ and no spin accumulation. Horizontal arrows represent elastic spin scattering. (b) Elastically relaxed state. Elastic relaxation produces spin accumulation $\mu_{sz}$ from energy accumulation $W$. The full/half-filled/empty circles represent occupied/partially filled/unoccupied states. The effect depends on the asymmetry between spin-resolved density of states $N_\uparrow(\varepsilon)$ and $N_\downarrow(\varepsilon)$, and is therefore absent in the normal state. }
  \end{figure}
%
Second, in superconductors the spin splitting in the Bogoliubov spectrum 
generated by FI through the magnetic proximity effect\cite{RevModPhys.90.041001,heikkila2019thermal,meservey1994spin,hijano2020coexistence} leads to the strong coupling between energy and spin degrees of freedom\cite{heikkila2019thermal}. 
 The mechanism of converting pumped quasiparticle energy to spin accumulation via elastic spin-relaxation processes is demonstrated in Fig.~\ref{Fig:SchematicWSzRelaxation} which shows non-equilibrium quasiparticle states on the spin-split Bogoliubov branches $E_p(p)$ for different momenta $p$. The spin quantization axis is determined by the induced Zeeman field $\bm h_0 = h_0 \bm z$, when the static magnetization direction is $\bm m_0 = \bm z$. 

  Energy  current $j_e$  generates spin-neutral energy accumulation $W$ by non-equilibrium quasiparticle states shown schematically by the filled circles in Fig.~\ref{Fig:SchematicWSzRelaxation}a. 
  The important feature of this distribution is that both spin-up and spin-down branches have the same number of occupied states. 
 Due to the spin splitting the spin-up and spin-down branches are filled up to  different energy levels.
The resulting population imbalance can relax due to the elastic spin scattering process. As a result, all spin-up and spin-down states with identical energies become equally populated. As one can see from  Fig.~\ref{Fig:SchematicWSzRelaxation}, in this state the net spin accumulation is non-zero because of the energy interval $\Delta- h_0< E_p< \Delta+ h_0$ where only the spin-down states exist.\cite{spinrelnote} 

\begin{figure*}
  \centering
  \includegraphics[width=1.0\linewidth]{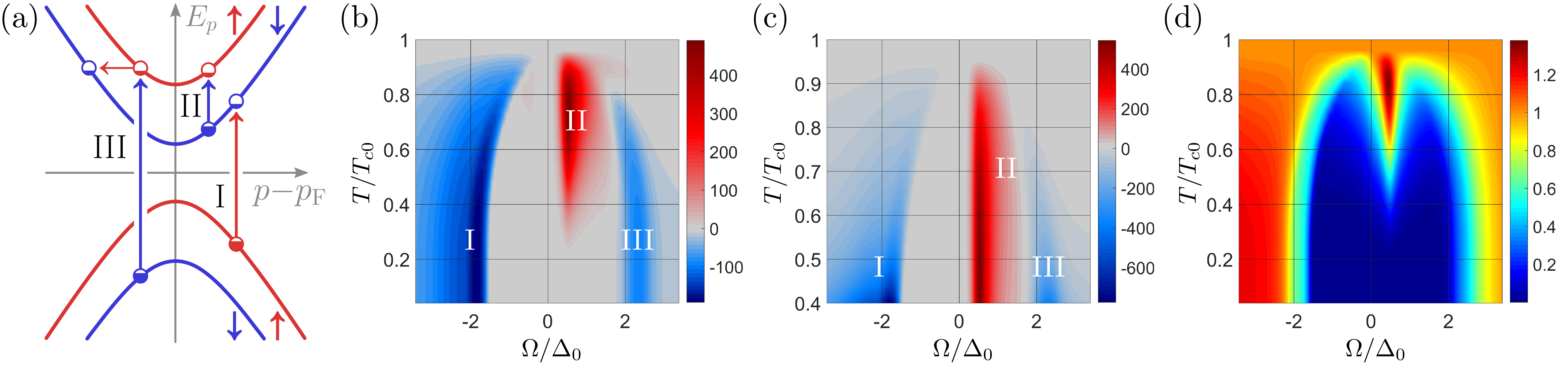} 
  \caption{\label{Fig:SzDfun} 
    (a) Quasiparticle excitation processes of the spin-split Bogoliubov spectrum. Vertical blue/red arrows are due to the absorption of a magnon with spin $\pm 1$. Horizontal arrows represent rapid spin relaxation. The filling of the circles shows the occupation of the states after spin relaxation. The corresponding peaks are labeled as I--III in the next panels.
 (b) Pumped spin accumulation
$( T_{c0}/h_\Omega^2)  \mu_z(T,\Omega)$,
(c) non-local voltage 
$ (e T_{c0}/h_\Omega^2)V_D(\Omega,T)$
and (d) Pumped energy accumulation $W(T,\Omega)/W(T_c,\Omega)$,
generated by the magnetization dynamics in the setup of Fig.~\ref{Fig:SchematicNonLocal}.
 The parameters used for (b)--(d) are $\Gamma/T_{c0}=10^{-3}$, $h_0/T_{c0}=0.528$ and  $\tau_{s}^{-1}/T_{c0} = 1.19$. For these parameters $T_c \approx 0.9 T_{c0}$, where $T_{c0}$ is the critical field at $h_0=\tau_{s}^{-1}=0$. 
  }
 \end{figure*}
 
 The energy-to spin conversion processes can be quantified using kinetic equations together with the collision integrals  corresponding to the spin-orbit or spin-flip scattering.
   Introducing the distribution functions $f_{\uparrow/\downarrow}$ and densities of states $N_{\uparrow/\downarrow}$ in spin-up/down subbands, we obtain\cite{SupplMat} the 
    spectral densities for spin and energy accumulations
        $f_s = N_\uparrow f_\uparrow - N_\downarrow f_\downarrow$
     and $f_e = \varepsilon ( N_\uparrow f_\uparrow + N_\downarrow f_\downarrow)$. 
     The elastic spin-scattering collision integral 
     is given by ${\cal I}_{s} = (f_s - \kappa_{se} f_e)/{\cal T}_1$, 
     where 
    ${\cal T}_1$
          is the longitudinal spin relaxation time in the superconducting state \cite{heikkila2019thermal} and spin-energy coupling coefficient 
$\kappa_{se} (\varepsilon)= 
(N_\uparrow - N_\downarrow)/[\varepsilon(N_\uparrow + N_\downarrow)]$. 
For weak spin splitting $h_0 \ll \Delta^2$, we can estimate $\kappa_{\rm se} \sim h_0/(\varepsilon \Delta)$, where $\Delta$ is the superconducting gap.
The spin-diffusion equation modified by the spin-energy coupling is given by 
\begin{subequations}
\begin{align} \label{Eq:Js}
&  \partial_x {\cal J}_{sz}   =  \frac{f_s(\varepsilon)  - \kappa_{se}  f_e }{{\cal T}_1},
\\ \label{Eq:Je}
& \partial_x {\cal J}_e = I_{e-ph} + \Gamma f_e,
\end{align}
\label{eq:spinenergydiffusion}%
\end{subequations}
where ${\cal J}_{sz} (\varepsilon)$ and ${\cal J}_e (\varepsilon)$ are the spectral densities of the time-independent spin  $ j_{sz} = \int d\varepsilon  {\cal J}_{sz}$ and energy $j_e = \int d\varepsilon  {\cal J}_e$  currents.
  The sources of these currents are determined by the boundary conditions at the FI interface with dynamical magnetization fixing the values of interfacial currents $ {\cal J}_e (x=0) \propto  \langle |\partial_t \bm m|^2 \rangle$ and  
  $ {\cal J}_{sz}(x=0)  \propto \bm z\cdot \langle \bm m \times \partial_t \bm m \rangle$. They are obtained generalizing the  theory of normal-state spin battery effect\cite{brataas2002spin}  for the superconducting case.\cite{SupplMat}
  In the limit of small SC film thickness $d$ the solution for spin accumulation $\mu_{sz} = (\bm \mu_s\cdot\bm z)$   is 
  $\mu_{sz}= -
  d^{-1}\int d\varepsilon  ( \Gamma^{-1} \kappa_{se} {\cal J}_e + {\cal T}_1 {\cal J}_{sz}  ) $.  The first term has a large prefactor $\Gamma^{-1}$ and provides the possibility of spin signal  enhancement  by the parameter $\kappa_{se}/(\Gamma {\cal T}_1)$  as compared to the normal state, where only the second term contributes.  The detailed calculation\cite{SupplMat} described below shows that both ${\cal J}_e$ and  ${\cal J}_{sz}$ are not dramatically smaller than their normal state magnitudes down to $T\approx 0.3 T_c$. Thus $\mu_{sz}$ is enhanced by the factor $\kappa_{se}/(\Gamma {\cal T}_1)$ at $T/T_c \approx 0.8 - 0.9$.

  The described effects are quantified using Keldysh-Usadel equation\cite{silaev2020large,silaev2020ff} 
  \begin{equation}\label{Eq:KeldyshUsadelFI}
  - \{\hat\tau_3\partial_t\commacirc \check g \} + 
  \partial_x( D\check g\circ \partial_x \check g)  = 
   [ \Delta \hat \tau_1 + \check \Gamma + \check \Sigma_{\rm so}\commacirc \check g\, ]
  \end{equation} 
for the quasiclassical Green's function (GF) $\check g$ in 8$\times$8 space consisting of Keldysh, Nambu and spin indices.\cite{heikkila2019thermal}
 The elastic spin relaxation is determined by the spin-orbit scattering self-energy $\check \Sigma_{\rm so}$   \cite{SupplMat}, while  $\check \Gamma$ describes the coupling to the normal reservoir to model the {\it inelastic relaxation} \cite{PhysRevB.103.024524}. 
  The  spin splitting and pumping induced by the electron scattering at the FI interface $x=0$
are modelled by the  {\it dynamical boundary conditions} \cite{silaev2020finite,Tokuyasu1988} 
 \begin{align} \label{Eq:BCFS}
  D \check g \circ \partial_x \check g (x=0) =
  iJ_{sd}
  [\hat\tau_3 \hat{\bm \sigma} \bm m \commacirc \hat g  ]  
  \,,
  \end{align}
   %
   where we denote $[A\commacirc B] (t_1,t_2) = \int dt A(t_1,t)B(t,t_2) - B(t_1,t) A(t,t_2)$ and similarly for the anticommutator $\{X\commacirc Y\}$. Here the interface is characterized by the effective exchange coupling\cite{ohnuma2014enhanced} $J_{sd}$.    
      Within the minimal  model of the FI\cite{Tokuyasu1988,millis1988quasiclassical} it can also be expressed through the spin-mixing angle \cite{silaev2020finite,silaev2020large}. 

    We assume the time-dependent magnetization is
    $\bm m_\perp (t) = m_\Omega(\cos(\Omega t), \sin(\Omega t), 0)$ consisting of the  left- and right-hand parts
     $\bm m_\perp (t) = m_{\Omega, l} e^{i\Omega t}(\bm x-i\bm y) + m_{-\Omega, r} e^{-i\Omega t} (\bm x+i\bm y)$ with 
     $m_{\Omega, l} = m_{-\Omega, r}= m_\Omega/2$. 
     In general, 
     solving Eqs.~(\ref{Eq:KeldyshUsadelFI}--\ref{Eq:BCFS}) to the second order in time-dependent field we obtain
      the stationary second-order correction to the Keldysh component of the GF
      $\hat g^K (\varepsilon) \propto m_\Omega^2$. It  
     consists of corrections to the spectral function
 analogous to those induced by the electromagnetic irradiation
 \cite{semenov2016coherent,
linder2016microwave} and of the  anomalous part
  \cite{
eliashberg1971inelastic, 
gor1975vortex, larkin1977non,
artemenko1979electric} $\hat g^a$ which determines the stationary spin accumulation and thereby the non-local voltage in Eq.~\eqref{Eq:VD}. 
      The calculation of $\hat g^a$ and its relation to the observables $W$, $ \mu_{sz}$  and the distribution functions $f_\uparrow$, $f_\downarrow$ is  presented in
      Supplementary Material \cite{SupplMat}.
     It provides the general expression for the spin accumulation 
                \begin{align} \label{Eq:SzGeneral}
    \mu_{sz} (\Omega, T ) = \chi_{lr} (\Omega,T) m_{l,\Omega} m_{r,-\Omega},
    \end{align}
where  $\chi_{lr}$ is the second-order spin response function.
    
  Here we consider a superconductor film with thickness $d_S \ll \ell_{sn},\xi_0$
        small compared to the spin relaxation and coherence lengths in the superconductor. 
       Then Eqs.~(\ref{Eq:KeldyshUsadelFI}--\ref{Eq:BCFS}) can be reduced\cite{SupplMat} to the coordinate-independent Usadel equation with an  
      effective Zeeman field
       $\bm h = J_{\rm sd} \bm m/d$ so that $h_0 =J_{\rm sd} /d $ and $h_\Omega = h_0 m_\Omega$. 
     
The calculated dependencies of pumped spin accumulation $\mu_{sz}$, non-local voltage $V_D$ and energy $W$  are shown in Fig.~\ref{Fig:SzDfun}(b--d). One can see the clear  correlation between these three quantities resulting from the strong spin-energy coupling in spin-split superconductors. The key feature of $\mu_{sz}(\Omega)$ and $V_D(\Omega)$ dependencies are the  sharp   
    peaks  labelled
by I and II as well as the  less pronounced peak labelled by III corresponding to the different spin excitation processes shown schematically on the energy level diagram Fig.~\ref{Fig:SzDfun}(a). 
The excitation processes I and II create nonequilibrium quasiparticle states on the spin-down branch at the energy interval $\Delta- h_0< E_p< \Delta+h_0$, which corresponds to the situation with spin-energy accumulation shown in Fig.~\ref{Fig:SchematicWSzRelaxation}.  Such states can relax only due to the slow energy relaxation which determines the large amplitude of the peaks I and II in Fig.~\ref{Fig:SzDfun}(b--c). 
The size of these peaks  scale as $ {\rm min} (\tau_{s}, \Delta^{-1}) h_0/ (\tau_{s} \Gamma) $ as demonstrated by the series of plots for different parameters\cite{SupplMat}. 
The process III is more complicated since it requires the existence of subgap spin-up states at $[\Delta-h_0, \Delta+ h_0]$ energy interval
which appear due to the broadening of spin subbands by the spin relaxation. The equilibration of spin-up and spin-down populations shown by the 
horizontal arrow leads to $f_\uparrow= f_\downarrow$ but the spin accumulation appears due to the DOS difference $N_\downarrow> N_\uparrow$.   

 Results in Fig.~\ref{Fig:SzDfun}(b--c) predict sizable spin and voltage signals 
  even for low frequencies $\Omega \ll \Delta_0$. They are especially pronounced  near the peak II associated with electron paramagnetic resonance frequency $\Omega \approx 2h_0$ usually reached in FMR experiments with resonance frequencies around several GHz. The excitation process  II in Fig.~\ref{Fig:SzDfun}(a) polarizes existing quasiparticles and therefore disappears at low temperatures $T\ll T_c$.  The processes I and III exist even at $T\to 0$ since they break Cooper pairs and create spin-polarized quasiparticles out from the vacuum state. As a result  peaks I and III become exponentially diverging in the voltage signal at low temperatures $T\ll T_c$ (not shown in Fig.~\ref{Fig:SzDfun}c) $ V_D\propto e^{\Delta/T}$ since the local conductance $G_F \propto e^{-\Delta/T}$ in Eq.~\ref{Eq:VD}. 
  
Because of energy conservation $W (\Omega) = \alpha (\Omega)  \Omega^2  m_\Omega^2/\Gamma $, where $\alpha (\Omega)$ is the frequency-dependent increase of Gilbert damping.  The plot of the ratio $W(\Omega,T)/W(\Omega,T_c) = \alpha (\Omega,T)/\alpha(\Omega,T_c)$
shows the presence of the superconducting gap since the damping is generally suppressed for $\Omega < 2\Delta_0$. For temperatures somewhat below $T_c$ there is a coherence peak \cite{kato2019microscopic,inoue2017spin,silaev2020finite,silaev2020large}  at around $\Omega \approx 2 h_0$. 
  
\begin{figure}
  \centering
  \includegraphics[width=\columnwidth]{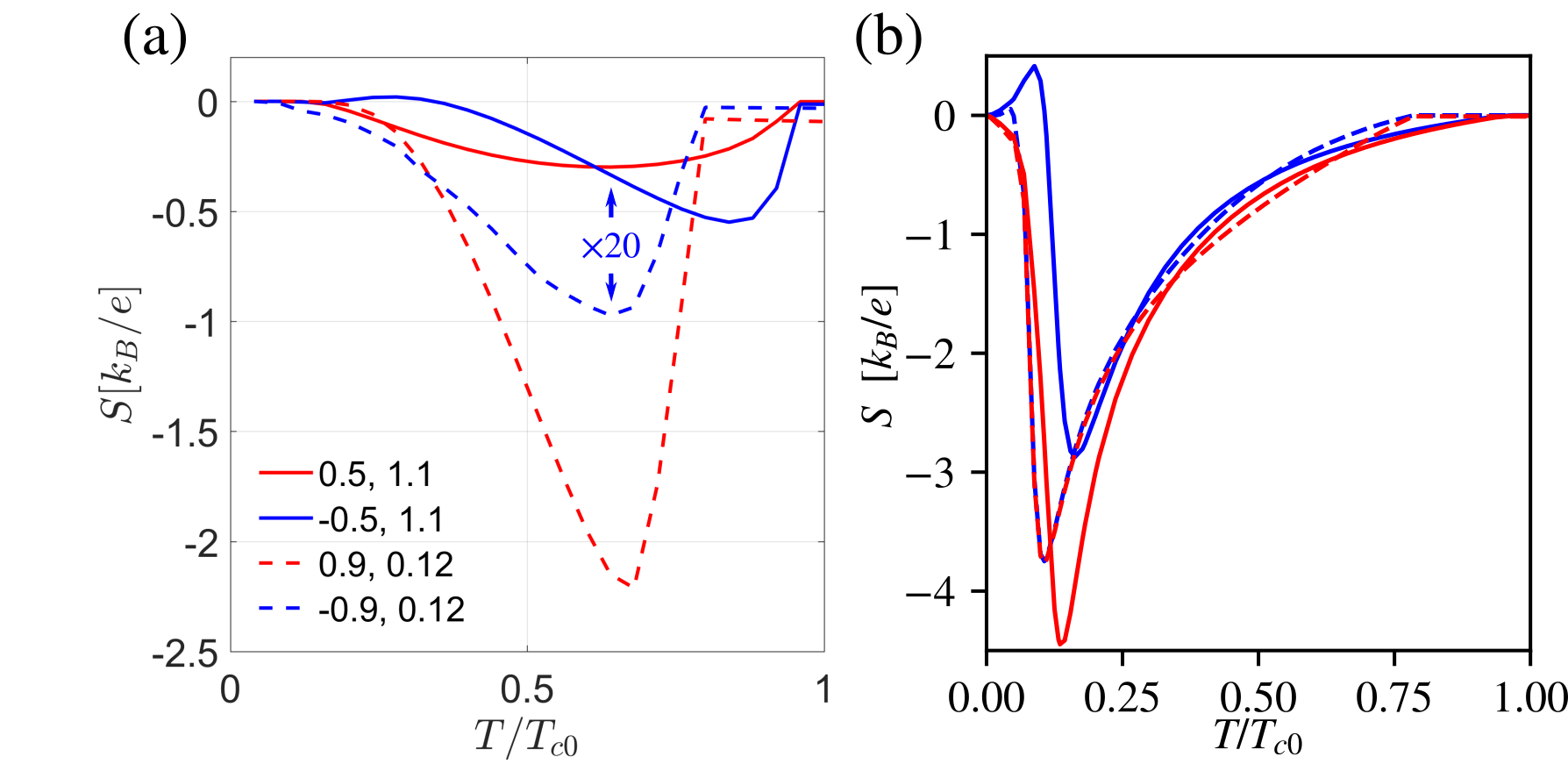}
  \caption{\label{Fig:SeebeckT}
 Magnon  Seebeck coefficient $S(T)$ in FI/SC/FM setup calculated using  
  (a) Model energy relaxation  (\ref{Eq:KeldyshUsadelFI}) with $\Gamma=10^{-3}T_{c0}$, scaled by the normal-state value $S^{(n)} = S (T_c)$; 
  (b) Quasiequilibrium model (\ref{Eq:Quasiequilibrium}) with electron-phonon relaxation.
   Red (blue) curves correspond to $(\bm h_{0}\cdot \bm m_0 )> (<) 0 $.
   Blue curves in (a) are multiplied by $20$ for clarity.
   Solid and dashed lines are for $|h_0| = 0.5T_{c0}$, $\tau_{so}^{-1}=1.1 T_{c0}$ (Nb) and $|h_0| = 0.9 T_{c0}$, $\tau_{so}^{-1}=0.12 T_{c0}$ (Al), respectively; 
  $P_D=0.5$, $\vec{P}_D\cdot\vec{h}_0>0$.
  }
\end{figure}

     Next, we consider the spin accumulation driven by the stochastic magnetization corresponding to the {\bf magnon thermal field} at temperature $T_m \neq T$ which can be  controlled with the help of electrical spin injection based on the spin Hall effect \cite{cornelissen2015long,Jeon2020giant}. 
    For that we find $\mu_{sz}$ by averaging Eq.~(\ref{Eq:SzGeneral})  
    over the fluctuations of magnetization. 
   This can be done \cite{SupplMat} 
   by 
   replacing the product of classical field components with the non-equilibrium Keldysh magnon propagator  $m_{l,\Omega} m_{r,-\Omega} \to  v_s \delta D^K(\Omega)$, where $v_s$ is the volume per spin, and summing over $\Omega$. 
   In the stationary case 
  $\delta D^{K}(\Omega) = D^{RA}(\Omega) \delta f_m (\Omega)$, where $D^{RA}(\Omega)$ and $f_m$ are the magnon density of states and the distribution function\cite{SupplMat}. For the thermally biased magnon state    $\delta f_m(\Omega) = n_B(\Omega/T_m)-n_B(\Omega/T)$, where $n_B (\Omega/T)= \coth (\Omega/2T)$. 
 This approach generalizes the calculation of the magnon-driven spin current\cite{PhysRevB.83.094410,adachi2013theory,Kato2019} to that of the magnon-driven spin accumulation. 
  For small magnon temperature bias this spin accumulation $\mu_{sz}\propto  (T-T_m) $  
 and the detector voltage (\ref{Eq:VD}) 
 can be expressed 
 through the linear Seebeck coefficient
 characterising the conversion of magnon temperature into the electric signal in FI/SC/FM non-local circuit $V_D= S(T-T_m) $  
    \begin{align} \label{Eq:SSE}
     S=
      P_D v_s m_M^{3/2} \frac{ G_{Fn}}{e G_{F}} \int_0^{\infty}
     \sqrt{\Omega}  \chi_{lr}\partial_T n_B d\Omega ,
        \end{align}
      where  $v_s$, the volume per unit spin in FI, determines the number of magnon modes. 
 For YIG,\cite{srivastava1987}  $m_M\approx
 \SI{1}{\electronvolt^{-1}\angstrom^{-2}}$ and\cite{cherepanov1993saga}  $v_s\approx
 \SI{500}{\angstrom\cubed}$. 
      Figure \ref{Fig:SeebeckT} shows $S(T)$ for  parameters qualitatively corresponding to the EuS/Al and YIG/Nb based FI/SC bilayers that have been studied recently \cite{Jeon2020giant,hijano2020coexistence}. 
  
  The spin signals are enhanced even more due to the energy dependence of the inelastic scattering rate when the relaxation is due to the electron-phonon coupling. This can be demonstrated in the 
  quasiequilibrium limit, assuming the rapid internal thermalization process that allows to parametrize the distribution function by temperature $T_S$ and the spin-dependent chemical potential shift $e V_s$. Then kinetic Eqs. \eqref{eq:spinenergydiffusion}  can be written as the following system describing energy, spin and charge currents at SC/FI and SC/FM interfaces
   \begin{align} \nonumber
      &  G_{e-ph}(T_S-T_{ph}) = G_{me} (T_m-T_S)
       \\ \label{Eq:Quasiequilibrium}
&    \mathcal{V}_S\nu eV_s/\tau_{sa}   =  
G_{ms} (T_m-T_S) 
\\ \nonumber
& (G_{F}/G_{Fn}) V_D =  P_D  V_s + \alpha_{th} (T_S-T_{F}).
   \end{align}
   Here $G_{e-ph}$ is the electron-phonon thermal conductance, $\alpha_{th}= e P_D\int_0^{\infty} (N_\uparrow -N_\downarrow ) \partial_T n_0 d\varepsilon  $ is the thermoelectric coefficient at the SC/FM interface, \cite{ozaeta2014predicted,RevModPhys.90.041001}
      $\mathcal{V}_S$ the superconductor volume,
   $\nu$ its density of states,
   and the energy-averaged spin relaxation rate is  $\tau_{sa}^{-1} = \int_0^\infty d\varepsilon\, \partial_\varepsilon n_0 {\cal T}^{-1}_1 N_\uparrow N_\downarrow / (N_\uparrow + N_\downarrow)$.
    The magnon-electron conductances for spin and heat, $G_{ms}$ and $G_{me}$, are 
  expressed\cite{SupplMat} through the linear spin susceptibility \cite{maki1973,silaev2020ff}, and were previously studied in the normal state\cite{bender2015interfacial,cornelissen2016magnon}. 
  Further we assume that the temperature of the phonon heat bath is equal to that of the ferromagnetic metal electrode $T_F=T_S$ to obtain the electric Seebeck coefficient  
  \begin{align}
      S =  \frac{G_{Fn}}{G_F}\left(\frac{P_D \tau_{sa} G_{ms}}{\mathcal{V}_S\nu} + \frac{\alpha_{th} G_{me}}{G_{me} + G_{e-ph}} \right).
  \end{align}
The second term is again due to the spin-energy mixing, and it provides the dominating contribution in the superconducting state. The Seebeck coefficient is plotted in Fig.~\ref{Fig:SeebeckT}b. \footnote{Fig. 4 assumes for simplicity that the electron-phonon coupling in Nb is the same as in Al.} Compared to the full nonequilibrium case, we find that due to the rapid decrease of the electron-phonon coupling with decreasing temperature, the signal persists to lower temperatures and is mainly limited by the Seebeck coefficient of the SF junction \cite{ozaeta2014predicted}.

The large value of the Seebeck coefficient converting the magnon temperature difference to an electrical voltage indicates that this device can be used as an ultrasensitive detector of propagating magnons,\cite{SupplMat} analogous to the thermoelectric detector suggested in Refs.~\onlinecite{heikkila2018thermoelectric,chakraborty2018thermoelectric}. The detector can have a very low noise equivalent power of the order of $NEP^2 \sim G_{\rm th} T^2$, limited by the weak thermal conductance $G_{\rm th} = G_{me}+G_{e-ph}$ from the superconductor to the relevant heat baths. Similar to the other nanoscale superconducting detectors \cite{govenius2014microwave,
govenius2016detection,
kokkoniemi2019nanobolometer}, they will also have a very good energy resolution $\Delta E = NEP \sqrt{\tau_{\rm eff}}$, provided that the thermal relaxation time $\tau_{\rm eff}$ is not too long. With suitable setting one can then approach even the detection of single propagating magnons with frequencies of a few tens of GHz.

     {\bf To conclude}
     we have shown  how the electron-hole symmetry breaking present in SC/FI bilayers mixes the spin and energy modes and leads to a giant enhancement of the spin battery effect. 
     This leads to the large magnon-driven Seebeck effect which can be considered as a very sensitive  detector of magnons.
     We expect this effect also to explain the giant spin-Hall signal measured in \cite{Jeon2020giant}, but its precise description would require appending the theory with the description of the spin-Hall angle\cite{bergeret2016manifestation, Tokatly2017, huang2018extrinsic}.

     The mechanism of producing giant spin signals does not necessarily require superconductors, but we expect similar effects in any system exhibiting strong spin-resolved electron-hole asymmetry, such as semimetals in the presence of large exchange fields or magnetic topological insulators\cite{otrokov2019prediction}. Such systems allow for an electrical access to the energy dissipation processes in ferromagnetic resonance, or detailed studies of the magnon spectra via the heat conductance $G_{me}$ between electrons and magnons.

      {\bf Acknowledgements}
  This work was supported by the Academy of Finland Projects 297439 and 317118, the European Union’s Horizon 2020 Research and Innovation Framework Programme under Grant No. 800923 (SUPERTED), and Jenny and Antti Wihuri Foundation.

\bibliographystyle{apsrev4-2} 
\bibliography{refs3} 

\nocite{Dynes1984,heikkila2019thermal,abrikosov1962spin,Tokuyasu1988,RevModPhys.77.1375,ohnuma2014enhanced,brataas2002spin,Silaev2015,maki1973,silaev2020ff,adachi2013theory,kamenev2011field,bender2015interfacial,cornelissen2016magnon,heikkila2018thermoelectric,chakraborty2018thermoelectric,ozaeta2014predicted}

\clearpage
\input{supplmat.tex}

\end{document}

%% file: supplmat.tex
 
\section{Supplementary material} 
  
  Here we provide technical details of the formalism,  verification of our approach by comparison with known results as well as the extended results of calculations for wide range of parameters.  
    In Sec.~\ref{SMSec:GeneralFormalism} we describe the general formalism of Keldysh-Usadel equation with the dynamical boundary conditions at the S/FI interface. 
    In Sec.~\ref{SMSec:SpinCurrent} we show that our general formalism yields the conventional expression for the spin current with pumped and backflow terms. 
    In Sec.~\ref{SMSec:SpinDiffusionNormal} we show that our formalism in the normal superconducting state yields the
    usual expression for static spin current and spin accumulation (spin battery effect). 
    
    The generalization of kinetic equations to describe spin accumulation in the superconducting spin sink are derived in Sec.~\ref{SMSec:KineticEq}. 
    
    In Sec.~\ref{SMSec:NumericalPerturbation} we describe the perturbation theory approach to solving Keldysh-Usadel equation to the second order of the driving Zeeman field 
    to calculate the stationary spin accumulation. Here we present the extended calculation results of spin and energy accumulation as well as the non-local voltage driven by the magnetization dynamics for a wide range of parameters.

 Beyond perturbation theory we have also developed the numerically exact solution of the non-stationary Keldysh-Usadel equation with the time-dependent Zeeman field. The method is described in Sec.~\ref{SMSec:HigherOrder}. 
 
 Using the results for spin accumulation driven by the deterministic magnetic signal we can treat the case of stochastic magnetization dynamics driven by the field of thermal magnons.
 The approach based on the calculation of electron-magnon collision integral is 
  presented in Sec.~\ref{SMSec:MagnonDerivation}.
   We verify our approach by deriving the known results for the magnon-driven spin and energy currents in the normal state of FI/metal bilayer. 
   
   In Sec.~\ref{SMsec:SpinEnergyPumping}
     We demonstrate that general relations between pumped spin and energy currents are valid in the superconducting state.  
  
  Section \ref{SMsec:detector} considers the FI/SC/FM system as a magnon detector and estimates the corresponding figures of merit.

   \subsection{General formalism}
   \label{SMSec:GeneralFormalism}
  
    We describe the superconducting film using Keldysh-Usadel equation \cite{heikkila2019thermal} with a gradient term and without external Zeeman field 
  \begin{equation}\label{SMEq:KeldyshUsadelFI}
  - \{\hat\tau_3\partial_t \commacirc \check g \} + 
  \partial_x \check{I}  = 
   [ \Delta \hat \tau_1 + \check \Gamma + \check \Sigma_{\rm so}\commacirc \check g ]. 
  \end{equation} 
Here  $\check g$ is the  quasiclassical Green's function (GF) in the 8$\times$8 space consisting of Keldysh, Nambu and spin indices, $D$ is the diffusion coefficient and $\check I = D(\check g \circ \partial_x \check g)$ is the matrix current in the  $x$-direction.\cite{heikkila2019thermal} We assume translation invariance in the $y{-}z$ plane. For double-time variables $\circ$ is a convolution product
\begin{equation}
   (A\circ B)(t_1,t_2) = \int dt A(t_1,t)B(t,t_2).
\end{equation}
For single-time variables it reduces to a product
\begin{equation}
\begin{split}
    (a\circ B)(t_1,t_2) = a(t_1)B(t_1,t_2),\\
    (B\circ a)(t_1,t_2) = B(t_1,t_2)a(t_2).
\end{split}
\end{equation}
The commutator is $ [X\commacirc Y] = X\circ Y-Y\circ X$, and the time-derivative acts as
\begin{align}
    \{\hat\tau_3 \partial_t\commacirc\check g\}(t_1,t_2)= \hat\tau_3 \partial_{t_1} \check g(t_1,t_2) + \partial_{t_2} \check g(t_1,t_2) \hat\tau_3.
\end{align}

The value of the superconducting order parameter $\Delta$ is determined from the self-consistency equation
\begin{equation}
    \Delta = \frac{\lambda}{16i}\int_{-\Omega_{\rm  D}}^{\Omega_{\rm  D}} d\varepsilon \Tr[\hat\tau_1 \hat g^K(\varepsilon)].
\end{equation}
We assume the weak-coupling limit, so that the coupling constant $\lambda$ and the high-energy cutoff $\Omega_{\rm D}$ can be eliminated in favor of the transition temperature $T_{c0}$ in the absence of pair-breaking effects.\cite{heikkila2019thermal}
We do not include the non-equilibrium correction to $\Delta$, as it only gives a spectral correction to the GFs and does not affect the second-order perturbation theory results for the non-local voltage $V_D$ or the energy accumulation $W$.

The coupling to the normal reservoir self-energy has spectral components  $\hat \Gamma^{R,A} = \pm  \Gamma \hat \tau_3 $ and the Keldysh component $\hat \Gamma^{K} = 2\Gamma \hat \tau_3 n_0 $ with the equilibrium distribution function in the Fourier representation $n_0(\varepsilon) =\tanh (\varepsilon /2T)$. The spectral components of this self-energy yield the frequently used Dynes\cite{Dynes1984} parameter which
     determines the smearing of the BCS density of states singularity. 
     In addition, this self-energy determines the relaxation of spin-independent non-equilibrium distribution functions.
     
Elastic spin relaxation in the ladder approximation is determined by the spin-orbit  scattering  
  self-energy  \cite{abrikosov1962spin} 
 \begin{align} \label{Eq:SpinOrbitalSelfEnergy}
  &  \check\Sigma_{\rm so} = 
  \hat{\bm\sigma}\cdot
  \check g \hat{\bm\sigma}/(6\tau_{\rm so}).
 \end{align}

The differential equation \eqref{Eq:KeldyshUsadelFI} is supplemented by dynamical boundary conditions at $x\,{=}\,0$ describing the 
spin splitting and pumping induced by the electron scattering at the FI interface with time-dependent magnetization.
 These boundary conditions are derived from the   spin-dependent  scattering matrix at the FI/SC interface\cite{Tokuyasu1988}
 \begin{align} \label{SMEq:BCFS}
  \check I(x=0) =
  iJ_{sd}[\hat{\bm \sigma} \bm m \hat\tau_3\commacirc \check g].
 \end{align}
The boundary condition determines the interfacial matrix current.
 
 \subsection{Boundary condition for spin current}
 \label{SMSec:SpinCurrent}

  To demonstrate how the spin pumping arises in this formalism, we now derive an expression for the energy-integrated spin current generated at the interface. The Fourier transform of the Keldysh part of the matrix current is
 \begin{equation}
 \!\!\hat I^K(\varepsilon,\varepsilon{-}\Omega)\!= iJ_{\rm sd} \!\!\int\!\!d\omega\!\left( \hat m_\omega g^K_{\varepsilon{-}\omega,\varepsilon{-}\Omega} {-} g^K_{\varepsilon,\varepsilon{-}\Omega{+}\omega} \hat m_\omega \right),
 \end{equation}
  where $\hat m_\omega \equiv \int dt\tau_3\bm\sigma\bm m(t) e^{-i \omega t}$. Fourier convention for double-time functions is 
  \begin{equation}
  f(\varepsilon_1,\varepsilon_2) = \int dt_1 dt_2 f(t_1,t_2) e^{-i\varepsilon_1 t_1 + i\varepsilon_2 t_2}.\label{eq:FTconvention}
  \end{equation}
  We extract the quasiclassical part of the energy-integrated current by imposing an energy cutoff $\Lambda$ satisfying $\Delta_0,T_{c0},\Omega\ll \Lambda \ll E_F$, so that
  \begin{align}
  \hat I^K(\Omega) &= \int_{-\Lambda}^{+\Lambda} \mkern-10mu d\varepsilon \hat I(\varepsilon+\Omega/2,\varepsilon-\Omega/2)\\
  &= iJ_{sd} \int d\omega \int_{-\Lambda}^{+\Lambda} \mkern-10mu d\varepsilon \big( \hat m_\omega g^K_{\varepsilon+\Omega/2-\omega,\varepsilon-\Omega/2}\nonumber\\ &\mkern140mu- g^K_{\varepsilon+\Omega/2,\varepsilon-\Omega/2+\omega} \hat m_\omega \big).\nonumber
  \end{align}
  In terms of a matrix $\hat n(\omega) = \int_{-\Lambda}^{+\Lambda}d\varepsilon g^K_{\varepsilon-\omega/2,\varepsilon+\omega/2}$, the current is
  \begin{equation}
  \hat I^K(\Omega) = i J_{sd}\left(\int d\omega (\hat m_\omega \hat n_{\Omega-\omega} {-} \hat n_{\Omega-\omega}\hat m_\omega) - 4 \Omega \bm\sigma{\bm m}_\Omega\right).
  \end{equation}
  The last term appears from energy shifts about the cutoff, where $g^K_{\varepsilon, \varepsilon{-}\omega} \approx 2\sgn(\varepsilon) \tau_3 \delta(\omega)$ does not depend on the state of the system. 

  Transforming to the time domain and extracting the spin-dependent part, we find the boundary condition for the spin current
  \begin{equation}
    \begin{split}
  \bm j_s(x=0) &= 2 J_{sd}\left[ \bm \mu_s(t)\times\bm m(t) \MSedit{-} \partial_t \bm m(t)\right].
    \end{split}\label{eq:spin_bc}
  \end{equation}
  Here the spin current and the quasiclassical part of the spin accumulation are defined as
    \begin{align}
      \bm j_s(t) &= 
       \MSedit{-}
      \frac{1}{8} \int_{-\infty}^\infty d\varepsilon \Tr[ \bm\sigma\hat I^K(\varepsilon, t) ],
      \\
      \bm\mu_s(t) &= \MSedit{-} \frac{1}{8} \int_{-\infty}^\infty d\varepsilon \Tr[\hat{\bm\sigma}\tau_3 \hat g^K(\varepsilon,t)],
  \end{align}
  respectively, with an implicit high-energy cutoff. Above, the center-of-mass time-coordinate $t$ is the Fourier transform of the frequency $\Omega=\varepsilon_1-\varepsilon_2$ and $\varepsilon=(\varepsilon_1+\varepsilon_2)/2$ as in Eq.~\eqref{eq:FTconvention}. In Eq.~\eqref{eq:spin_bc}, the latter term is the pumped spin current and the former term is the back-flow current due to spin accumulation. This expression corresponds to the general one\cite{RevModPhys.77.1375} 
      with a 
 purely imaginary  spin-mixing conductance  associated with the interfacial exchange constant \cite{ohnuma2014enhanced}  $g_{\uparrow\downarrow} = - 2i J_{\rm sd}/\nu $.
  
  
   \subsection{Spin diffusion in the normal state}
   \label{SMSec:SpinDiffusionNormal}
   Consider a normal metal (N) in contact with FI with time-dependent magnetization $\bm m(t)$. In the normal state Eq.~(\ref{SMEq:KeldyshUsadelFI}) is greatly simplified because we know the spectral functions 
   $\hat g^{R/A}  = \pm \tau_3$ and they are not perturbed by the time-dependent boundary conditions (\ref{Eq:BCFS}). 
   The Keldysh function is given by  $\hat g^K = 2\tau_3 \hat f $ where $\hat f = f_L + \bm f \hat{\bm \sigma}$. The distribution functions $f_L$ and $\bm f$ parametrize the energy and spin accumulations, respectively.
    The spin accumulation is given by 
   \begin{align} \label{Eq:SpinNormal}
    & \bm \mu_s (t)= - \int_{-\infty}^{\infty} d\varepsilon \bm f (t, \varepsilon)
    \\ \label{SMEq:jsNormal}
    & \bm j_s = D \partial_x \bm \mu_s 
    \end{align}
    
    Spin diffusion equation in the normal metal, obtained from the Keldysh part of Eq.~(\ref{SMEq:KeldyshUsadelFI}), is
 \begin{align}\label{Eq:SpinDiffusion}
  \partial_x \bm j_s  =  \partial_t \bm \mu_s + \bm \mu_s / \tau_{s},
     \end{align}    
    where $\tau_{s}^{-1}$ is the spin relaxation rate in the normal state. In the normal-state we can use the energy-integrated Eq.~\eqref{eq:spin_bc} as the  boundary condition for the spin current generated at the FI/N interface at $x=0$. The length of the normal metal is $d$ and the other interface is to vacuum so that the current vanishes at $x=d$.
    
    From the diffusion equation \eqref{Eq:SpinDiffusion}, we find that the spin accumulation at frequency $\omega$ is determined by the spin current at $x=0$ at the same frequency,
    \begin{align} 
        \bm \mu_s(x,\omega) = 
        - \bm j_s(x=0,\omega) \frac{\cosh [\kappa(x-d)]}{D \kappa \sinh (\kappa d)},\label{eq:SpinDiff_solution}
    \end{align}
and the boundary condition mixes the harmonics. Here the wavevector is $\kappa = \sqrt{1+i\omega \tau_{s}}/\lambda_s$ with the spin diffusion length $\lambda_s= \sqrt{D\tau_s}$.
We assume the FI magnetization $\bm m$ has a circularly polarized alternating component $\bm m_{\perp}(t) = \Re[m_\Omega(\bm x+i\bm y) e^{i\Omega t}]$, with $m_\Omega^*=m_\Omega$.
It drives the alternating spin accumulation (\ref{eq:SpinDiff_solution}) with frequency $\omega= \pm \Omega$ and  for static spin accumulation with $\omega=0$.
The latter one determines the  spin battery effect \cite{brataas2002spin, RevModPhys.77.1375}  
             \begin{equation}
                \langle \bm \mu_s\rangle (x) = - \langle \bm j_s\rangle ({x{=}0}) \frac{\lambda_{s} \cosh [(x-d)/\lambda_{s}]}{D  \sinh ( d/\lambda_{s})},
           \end{equation}
           where $\langle .. \rangle $ denotes the time averaging.

We assume the FI magnetization $\bm m$ has a static component $\bm m_0 = \bm z$ and a circularly polarized alternating component $\bm m_{\perp}(t) = \Re[m_\Omega(\bm x+i\bm y) e^{i\Omega t}]$, with $m_\Omega^*=m_\Omega$.
  Solving Eqs.~\eqref{eq:spin_bc} and \eqref{eq:SpinDiff_solution}, we find the linear response spin accumulation at the interface,
  \begin{align}
    & \bm  \mu_{s}(x=0,\Omega) = \chi_l(\Omega) m_\Omega(\bm x+i\bm y), 
    \\ \label{SMEq:chil}
    & \chi_l(\Omega) = - \frac{\Omega}{1- i (D\kappa/2J_{sd}) \tanh(\kappa d)}.
  \end{align}
     In time-domain, $\bm\mu_s(t)$ is real, so the negative frequency is given by $\bm  \mu_{s}(x=0,-\Omega) = \bm  \mu_{s}(x=0,\Omega)^*$.

At the second order in $\bm m_\perp(t)$, the static spin current at the interface is given by the boundary condition \eqref{eq:spin_bc} as
\begin{equation} \label{SMEq:jsAverage}
    \begin{split}
   & \langle \bm j_s(x=0) \rangle = J_{sd} \bm\mu_s(x=0,\Omega)^*\times \bm m(\Omega) + \text{c.c.}      \\
    & = - 2J_{sd} \Im [\chi_l(\Omega)]m_\Omega^2 \bm z.
    \end{split}
\end{equation}

In the low-frequency limit $\omega\tau_s \ll 1$  we can put $\kappa = \lambda_s^{-1}$ so that combining Eqs.~(\ref{SMEq:chil}--\ref{SMEq:jsAverage}) we get the constant spin current in the conventional form \cite{brataas2002spin, RevModPhys.77.1375}  
 \begin{align}
  &   \nu \langle \bm j_s(x=0) \rangle = {\rm Re} A_{\rm eff}^{\uparrow\downarrow} \langle \bm m \times \partial_t \bm m   \rangle  
  \\ \label{SMEq:Aeff}
  & \frac{1}{A_{\rm eff}^{\uparrow\downarrow}} =
  -\frac{\nu}{2iJ_{sd}} + \frac{\nu D}{\lambda_s} \frac{1}{\tanh(d/\lambda_s)}
 \end{align}
 Here the first term is the inverse of pure imaginary spin-mixing conductance of the FI interface $- 2i J_{sd}/\nu $ while the second term is the usual contribution from the spin relaxation in the spin sink\cite{RevModPhys.77.1375} with $\nu D/\lambda_s$ is the dimensional resistance of the normal metal layer of the thickness $\lambda_s$. 
 
 In the thin-film limit $d\ll \lambda_{\rm sd}$ and beyond the small 
 frequency limit 
 the susceptibility (\ref{SMEq:chil})
 becomes 
 \begin{align}
  \chi_{l}(\Omega) &= 2h_0\Omega/(\Omega-2h_0 - i/\tau_{s}),
  \end{align}
  with the effective field $h_0 = J_{sd}/d$. 
  It corresponds to the Bloch equation 
  \begin{align} \label{SMEq:NormalThinFilm}
      \partial_t \bm \mu_s + 2 \bm \mu_s \times \bm h + \bm \mu_s/\tau_{\rm sn} = 2 \partial_t\bm h
  \end{align}
 with $\bm h = h_0 \bm m$. 
 Here we have electron paramagnetic resonance at $\Omega = 2 h_0$.
 
 
  \subsection { Derivation of kinetic equations}
  \label{SMSec:KineticEq}
 
  The stationary non-equilibrium Keldysh function can be presented in the form 
 \begin{align}  \label{Eq:Keldysh2order}
    & \hat g^K_{hh} =  n_0(\varepsilon) 
     (\hat g^R_{hh} - \hat g^A_{hh}) +
      \hat g^a_{hh},
 \end{align}
 where $\hat g^{R,A}_{hh}$ are the corrections to the spectral function and $\hat g^a_{hh}$ is the anomalous part which contains the information about non-equilibrium quasiparticles. The anomalous and spectral parts can be calculated separately.
 
 In general, the  corrections to GF satisfy the relation coming from the normalization condition  \begin{align}
        g_0^R  g_{hh}^a +
       g_{hh}^a g_0^A = -  g_h^R\circ g_h^a -
       g_h^a\circ g_h^A,
 \end{align} 
 where $g_h^{R/A/a}$
 are the first-order corrections and $g_{hh}^{R/A/a}$ 
 are the second-order corrections.
 To derive the simplified description in terms of the stationary kinetic equations 
 one can use a parametrization in terms of the spin-dependent distribution functions 
  \begin{align} \label{SMEq:gaDF}
 \hat g^a_{hh} = (\hat g^R_0 - \hat g^A_0) ( f_L \tau_0 + f_{T3} \sigma_z ).
 \end{align}
 This parametrization implies that $\hat g_0^R  \hat g_{hh}^a +
       \hat g_{hh}^a \hat g_0^A =0$ and therefore it is not exact.
       It neglects the contribution $\hat g_h^R\circ  \hat f_{h} -
       \hat f_{h}\circ \hat g_h^A $ to the second-order correction to the anomalous function, where $\hat f_{h}$ is the first-order correction to the distribution function. However this contribution 
      does not contain large parts determined by the inelastic relaxation. Therefore 
        by comparing the results given by this parametrization (\ref{SMEq:gaDF}) with the general form of $g_{hh}^a$ 
       we find that they coincide with good accuracy 
       for not very small spin relaxation, that is when $\Gamma \tau_s \ll 1$.
       All numerical results in paper are obtained with general $\hat g^a_{hh}$
       as explained in Sec.~\ref{SMSec:NumericalPerturbation}
   With good accuracy 
 we can thus parametrize the stationary anomalous GF with the help of the distribution functions.

  Distribution functions satisfy stationary  kinetic equations 
 \begin{align}
  &  \partial_x {\cal J}_e  + 
 \Gamma \varepsilon (N f_L + N_z f_z) + I_{ph}  =0
 \\ 
 & \partial_x {\cal J}_s  +  {\cal I}_{so} =0,
 \end{align}
 where the spin-relaxation collision integral is given by
    \begin{align} \label{SM:Iso0}
    {\cal I}_{so} =  {\rm Tr} ( \sigma_z [\Sigma_{so}, \hat g]^K)/4 =  \tau_{so}^{-1} N f_{T3}
       \end{align} 
 with spin relaxation time given by 
$\tau_{so}^{-1}  = (2/3N\tau_{s}) {\rm Tr} [(g_s^{RA})^2 - (g_t^{RA})^2 ] $ with spin-singlet $g_s^{RA}$ and spin-triplet $g_t^{RA}$ parts of the difference $\hat g^{RA} = \hat g^R - \hat g^A$.  
The spectral densities of currents are given by 
\begin{align}
 &{\cal J}_{sz} = D_{T3} \partial_x f_L + 
 D_{L} \partial_x f_{T3}
 \\
 & {\cal J}_e = \varepsilon 
  ( D_L \partial_x f_L + D_{T3} \partial_x f_{T3})
\end{align}
 with diffusion coefficients found in \cite{Silaev2015}. 
 These kinetic equations can be rewritten in terms of the spin-up and spin-down distribution functions $f_{\uparrow/\downarrow} = f_L \pm  f_{T3}$. 
  Then we obtain the 
    spectral densities for spin and energy accumulations
        $f_s = N_\uparrow f_\uparrow - N_\downarrow f_\downarrow$
     and $f_e = \varepsilon ( N_\uparrow f_\uparrow + N_\downarrow f_\downarrow)$. 
     In this representation the spin-orbit scattering collision integral (\ref{SM:Iso1}) 
     is given by 
      \begin{align} \label{SM:Iso1}
      {\cal I}_{so} = 
     \frac{f_s - \kappa_{se} f_e}{{\cal T}_1}    
      \end{align}
      where 
     { ${\cal T}_1 = 
     \tau_{so} N_\uparrow N_\downarrow /N^2 $ 
     }          is the longitudinal spin relaxation time and spin-energy coupling is quantified by the coefficient
      \begin{align} \label{SMEq:kappa_sh}
       \kappa_{se} (\varepsilon)= 
       \frac{1}{\varepsilon } 
       \frac{N_\uparrow - N_\downarrow}{N_\uparrow + N_\downarrow}.    \end{align}  
 
Note that $\kappa_{se}(\varepsilon) \neq 0$ requires the description of the static magnetic proximity effect, i.e., the generation of the spin splitting $\bm h_0$ in the superconductor.
      The spin-diffusion equation modified by the spin-energy coupling is given by 
      \begin{subequations}
    \begin{align} \label{SMEq:Js}
  &  \partial_x {\cal J}_{sz}   =  \frac{f_s - \kappa_{se}  f_e}{{\cal T}_1}  
 \\ \label{SMEq:Je}
  & \partial_x {\cal J}_e = I_{e-ph} + \Gamma f_e,
 \end{align}
 \label{SMeq:spinenergydiffusion}
 \end{subequations}
  where ${\cal J}_{sz} (\varepsilon)$ and ${\cal J}_e (\varepsilon)$ are the spectral densities of the time-independent spin and energy currents. 
  The sources in Eqs.~(\ref{SMEq:Js}--\ref{SMEq:Je})  are determined by the boundary conditions for these currents at the FI/SC interface, generated by the magnetization dynamics. 
  We obtain it from the general boundary conditions (\ref{SMbc:Js}--\ref{SMbc:Je}) by 
  leaving only the anomalous part of the sources 
   \begin{align} \label{SMEq:Jas}
 {\cal J}^{(a)}_{sz} (\varepsilon) = \frac{i J_{sd}}{8}  
 {\rm Tr} ( {\sigma_z}  [ {\bm\sigma\bm m} \hat\tau_3\commacirc \hat g_h]^a ) (\varepsilon)
 \\ \label{SMEq:Jae}
  {\cal J}^{(a)}_e (\varepsilon) = \frac{i J_{sd} }{4} 
 \varepsilon {\rm Tr} (   [ {\bm\sigma\bm m} \hat\tau_3\commacirc \hat g_h]^a ) (\varepsilon).
  \end{align}

 The anomalous part of the boundary conditions can be calculated by subtracting the spectral part, i.e., the part independent of the nonequilibrium state of the system, from the full Keldysh component of the currents.
 We assume the magnetization dynamics given by $\bm m(t) = m_0\bm z + \bm m_\perp (t)$ with 
     rotating time-dependent component given by $ \bm m_\perp (t) = m_\Omega(\cos(\Omega t), \sin(\Omega t), 0)$. It can be represented as the sum of 
     left-hand and right-hand components
     with the same amplitudes 
     $\bm m_\perp (t) = m_{\Omega, l} e^{i\Omega t}(\bm x-i\bm y) + m_{-\Omega, r} e^{-i\Omega t} (\bm x+i\bm y)$ where 
     $m_{\Omega, l}= m_{-\Omega, r} = m_\Omega/2$. 
     This signal can be induced in the standard ferromagnetic resonance setup.  The frequency $\Omega$ can be tuned by the external magnetic field. 
 
The spectral part of the boundary conditions reads
  \begin{align}
  & \frac{i}{4}  {\rm Tr} ( \hat\tau_3 \sigma_z  [ {\bm\sigma\bm m}\commacirc \hat g_h]^{sp} ) (\varepsilon)= 
     m_{l,\Omega} m_{r,-\Omega}  n_0(\varepsilon) \times
    \\   \nonumber
  &  [  \chi^{RA}_r
  (\varepsilon-\Omega,\varepsilon) 
   -  \chi^{RA}_l (\varepsilon,\varepsilon-\Omega) 
    -
      \\ \nonumber
  & \chi^{RA}_l
  (\varepsilon+\Omega,\varepsilon) 
   +  \chi^{RA}_r (\varepsilon,\varepsilon+\Omega) ]
     \end{align}

\begin{figure}[htb!]
 \centerline{
 $  \begin{array}{c}
 \includegraphics[width=0.5\linewidth]
 {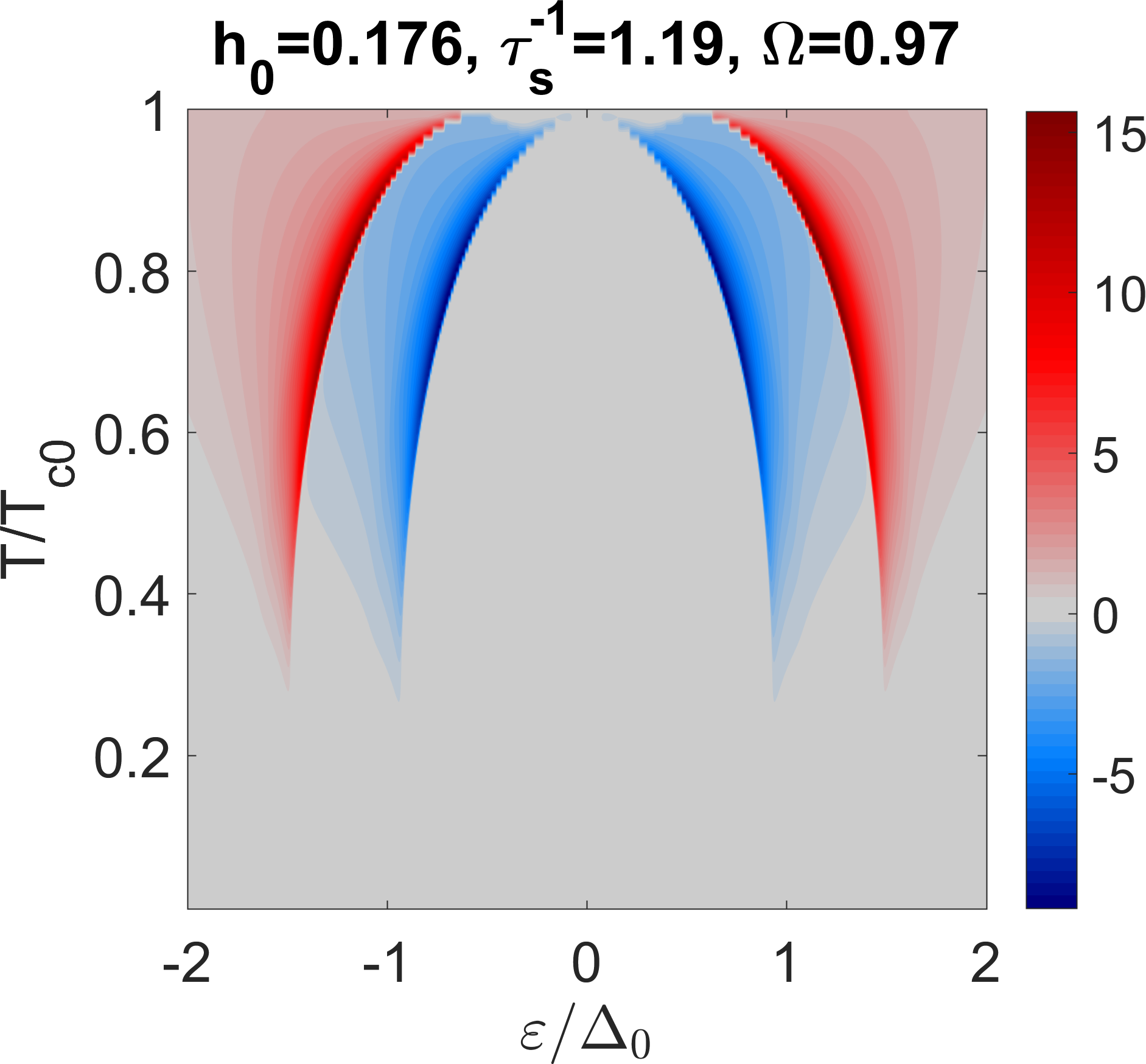} 
 \includegraphics[width=0.5\linewidth]
 {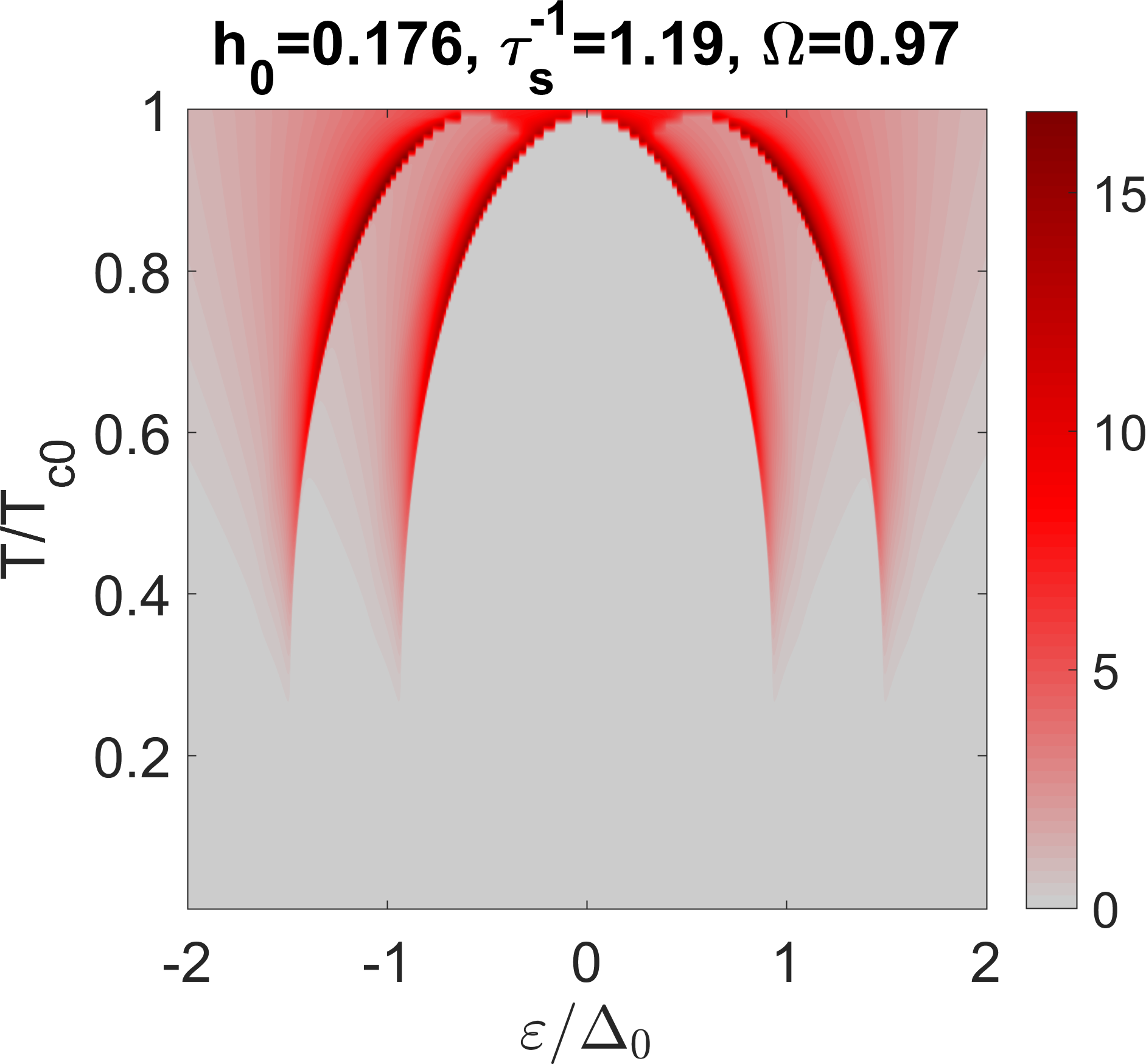} 
 \\
  \includegraphics[width=0.5\linewidth]
 {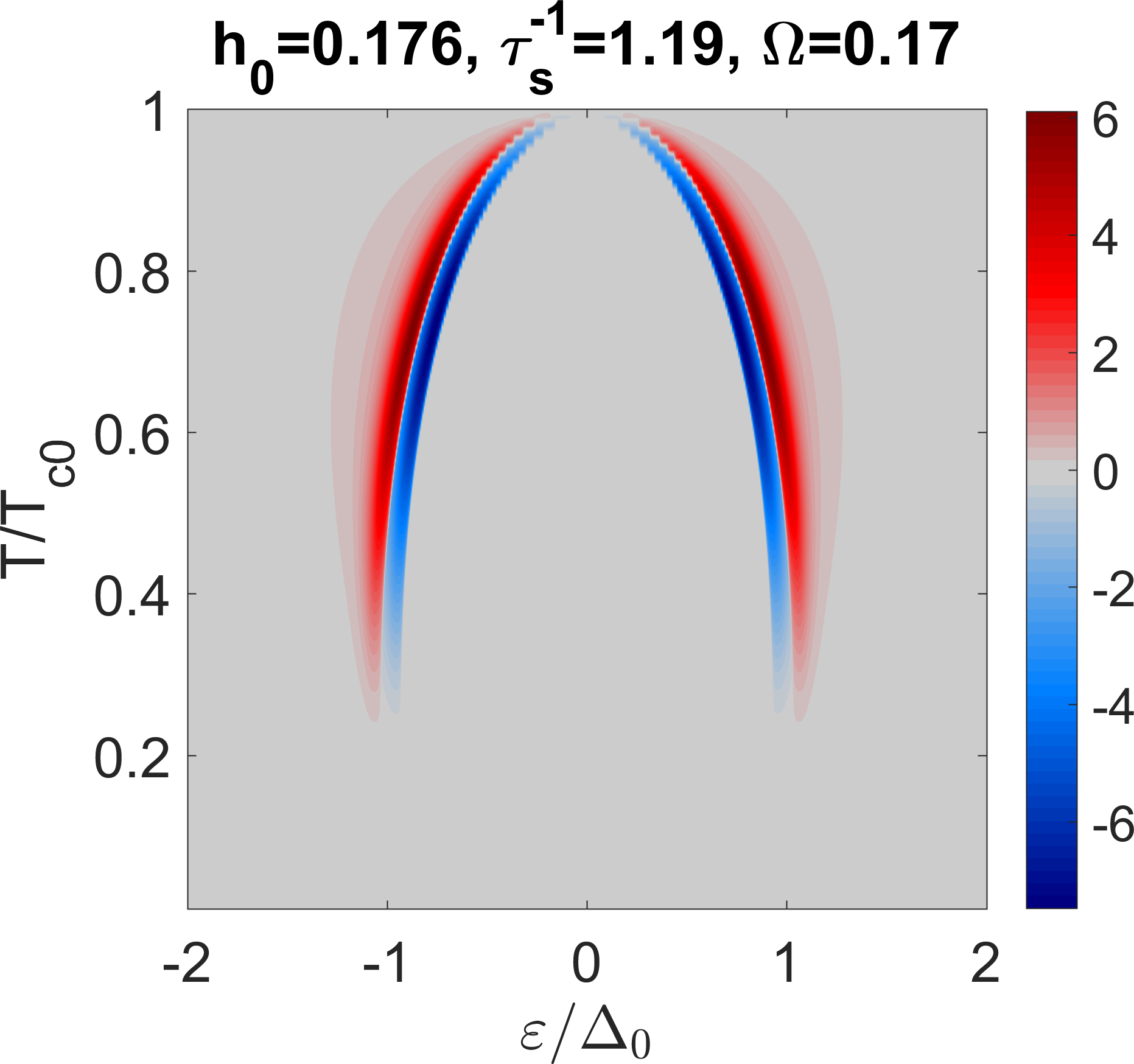} 
 \includegraphics[width=0.5\linewidth]
 {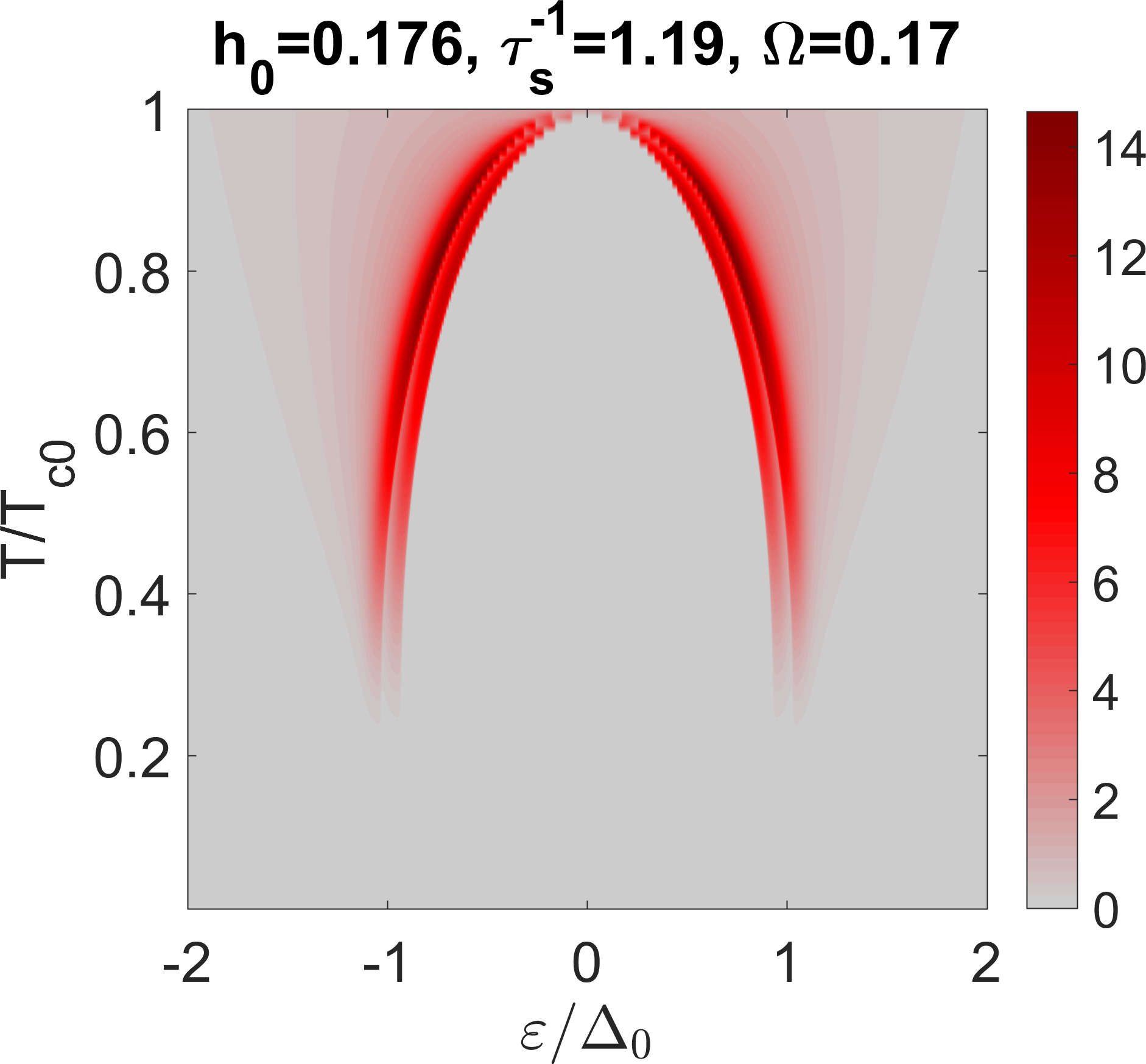} 
   \end{array}$}
  \caption{\label{Fig:jsQ}
  Spectral densities of interfacial currents generated by the magnetization dynamics.
  (Left column): Energy current ${\cal J}_e(\varepsilon,T)$. (Right column): Spin current ${\cal J}_{sz}(\varepsilon,T)$. 
  The energy-like quantities $h_0,\tau_{s}^{-1},\Omega$ are given in terms of $T_{c0}$.
    }
 \end{figure}

  With that we obtain the anomalous part
   \begin{align}
  & \frac{i}{4}  {\rm Tr} ( \hat\tau_3 \sigma_z  [ {\bm\sigma\bm m} \commacirc \hat g_h]^a ) (\varepsilon) = 
    \\ \nonumber
  &  m_{l,\Omega} [\chi^K_r
  (\varepsilon-\Omega,\varepsilon) - 
    \chi^{RA}_r(\varepsilon-\Omega,\varepsilon) n_0(\varepsilon)]
  \\   \nonumber
  & - m_{r,-\Omega} [\chi^K_l (\varepsilon,\varepsilon-\Omega) 
   - 
    \chi^{RA}_l(\varepsilon,\varepsilon-\Omega) n_0(\varepsilon)] 
    +
      \\ \nonumber
  &  m_{r,-\Omega} [\chi^K_l
  (\varepsilon+\Omega,\varepsilon) -
    \chi^{RA}_l(\varepsilon+\Omega,\varepsilon) n_0(\varepsilon)]
  \\   \nonumber
  & - m_{l,\Omega} [\chi^K_r (\varepsilon,\varepsilon+\Omega) 
   - 
    \chi^{RA}_r(\varepsilon,\varepsilon+\Omega) n_0(\varepsilon)] 
   \end{align}
   Substituting these expressions to (\ref{SMEq:Jas},\ref{SMEq:Jae}) we get the spectral densities of the anomalous parts of spin and energy currents.  
   The examples of ${\cal J}^{(a)}_{sz} (\varepsilon, T)$, 
   ${\cal J}^{(a)}_{e} (\varepsilon, T)$ functions at a given frequency of the driving magnetization are shown in Fig.~\ref{Fig:jsQ}.

   \subsection{Numerical perturbation calculations in the thin film limit }
   \label{SMSec:NumericalPerturbation}
  In this section we develop the perturbation theory which allows for calculating the corrections to spectral and anomalous parts of the GF in FI/SC system to the second order of the  time-dependent magnetization. 
    This calculation yields the boundary conditions for spin and energy currents which are the sources for kinetic equations discussed in Sec.~\ref{SMSec:KineticEq}. 
    Besides that here we consider a general form  of the anomalous function and hence can go beyond the approximation used for deriving the kinetic equations. This allows to study the limit of vanishing spin relaxation. 

  Integrating Eq.~\eqref{SMEq:KeldyshUsadelFI} by thickness using boundary conditions (\ref{SMEq:BCFS}) we get the 
   time-dependent Eilenberger equation 
  \begin{equation}\label{SMEq:KeldyshUsadelT}
  - \{\hat\tau_3\partial_t\commacirc \check g \}   = 
   [\hat{\bm \sigma} \bm h \hat\tau_3 + \Delta \hat \tau_1 + \check \Gamma + \check \Sigma_{\rm so} \commacirc \check g ] 
  \end{equation} 
where $\hat\sigma_k,\hat\tau_k$, $k=0,1,2,3$ are Pauli matrices. 
The effective Zeeman field is $\bm h = (J_{sd}/d )\bm m$.
  In the presence of both the non-zero spin-splitting field $\bm h$ and spin relaxation we can solve Eq.~\eqref{SMEq:KeldyshUsadelT}
 only numerically. 
 Let us write the iteration scheme for 
 Eq.~\eqref{SMEq:KeldyshUsadelT} considering the time-dependent Zeeman field $\bm h(t)$
 as a perturbation. 
  \begin{equation}
   -\{\hat\tau_3\partial_t\commacirc \check g \}  
  =
   i [\title{\bm h}\bm \sigma \tau_3\commacirc \check g]
   + 
   [ h_{0} \sigma_z \tau_3 + \Delta \hat \tau_1 + 
   \check \Gamma + \check \Sigma_{s}\commacirc 
   \check g ]  . 
  \end{equation}  
  Zeroth order solution is found in the form $\hat g_0(t_1,t_2) = \hat g_0 (\varepsilon)e^{i\varepsilon (t_1-t_2)}$, with the Keldysh component $\hat g_0^K (\varepsilon) = g^{RA} n_0(\varepsilon) $. The spectral components satisfy the stationary equilibrium Eilenberger equation 
   \begin{align}
   [ (h_{0} \sigma_z + \varepsilon \pm  i \Gamma) \tau_3 + 
   \Delta \hat \tau_1 + \check \Sigma_{\rm so} , 
    g^{R,A}_0 ] =0
   \end{align}
   
  The first-order perturbation solutions $g_h(12) e^{i\varepsilon_1 t_1 - i\varepsilon_2 t_2} $ 
  and 
  $g_h(21) e^{i\varepsilon_2 t_1 - i\varepsilon_1 t_2} $ 
  where $\varepsilon_1 = \varepsilon_2 + \Omega $
   are determined by  
   \begin{align} \nonumber
   & [ (i \varepsilon_1 + \Gamma_{\varepsilon_1}  )
   \tau_3  + \Lambda + \check \Sigma_{0}(1) ]    
   g_h (12)
   - 
   \\ \label{Eq:Perturbaition1order}
   &  g_h (12)
    [ \tau_3 
   (i\varepsilon_2 + \Gamma_{\varepsilon_2} ) + \Lambda + 
   \check \Sigma_{0}(2) ]
    + 
  \\ \nonumber
  &   \Sigma_{h}(12) g_0(2) 
   - g_0(1) \Sigma_{h}(12) 
        =
   \\ \nonumber
  &  i [
    \check g_0(1) 
  {\bm h}_\Omega \bm \sigma \tau_3
  - 
  {\bm h}_\Omega \bm \sigma \tau_3 
  \check g_0(2)
  ]
   \end{align}
   
  \begin{align} \nonumber
   & [ (i \varepsilon_2 + \Gamma_{\varepsilon_2}  )
   \tau_3  + \Lambda + \check \Sigma_{0}(2) ]    
   g_{h} (21) 
   - 
   \\
   &  g_h(21)
    [ \tau_3 
   (i\varepsilon_1 + \Gamma_{\varepsilon_1} ) + 
   \Lambda + 
   \check \Sigma_{0}(1) ]
    + 
  \\ \nonumber
  &   \Sigma_{h}(21) g_0(1) 
   - g_0(2)  \Sigma_{h} (21) 
        =
   \\ \nonumber
  &  i [
    \check g_0(2) 
  {\bm h}_{-\Omega} \bm \sigma \tau_3
  - 
  {\bm h}_{-\Omega} \bm \sigma \tau_3 
  \check g_0(1)
  ].
   \end{align}
  Here we also denote $\Lambda = h_{0} \sigma_z \tau_3 + 
   \Delta \hat \tau_1$. 

   The second order perturbation yields the stationary correction from
   $ g_{hh} (\varepsilon)e^{i\varepsilon (t_1-t_2)}$
   \begin{align} \nonumber
   &     [ (i \varepsilon + \check\Gamma )
   \tau_3 +   \Lambda + \check \Sigma_{0} , 
     g_{hh} ] 
  + 
   [  \Sigma_{hh},  g_0 ]
       =
   \\ \nonumber
  & i [
  \check g_h(12) {\bm h}_{-\Omega} 
  \bm \sigma \tau_3
  -
  {\bm h}_{-\Omega} \bm \sigma \tau_3 
  \check g_{h} (31)
  ] +
  \\ \nonumber
  & i [
  \check g_h (13) {\bm h}_{\Omega} 
  \bm \sigma \tau_3
  -
  {\bm h}_{\Omega} \bm \sigma \tau_3 
  \check g_h(21)
    ]
  -
        \\ \nonumber
  & [ \check \Sigma_{h} (12)
   g_{h}(21) 
   - 
   g_{h} (13)  \check \Sigma_{h}(31) +
      \\ \label{Eq:Perturbaition2order}
  &  \check \Sigma_{h}(13)  g_{h} (31) 
   - 
    g_{h}(12) \check \Sigma_{h} (21) ].
       \end{align}   
 
 Here the spin-orbit terms are
  \begin{align}
  & \check \Sigma_{h}(12) = \bm \sigma \hat g_{h}(12)  \bm\sigma/6\tau_{\rm so}  
 \\
 & \Sigma_0 = \bm \sigma \tilde g_0 \bm\sigma/6\tau_{\rm so}.  
 \end{align}
 
  The linear spin response is diagonal in the circular basis. Provided that there are components $h_{l,\Omega}$, $h_{r, -\Omega}$ we can write 
  $\bm \mu_s(\Omega) = \chi_l  h_{l, \Omega} (\bm x + i\bm y)/h_0  $ and $\bm \mu_s(- \Omega) = \chi_r h_{r, -\Omega} (\bm x - i\bm y) /h_0 $.
For left-hand field, $\vec{h}_\Omega=(1,i,0)h_{l,\Omega}/2$,
the first-order (imaginary-time) Green function solution reads
 \begin{align}
     g_h(12)
     =
     ih_{l,\Omega}\sigma_l
     \frac{
     \tau_3 - g_{+}(1) \tau_3 g_{-}(2)
     }{s_+(1) + s_-(2) + \frac{2}{3\tau_{\rm so}}}
     \,.
\end{align}
Here $g_0(\varepsilon)=\frac{1+\sigma_z}{2}g_+(\varepsilon)+\frac{1-\sigma_z}{2}g_{-}(\varepsilon)$, and $g_{\pm}=(\omega_\pm\tau_3+\Delta_\pm\tau_1)/s_\pm$,
$\omega_\pm=-i\varepsilon\pm{}ih_0+\frac{\omega_\mp}{3\tau_{\rm so} s_\mp}$,
$\Delta_\pm=\Delta+\frac{\Delta_\mp}{3\tau_{\rm so} s_\mp}$,
$s_\pm=\sqrt{\omega_\pm^2+\Delta_\pm^2}$.
The algebraic equations for $\omega_\pm$, $\Delta_\pm$ need
to be solved numerically.
The $R$, $A$ components are obtained via $g_h^{R,A}(12)=g_h(1^{R,A},2^{R,A})$ where
$\varepsilon^{R,A}=\varepsilon\pm i\Gamma$.
The spin susceptibility is conveniently obtained via the analytic
continuation,
\begin{align}
    \chi^K_l(1,2)
    &=
    \chi_l(1^R,2^R)\tanh\frac{\varepsilon_2}{2T}
    -
    \chi_l(1^A,2^A)\tanh\frac{\varepsilon_1}{2T}
    \notag
    \\&
    +
    \chi_l(1^R,2^A)
    [\tanh\frac{\varepsilon_1}{2T}-\tanh\frac{\varepsilon_2}{2T}]
    \,,
\end{align}
where $\chi_l(12)=\frac{h_0}{8h_{l,\Omega}}\tr\tau_3\sigma_rg_h(12)$. The calculation for $\chi$ has been previously discussed in Refs.~\onlinecite{maki1973,silaev2020ff}.

 
  Now let us consider the equation for the rectified spin polarization  $\mu_{sz} \propto \bm z\cdot( \bm m_\Omega \times \bm m_{-\Omega} )$  which is given by the second-order non-linear spin response of the superconductor. We search for the correction to Keldysh function in the form 
 \begin{align}  \label{Eq:Keldysh2order2}
    & \hat g^K_{hh} =  n_0(\varepsilon) 
     (\hat g^R_{hh} - \hat g^A_{hh}) +
      \hat g^a_{hh}
 \end{align}
 where $\hat g^{R,A}_{hh}$ are the corrections to the spectral function and $\hat g^a_{hh}$ is the anomalous part which contains the information about non-equilibrium quasiparticles. The anomalous and spectral parts can be calculated separately  from  Eq.~\eqref{Eq:Perturbaition2order}.
  We are interested in the anomalous part since it determined the non-equilibrium spin accumulation and thereby the non-local voltage in Eq.~\eqref{Eq:VD}
 \begin{align} \label{SMEq:SpinPolarizationGen}
  & \bm \mu_s = 
  \MSedit{-} 
  \int_{-\infty}^{\infty} d\varepsilon {\rm Tr}\; [\hat\tau_3 \hat{\bm \sigma} g^a_{hh} (\varepsilon)] /8
   \\
  & W= \MSedit{-}  
  \int_{-\infty}^{\infty} d\varepsilon {\rm Tr}[\hat \tau_3 \hat g^a_{hh}(\varepsilon)]/4
   \end{align}
  
\begin{figure*}[htb!]
 \centerline{
 $  \begin{array}{c}
 \includegraphics[width=0.20\linewidth]
 {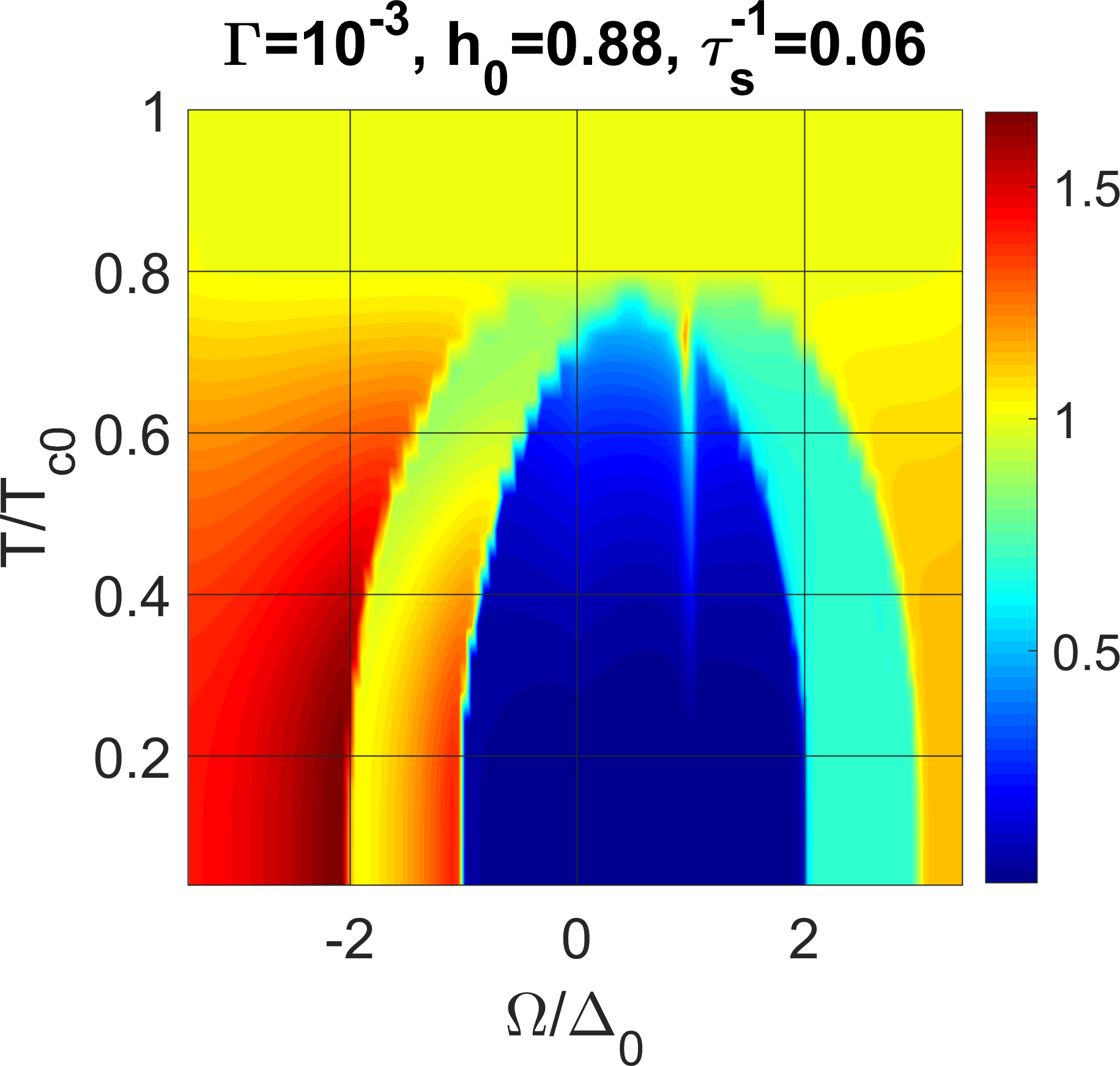} 
 \includegraphics[width=0.20\linewidth]
 {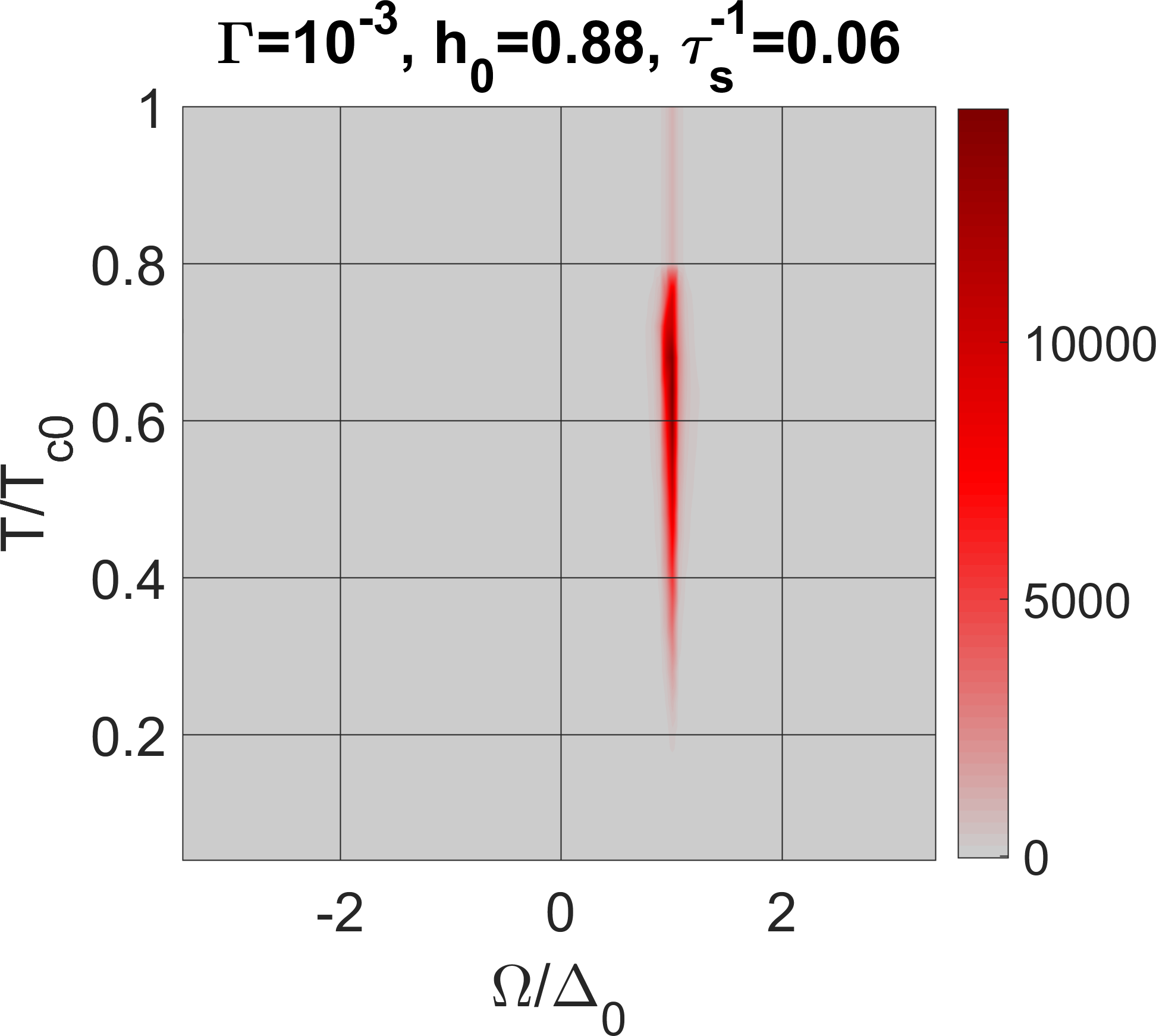} 
 \includegraphics[width=0.20\linewidth]
 {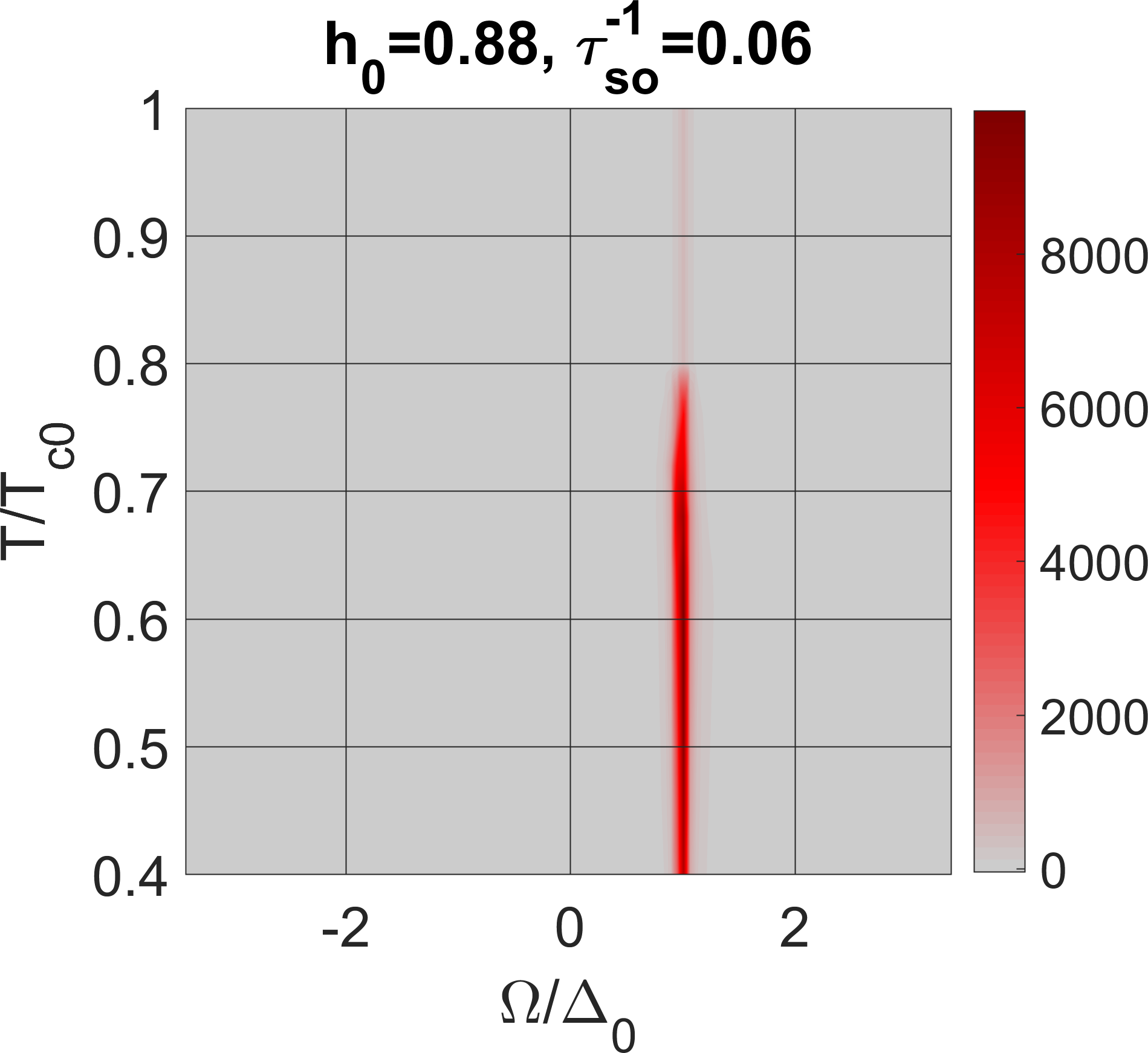}
 \includegraphics[width=0.18\linewidth]
 {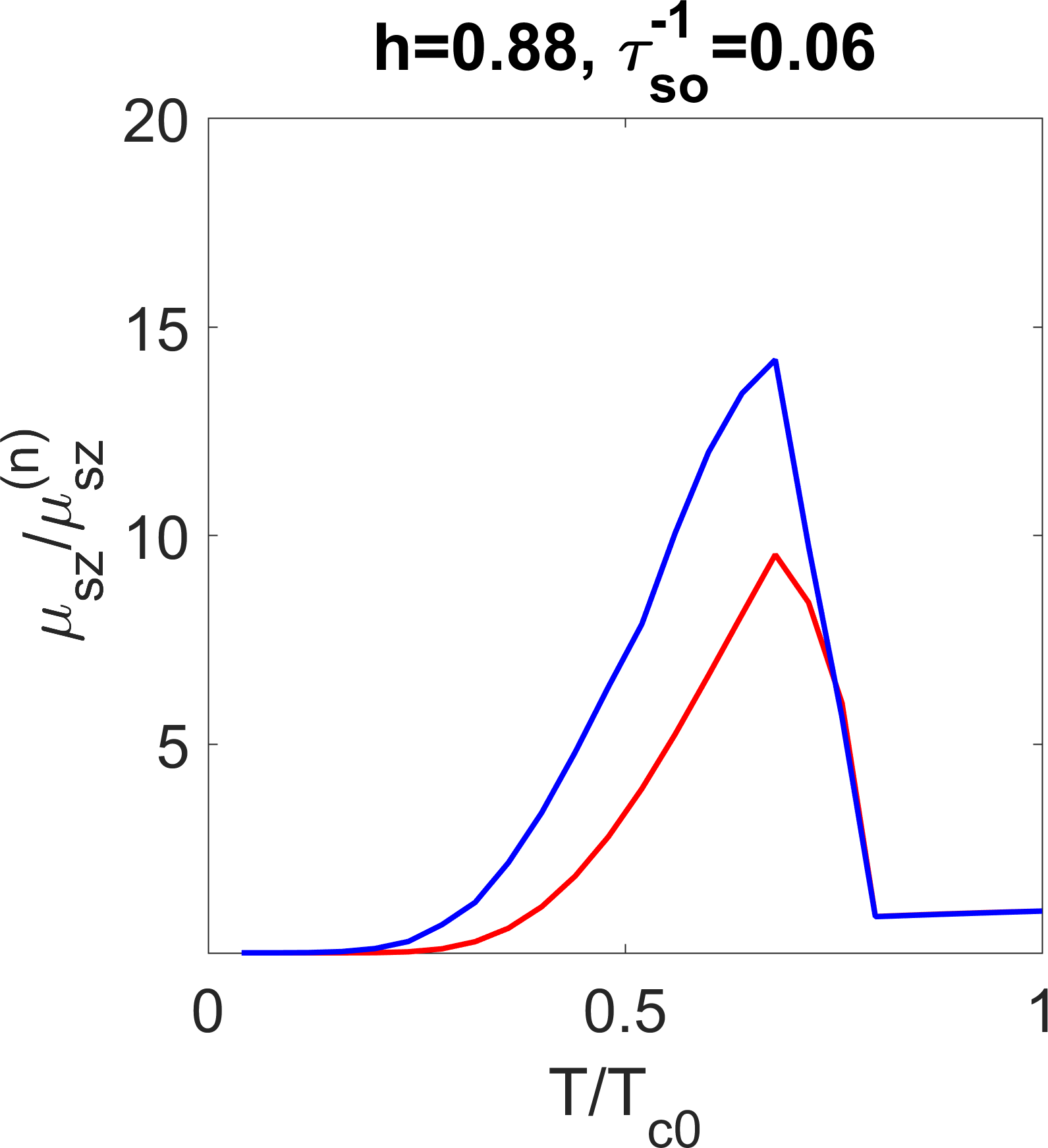} 
  \includegraphics[width=0.175\linewidth]
 {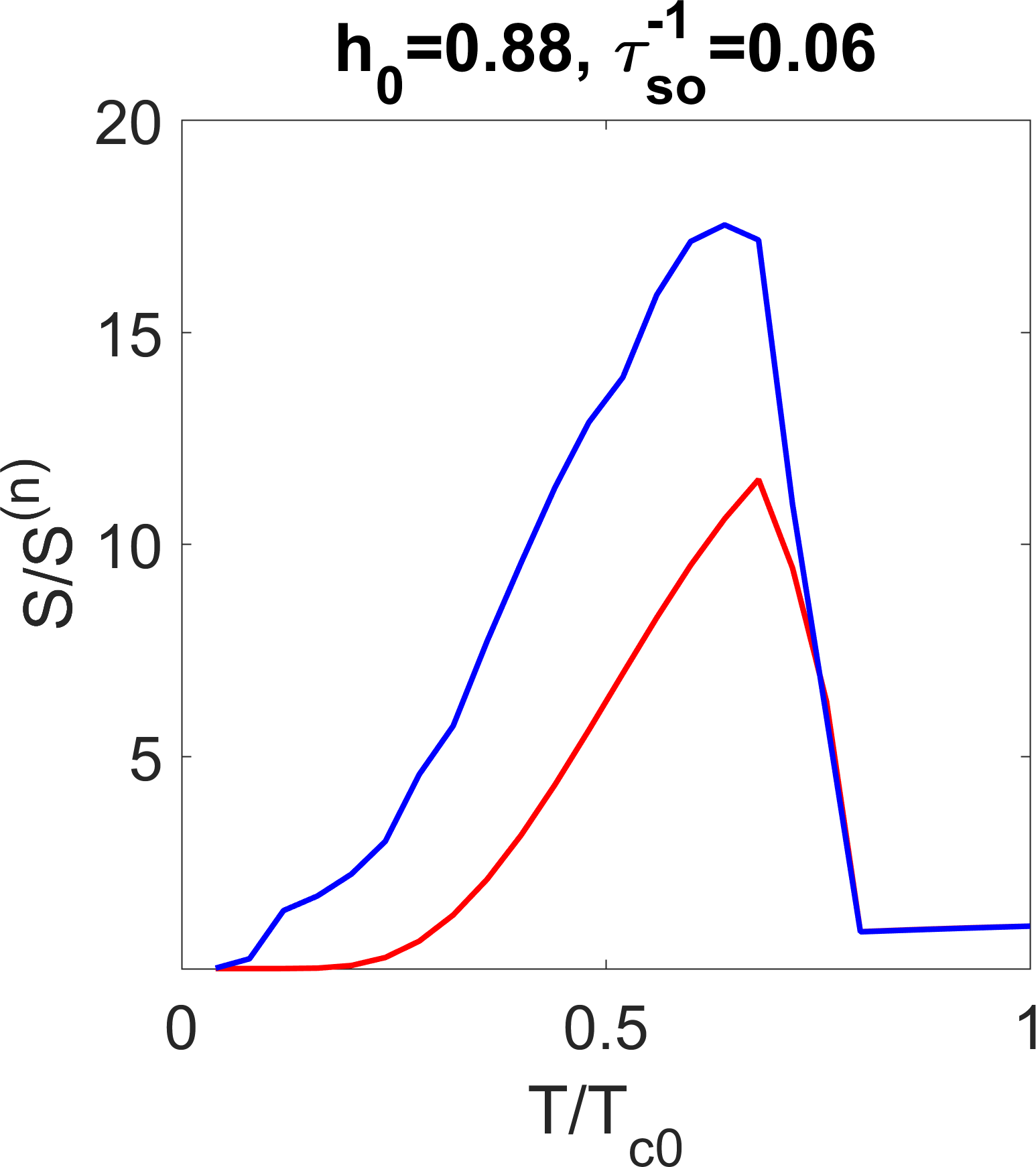} 
 \\
 \includegraphics[width=0.20\linewidth]
  {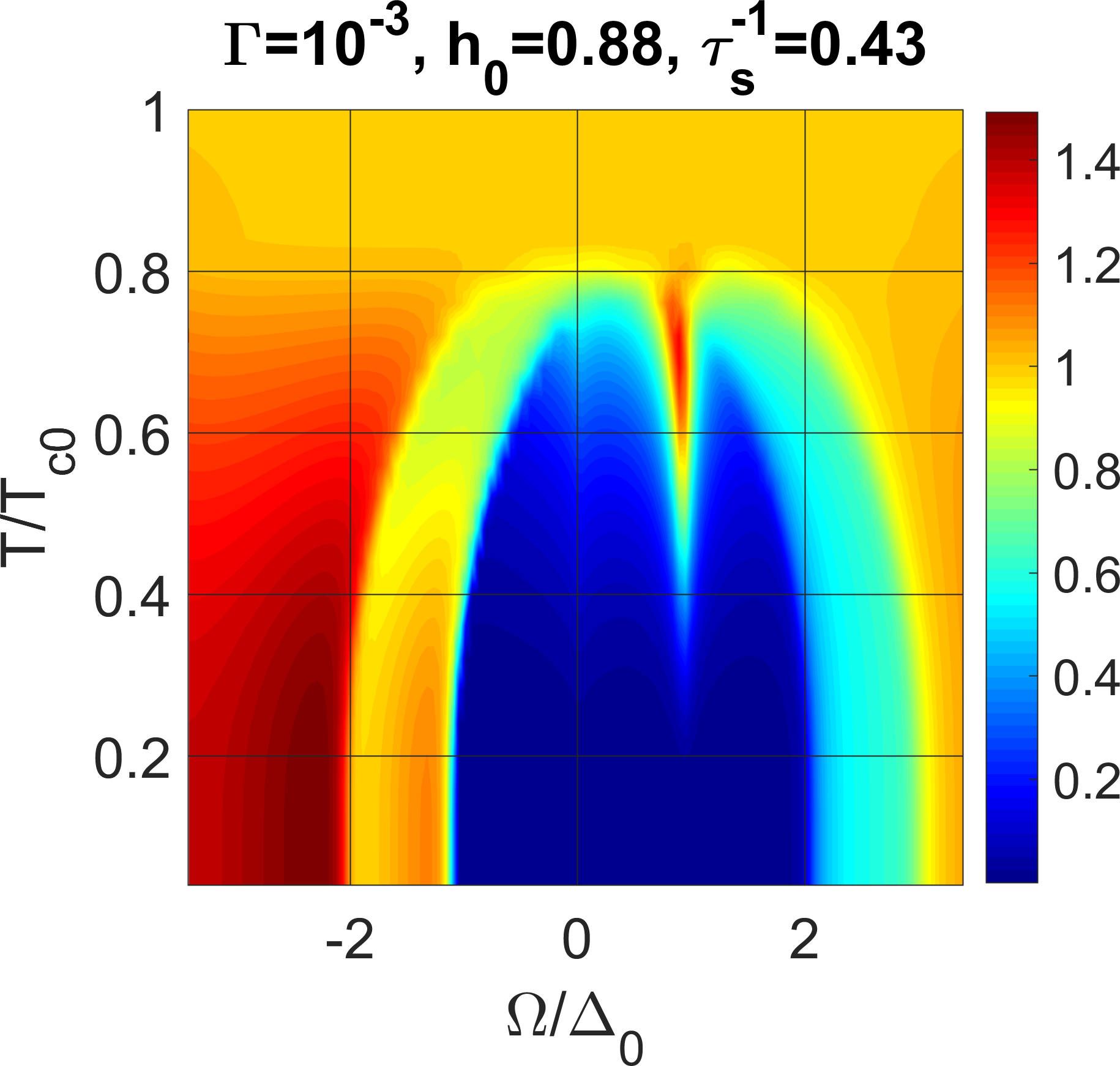}   %
  \includegraphics[width=0.20\linewidth]
 {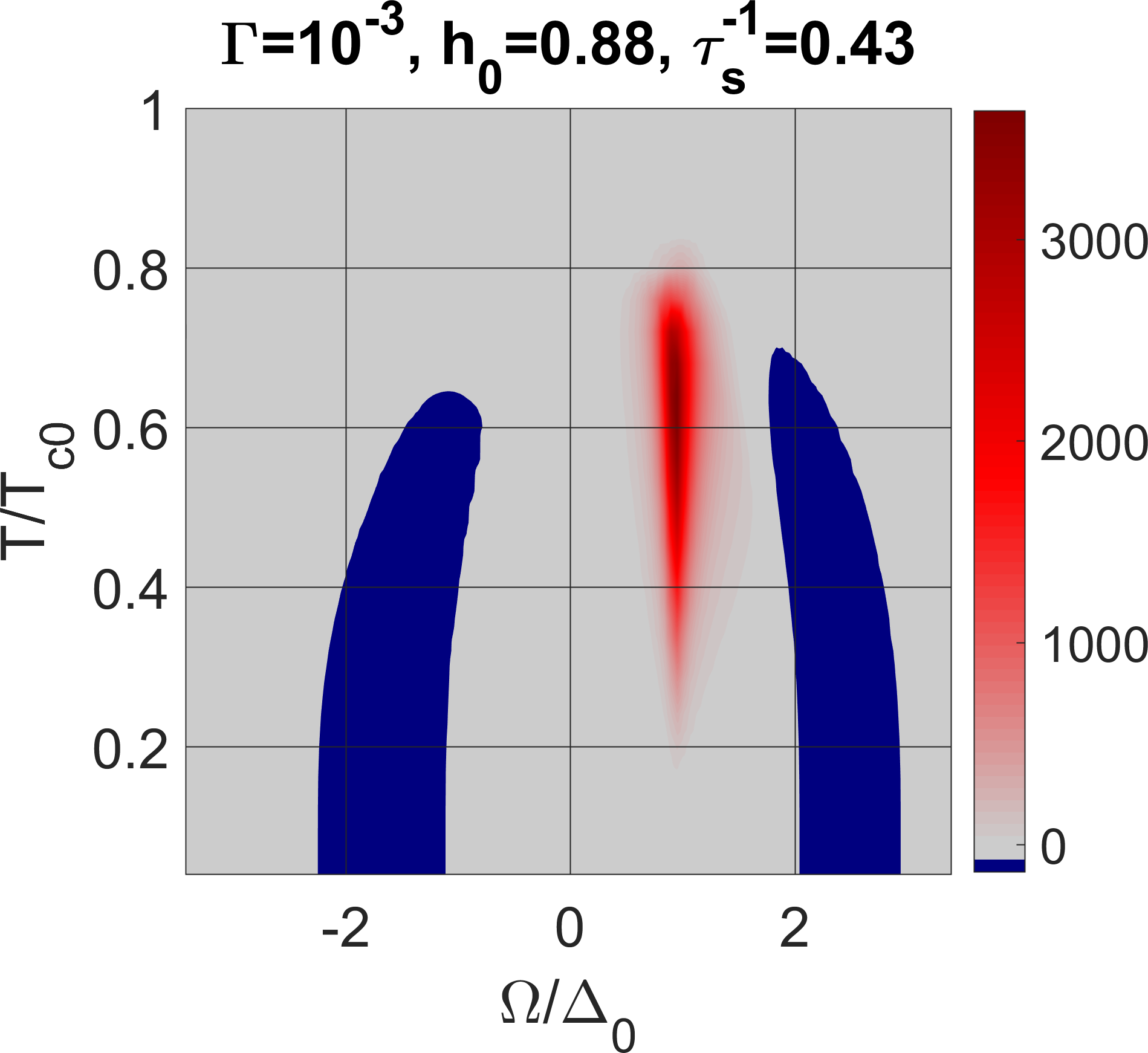} 
 \includegraphics[width=0.20\linewidth]
 {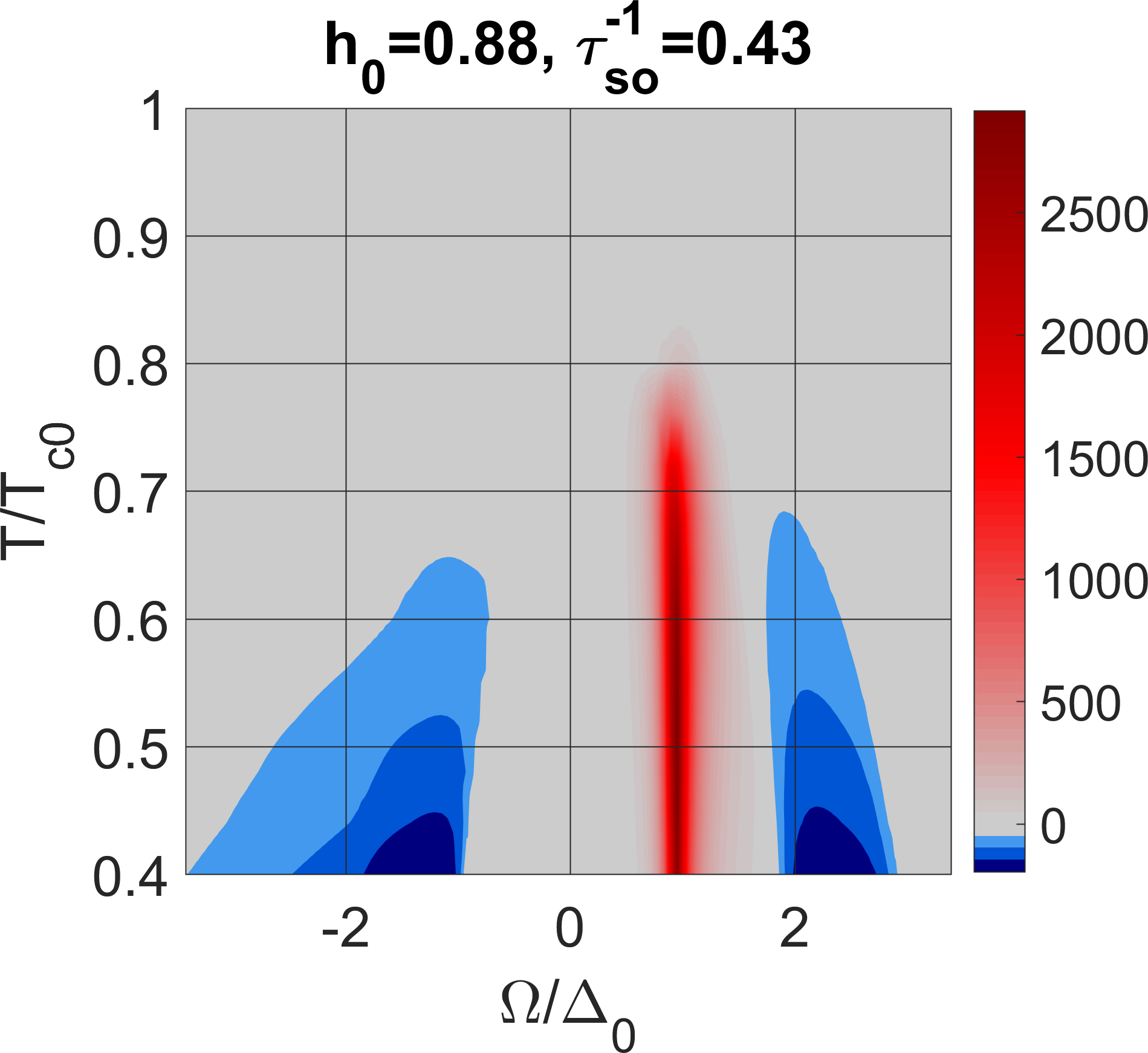}
 \includegraphics[width=0.18\linewidth]
 {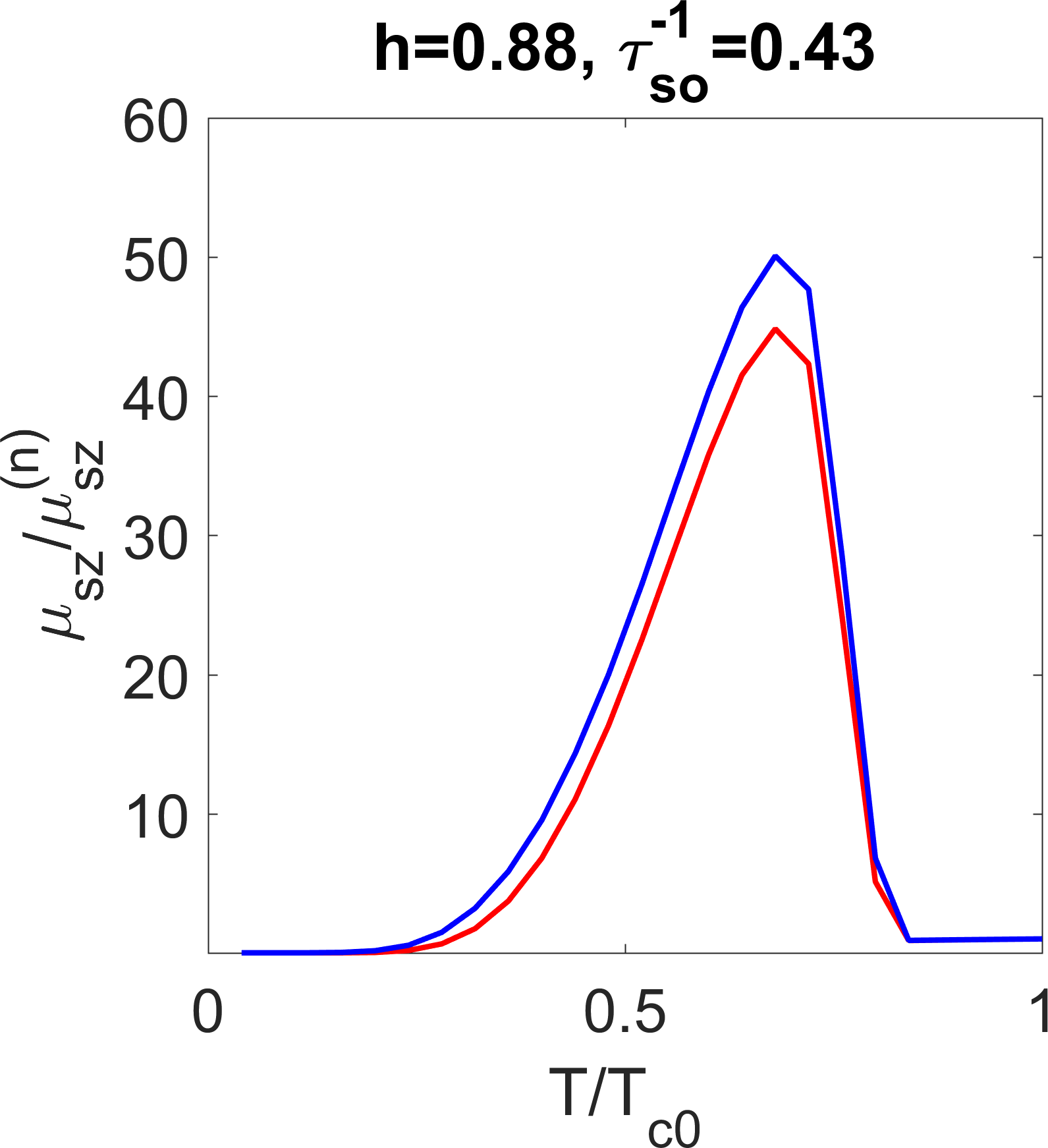} 
  \includegraphics[width=0.175\linewidth]
 {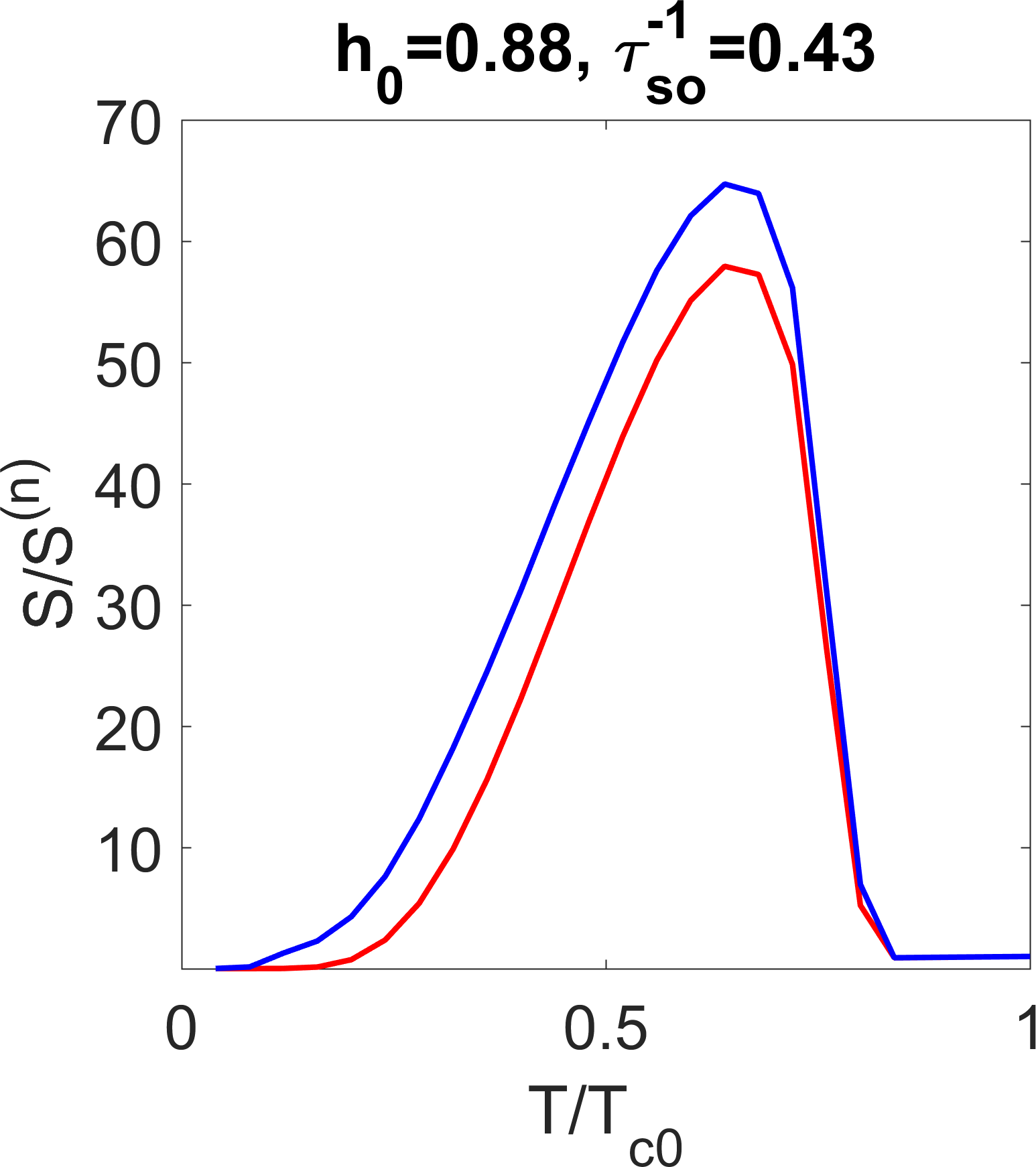} 
  \\
 \includegraphics[width=0.20\linewidth]
  {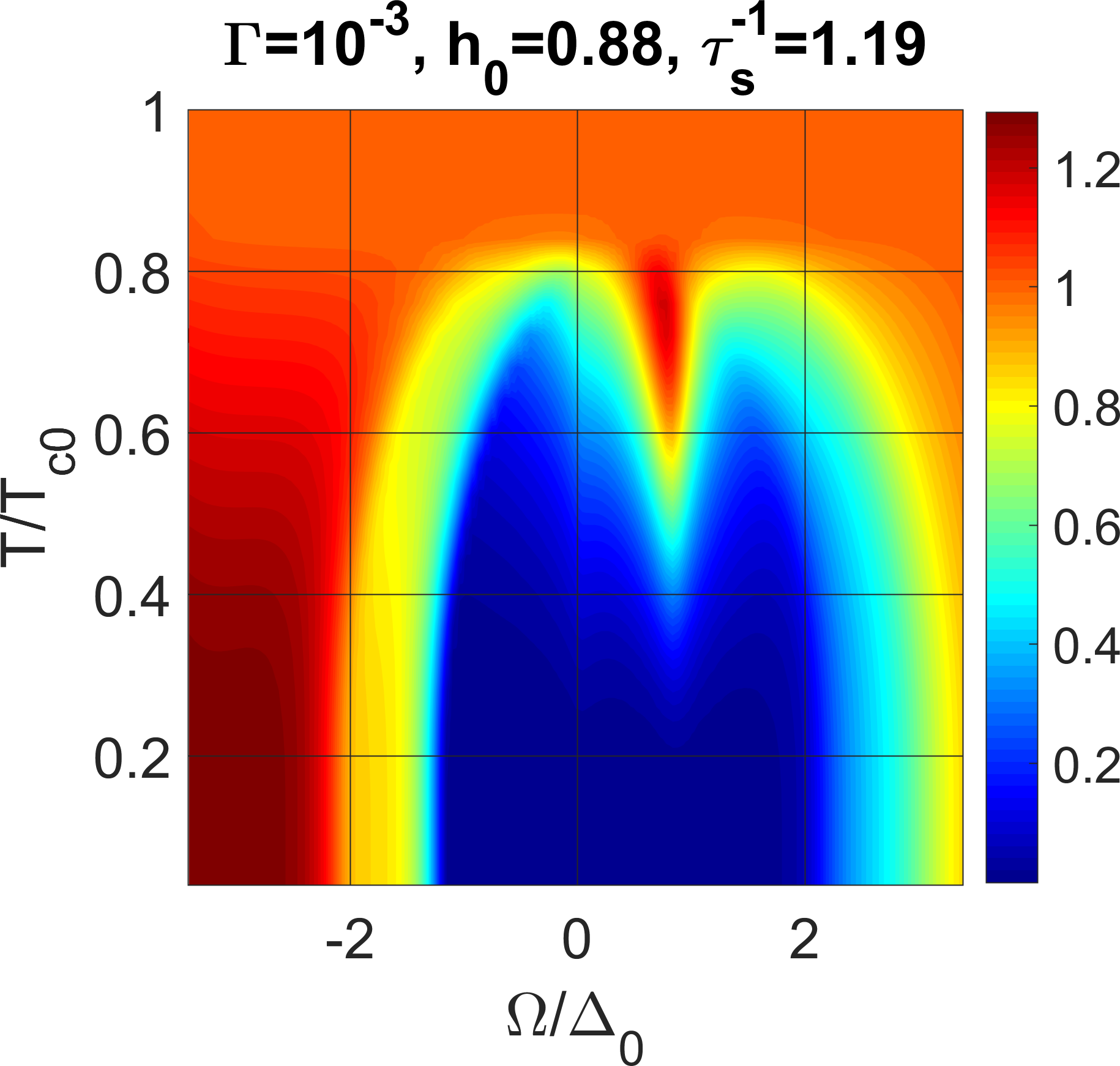}   %
   \includegraphics[width=0.20\linewidth]
 {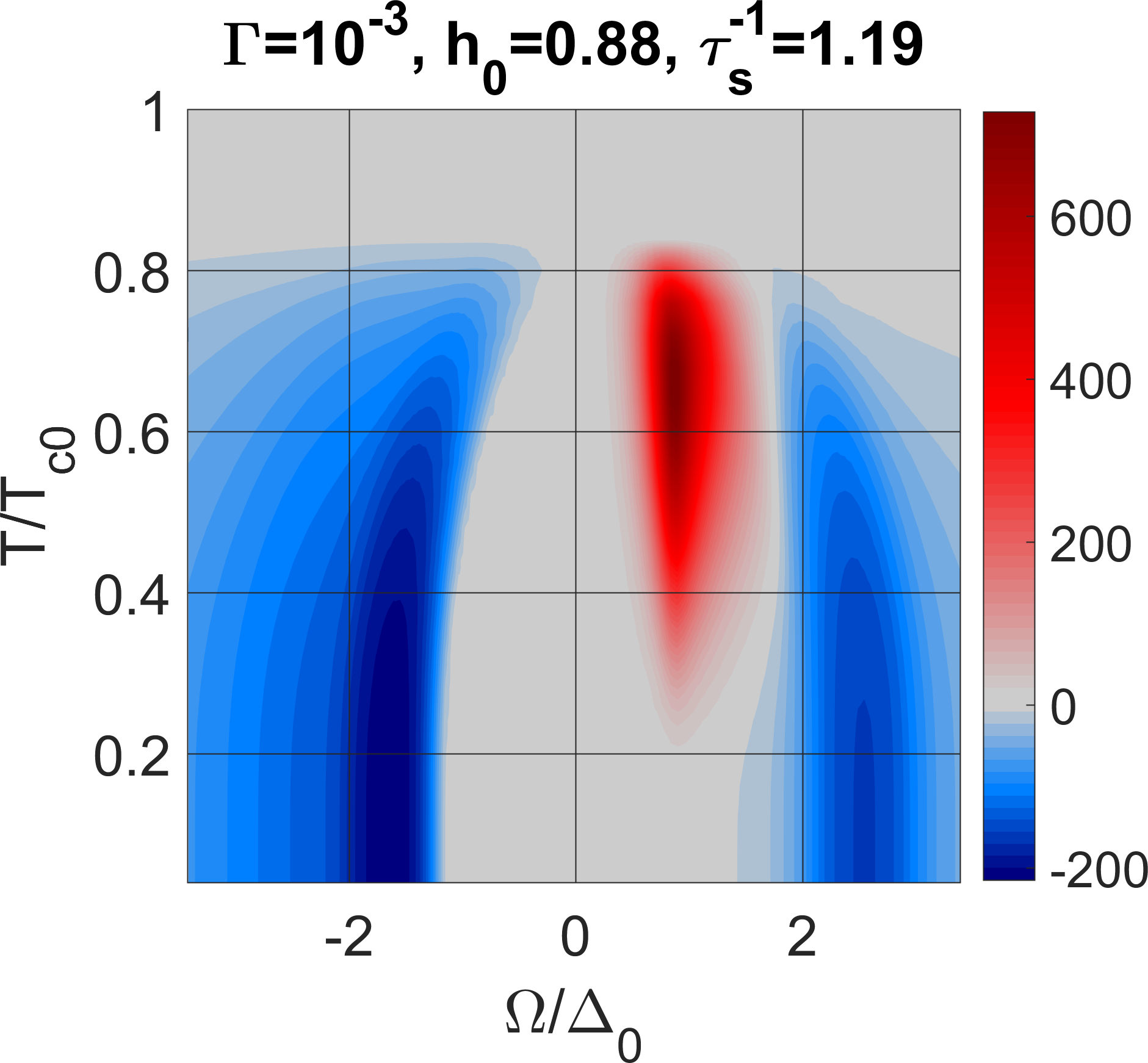}
 \includegraphics[width=0.20\linewidth]
 {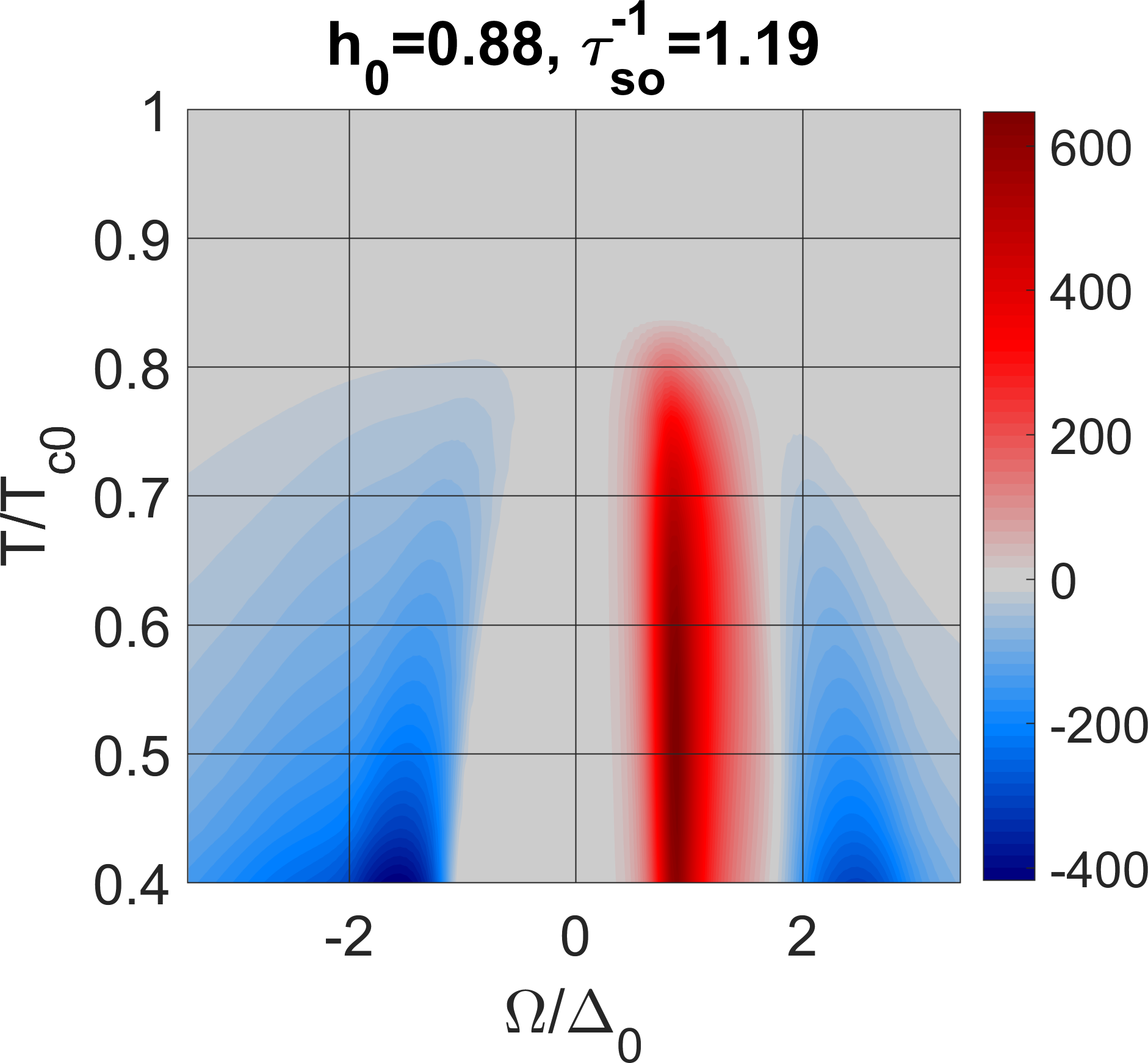}
 \includegraphics[width=0.18\linewidth]
 {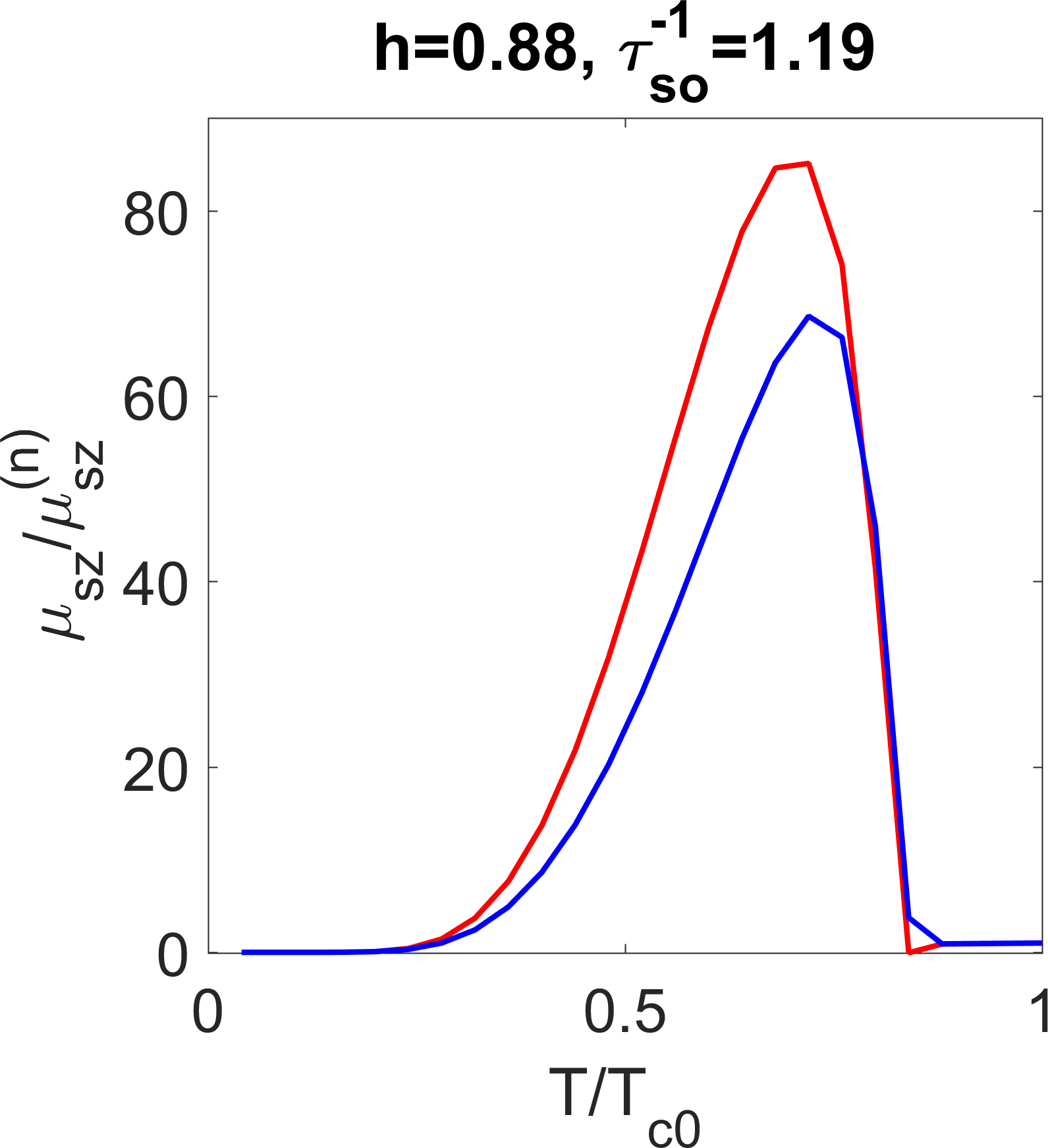}
  \includegraphics[width=0.175\linewidth]
 {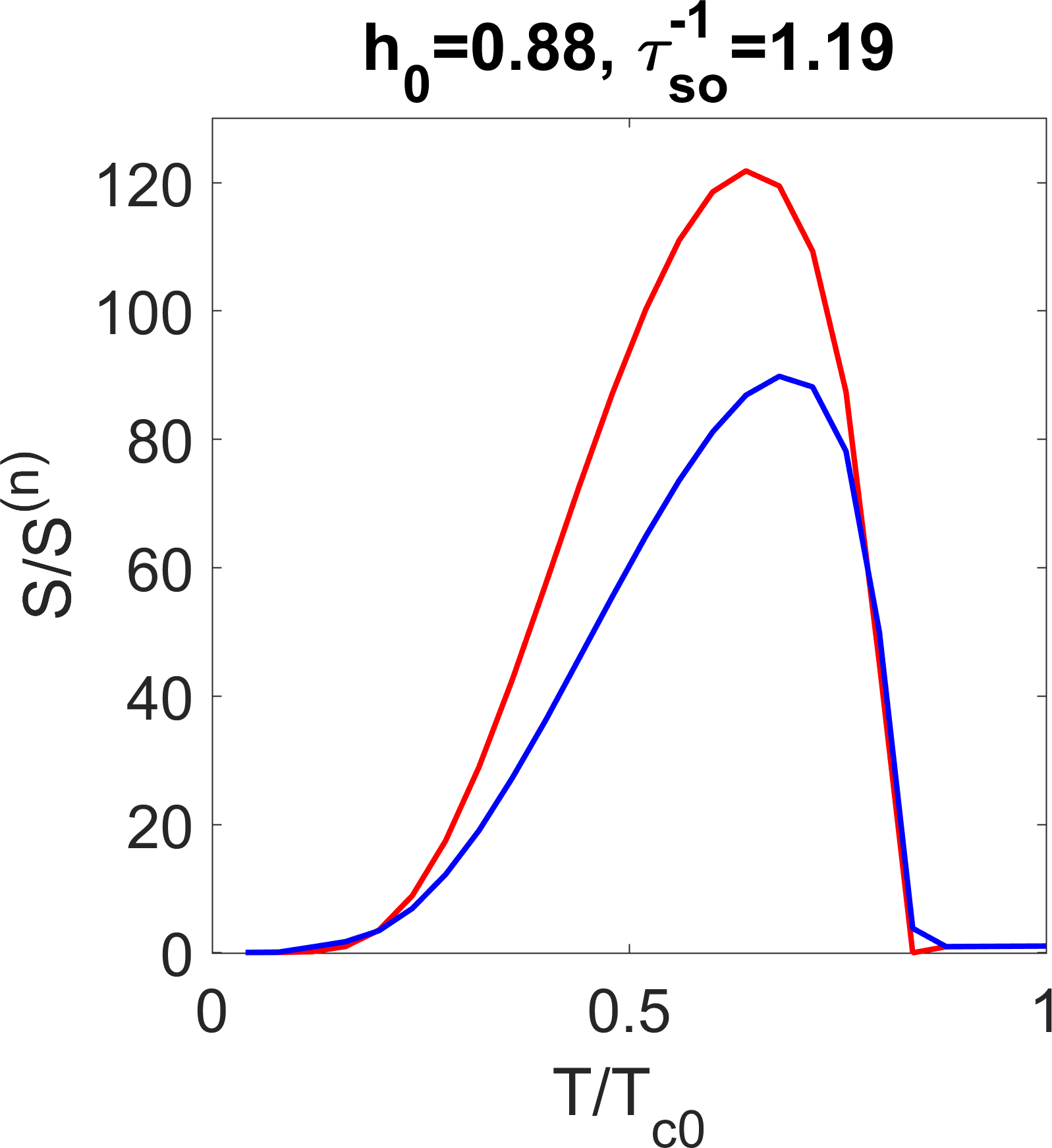}
  \\
 \includegraphics[width=0.20\linewidth]
  {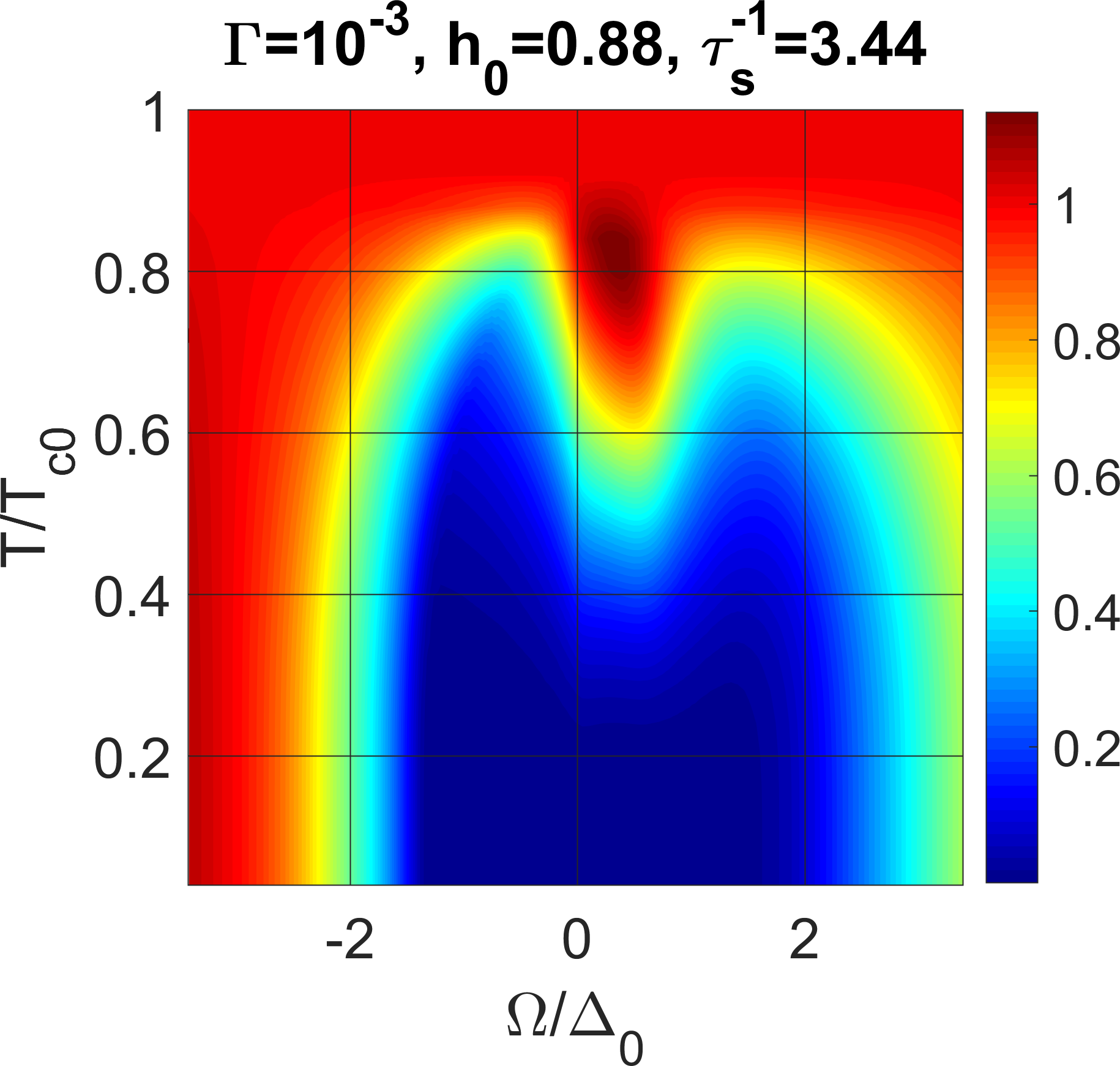}   %
  \includegraphics[width=0.20\linewidth]
 {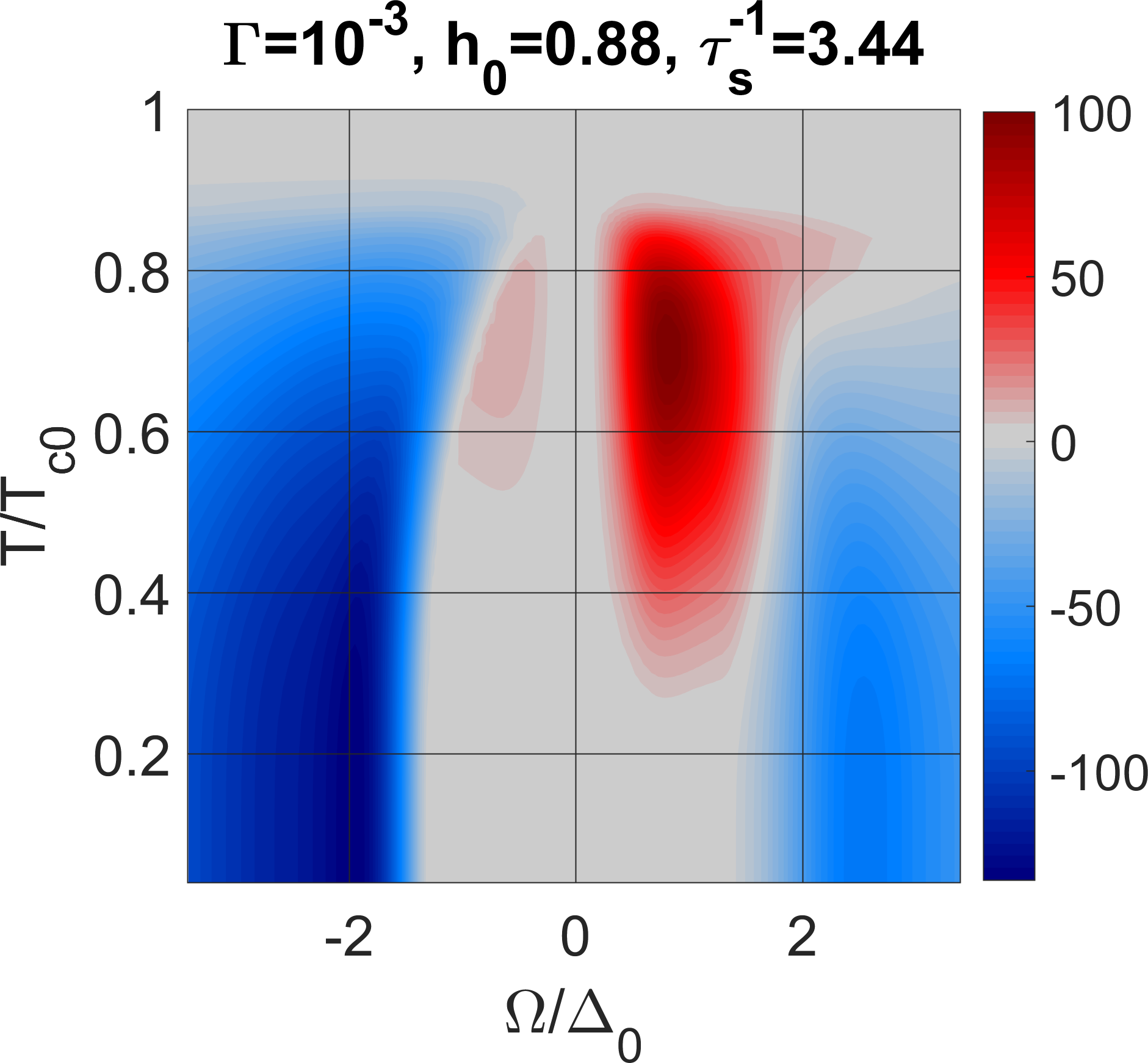} 
 \includegraphics[width=0.20\linewidth]
 {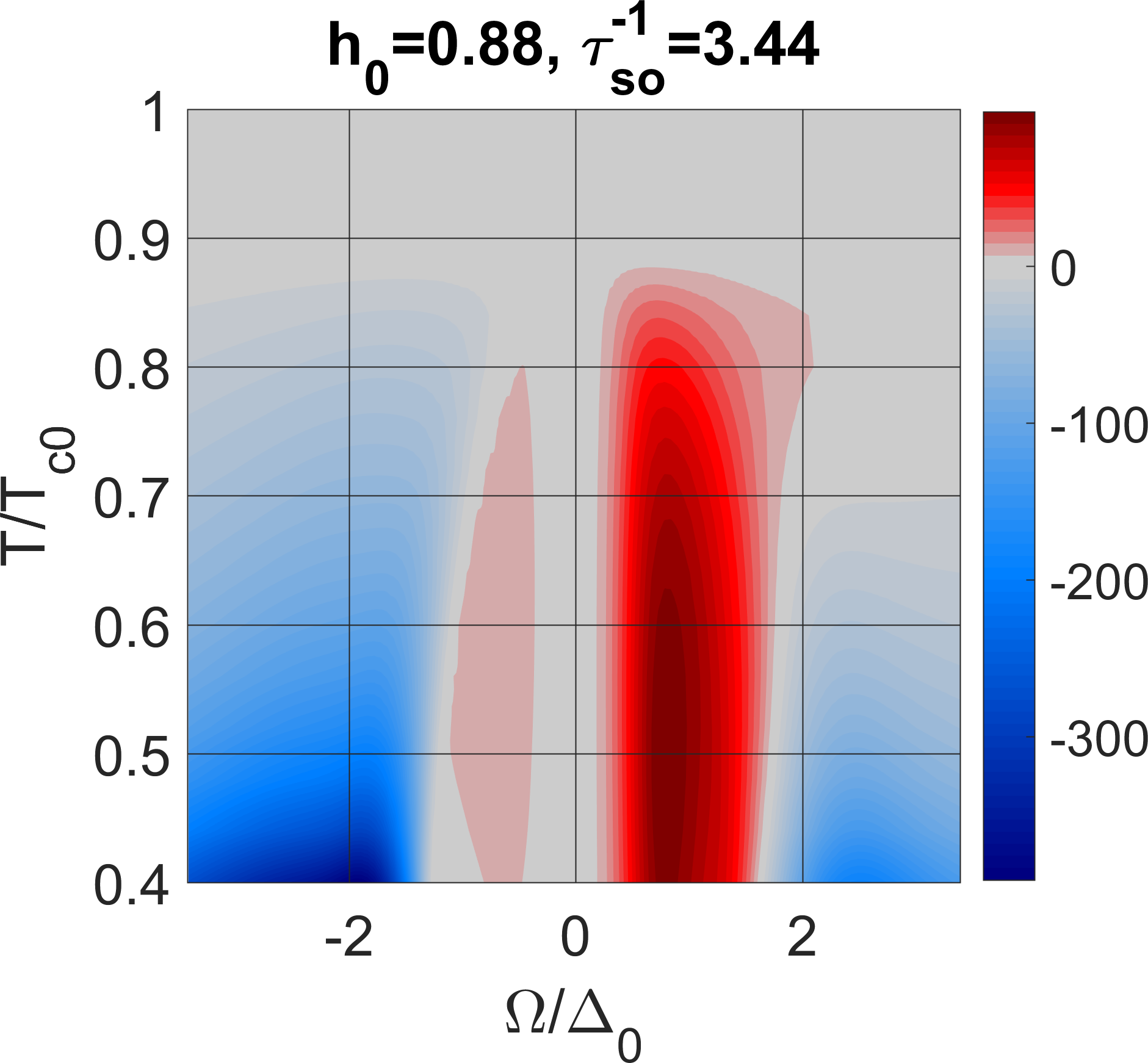}
 \includegraphics[width=0.18\linewidth]
 {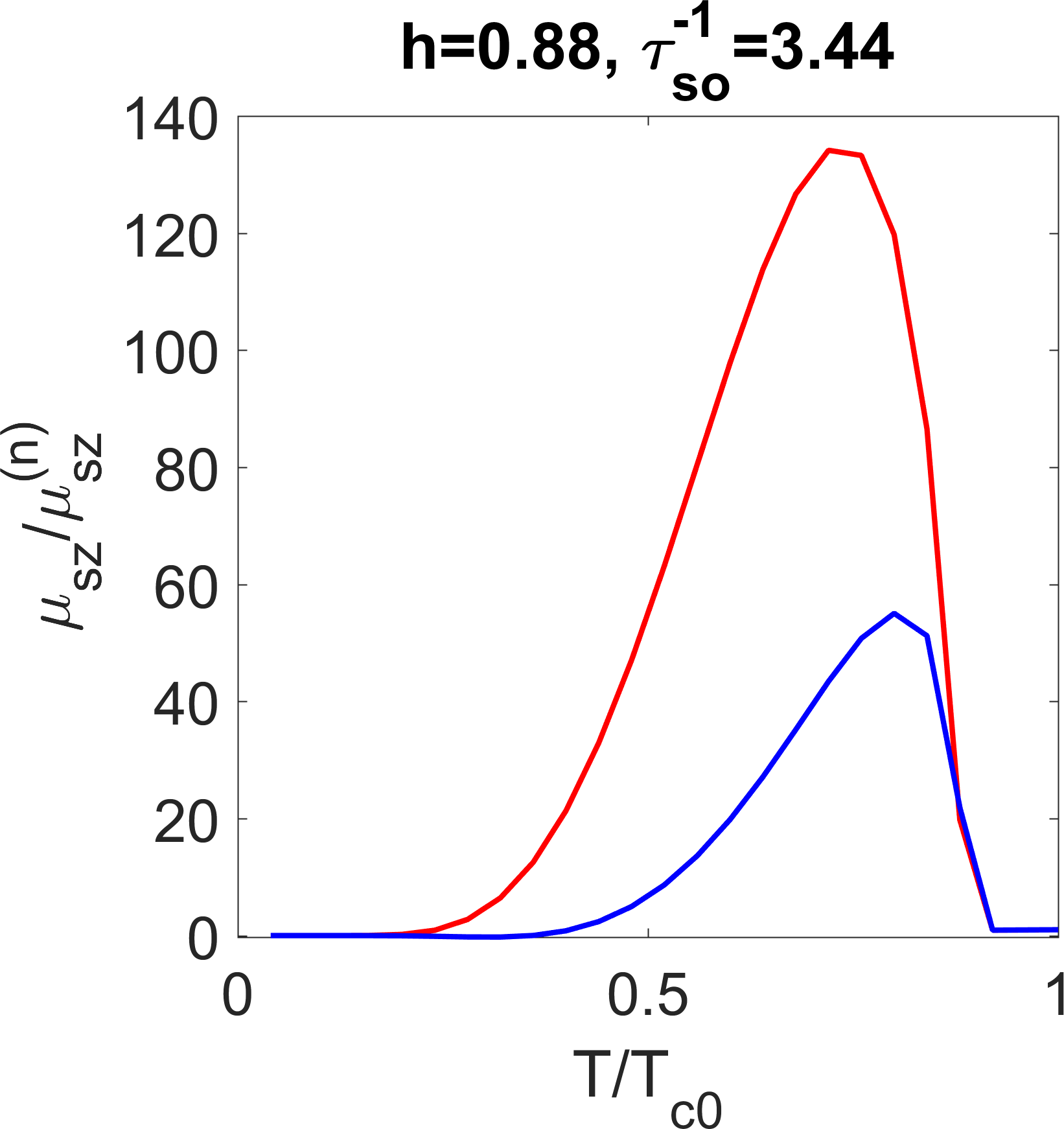} 
   \includegraphics[width=0.175\linewidth]
 {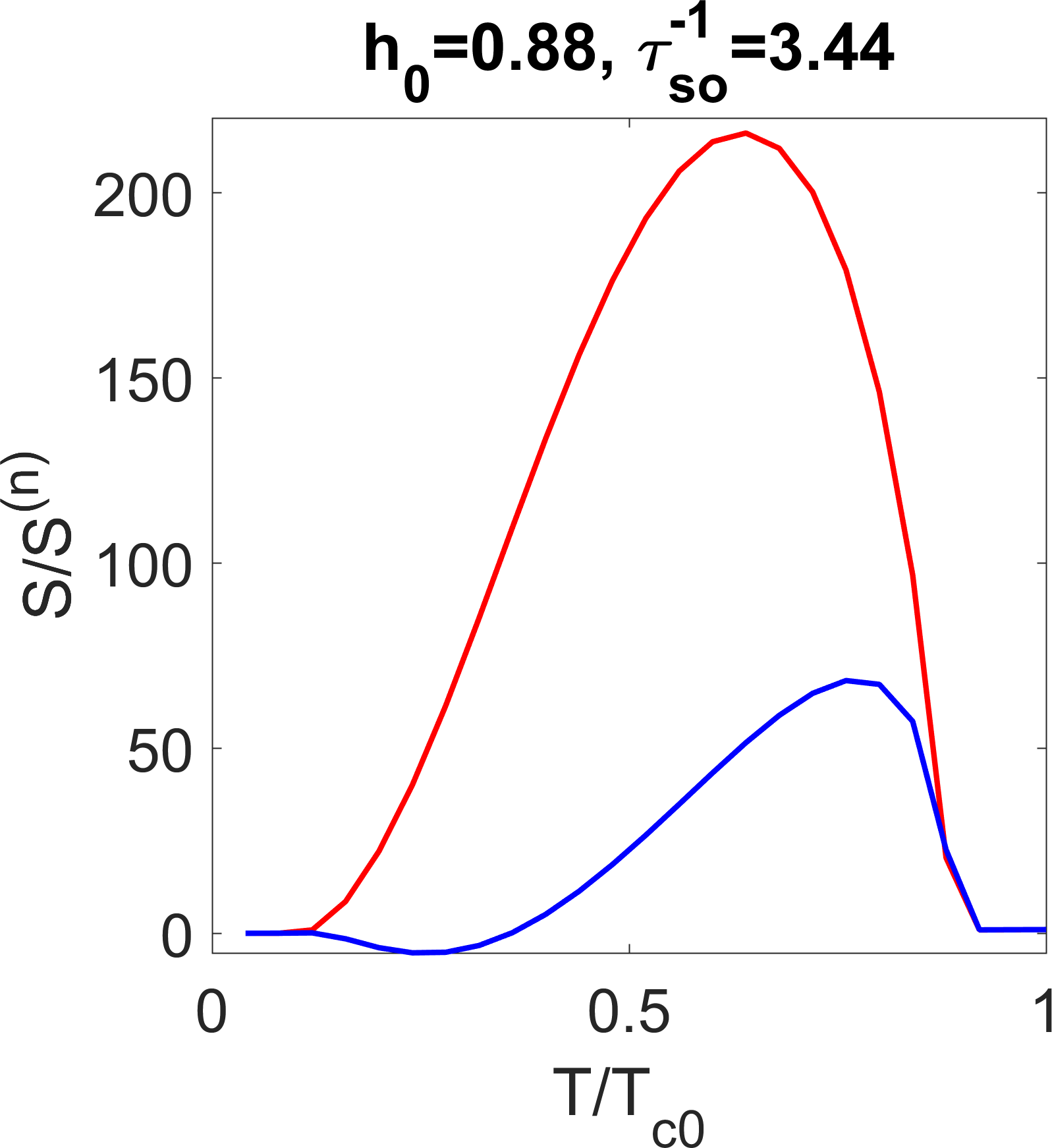} 
   \\
 \includegraphics[width=0.20\linewidth]
  {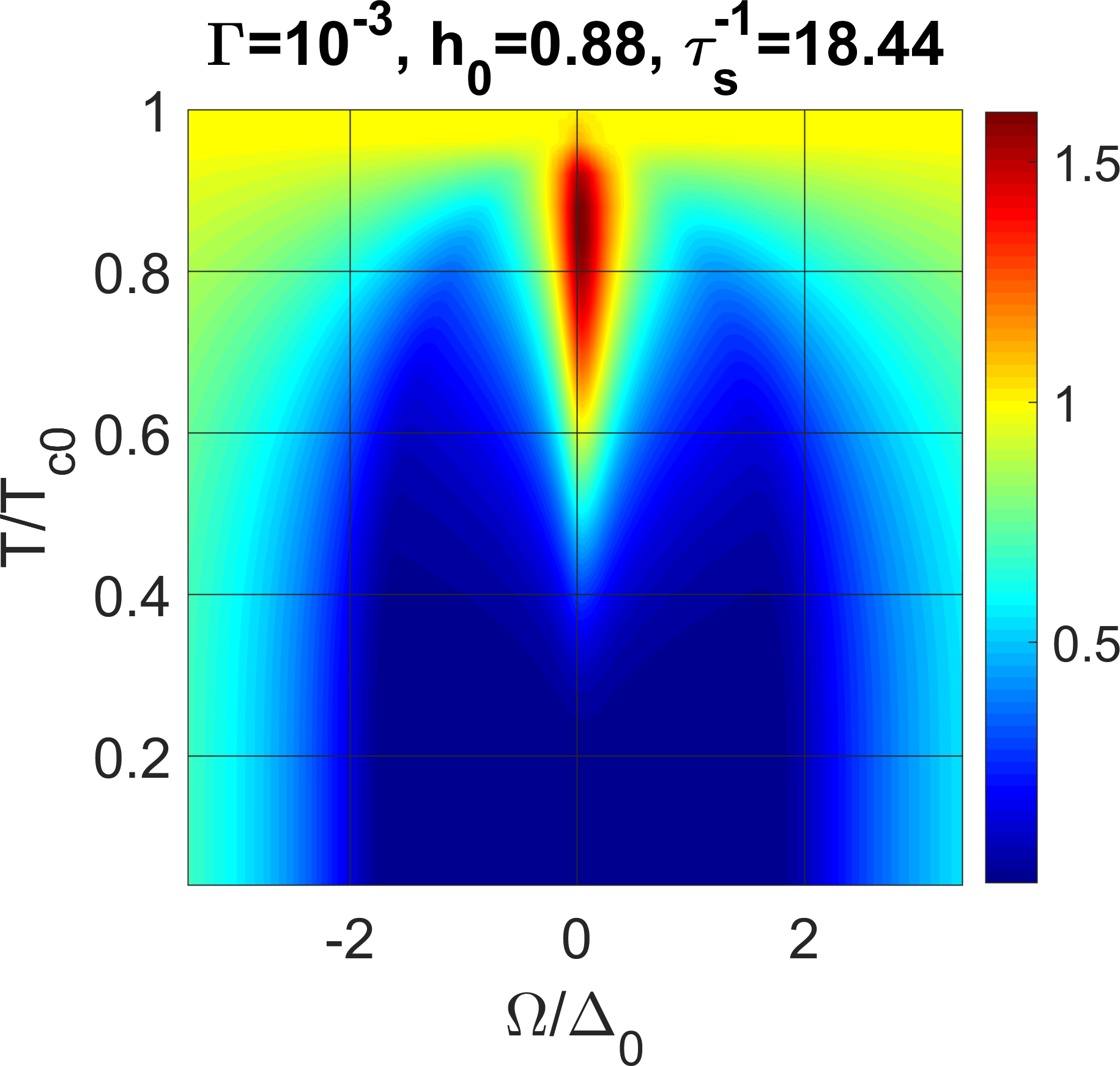}   %
  \includegraphics[width=0.20\linewidth]
 {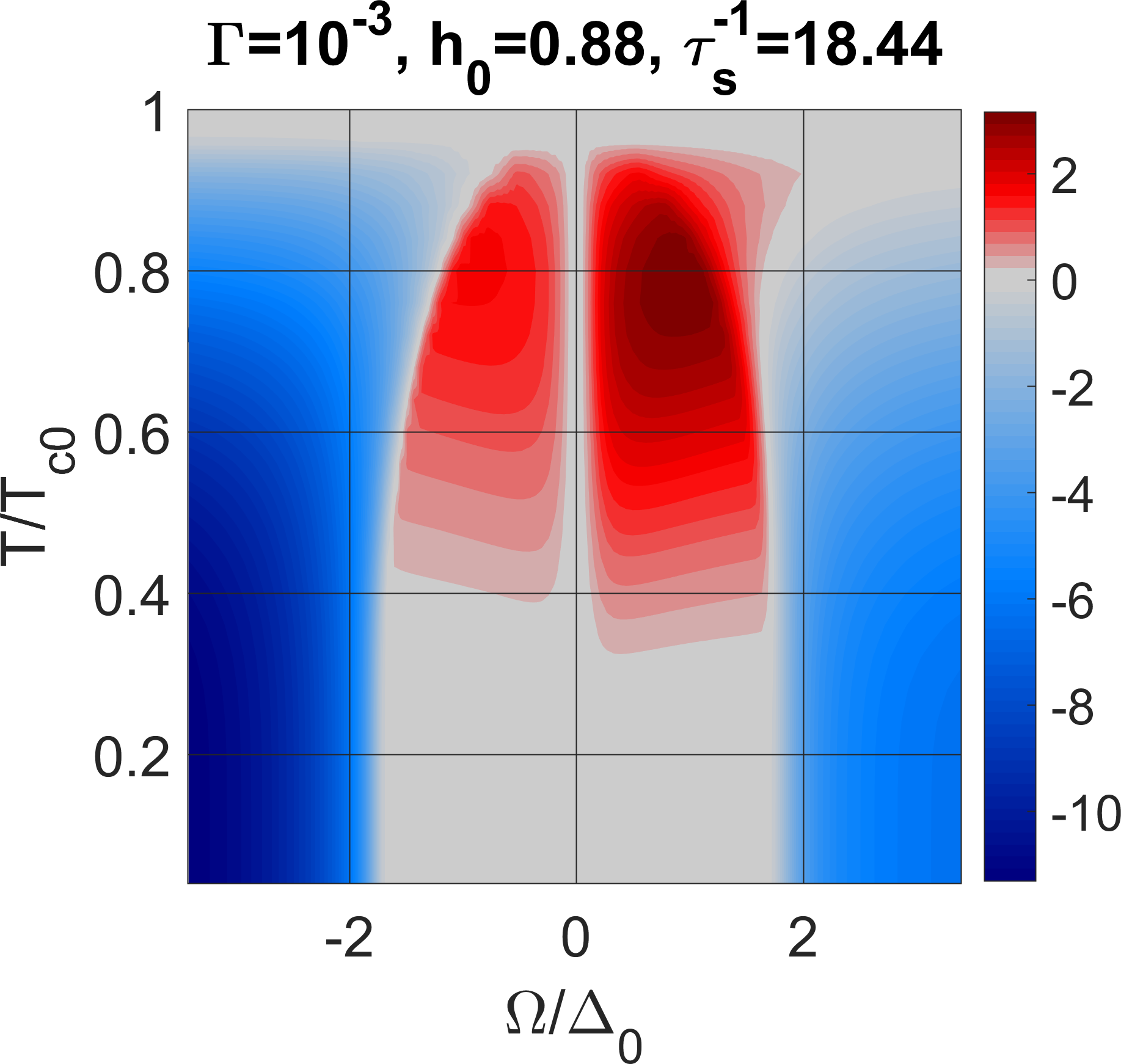} 
 \includegraphics[width=0.20\linewidth]
 {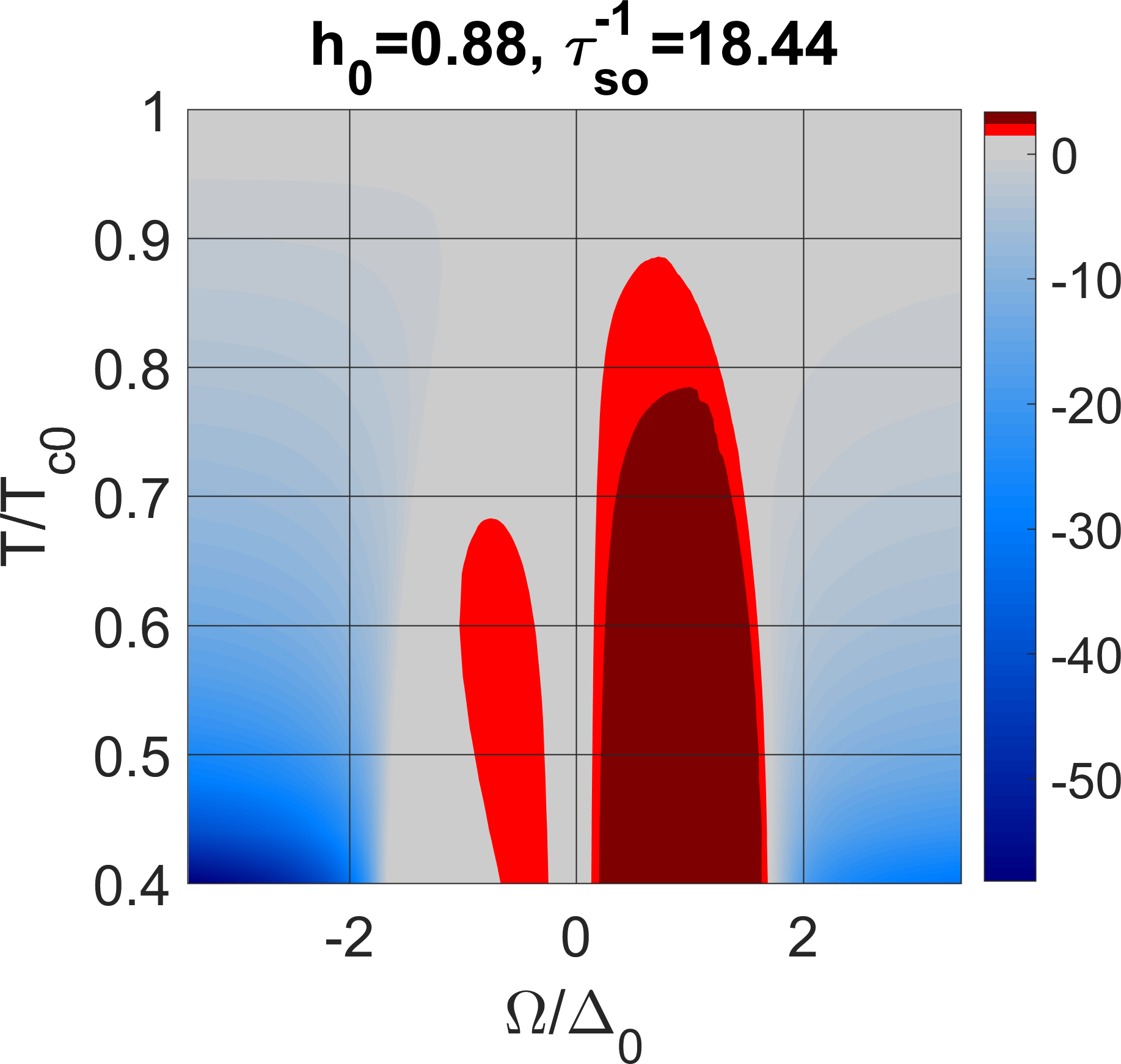}
 \includegraphics[width=0.18\linewidth]
 {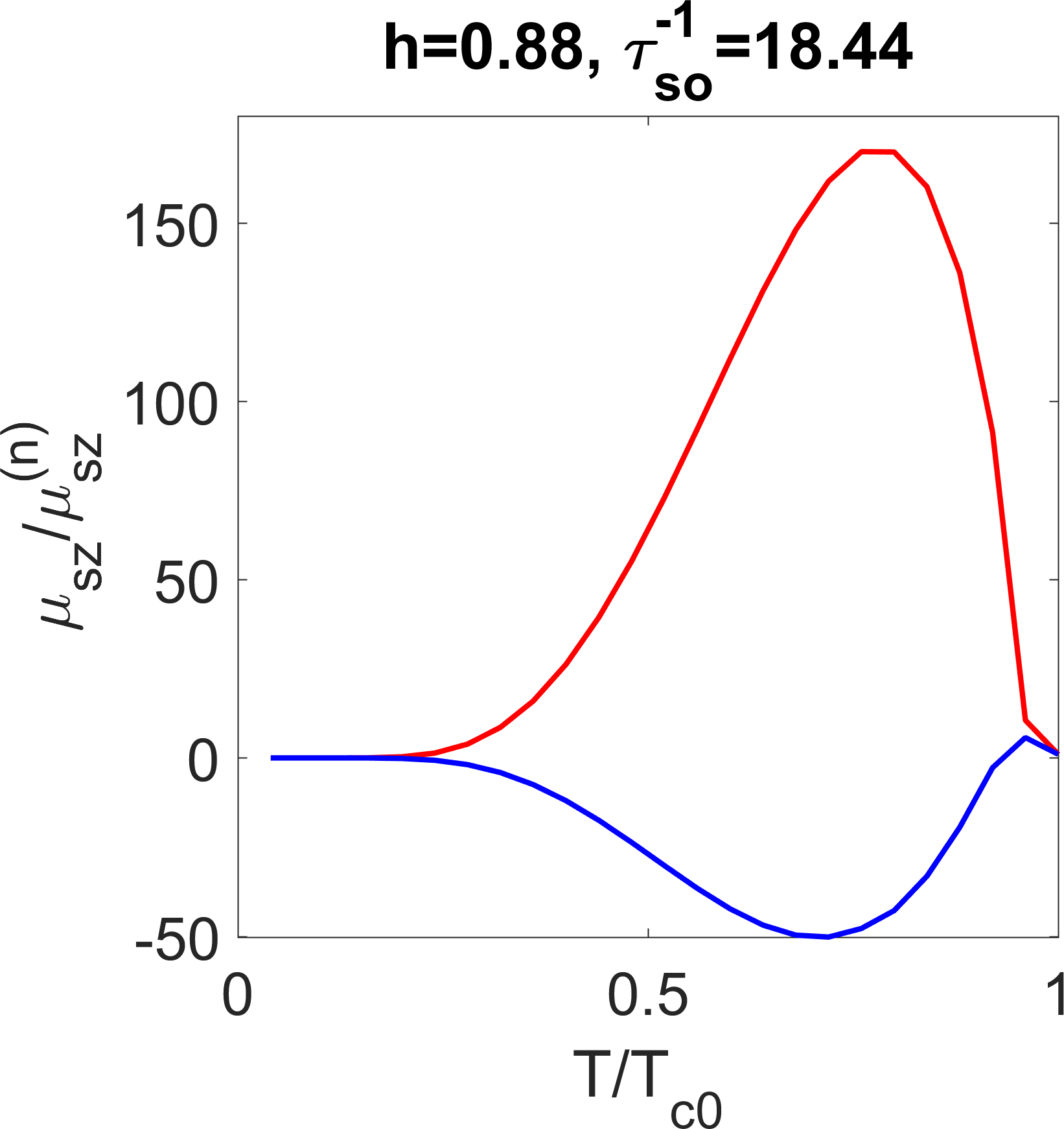} 
  \includegraphics[width=0.175\linewidth]
 {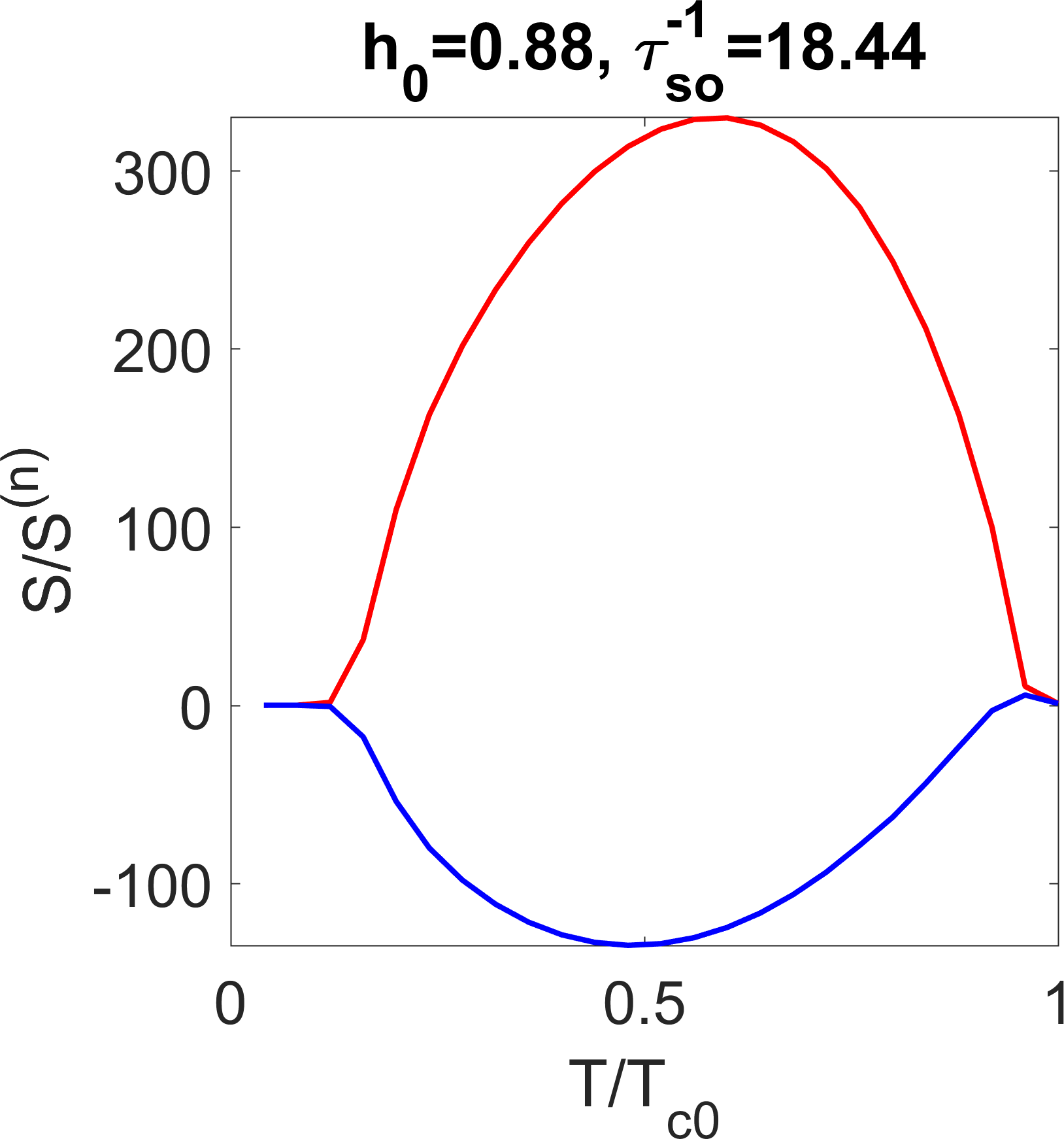} 
   \end{array}$}
  \caption{\label{Fig:EnergyPolarWT-so-scan}
(1st column): Pumped energy of the electronic system $ W (T,\Omega)/ W (T_c,\Omega)$.
(2nd column): Pumped spin accumulation $( T_{c0}/h_\Omega^2)  \mu_z(T,\Omega)$. 
(3rd column): Non-local voltage generated by the pumped spin accumulation $ (e T_{c0}/h_\Omega^2)V(\Omega,T)$. 
(4th column): Magnon-induced spin accumulation.. 
(5th column):  nonlocal Seebeck coefficient in the FI/SC/FM bilayer.
 Parameters are $h_0/T_{c0} =0.88$, energy relaxation rate $\Gamma /T_{c0} =10^{-3}$.  We consider circular polarization  $h_{l,\Omega},h_{r,-\Omega} \neq 0$.  
Scan over $\Omega$, $T$, different values of spin relaxation. 
  }
 \end{figure*}

\begin{figure*}[htb!]
 \centerline{
 $  \begin{array}{c}
 \includegraphics[width=0.20\linewidth]
 {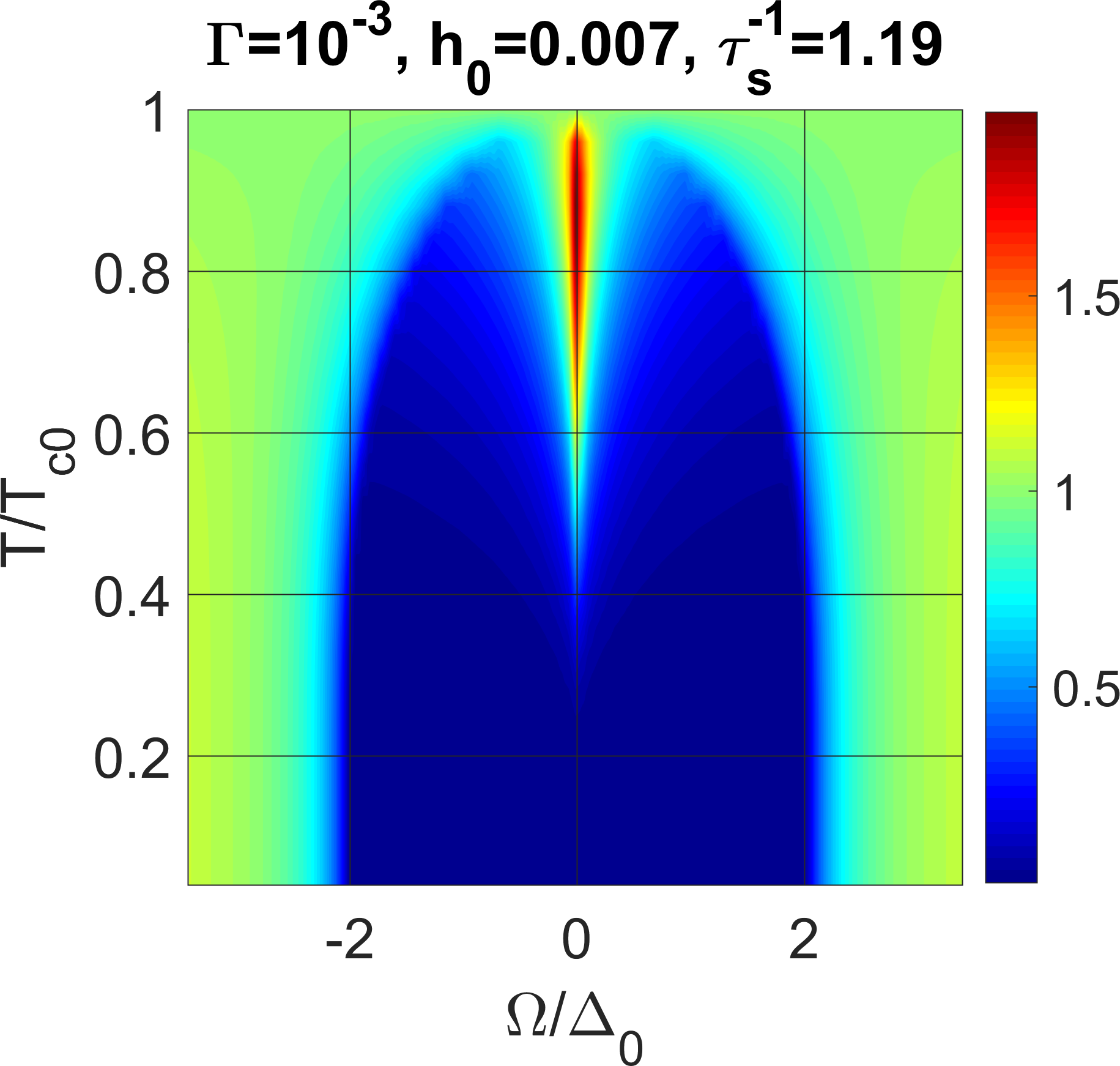} 
 \includegraphics[width=0.20\linewidth]
 {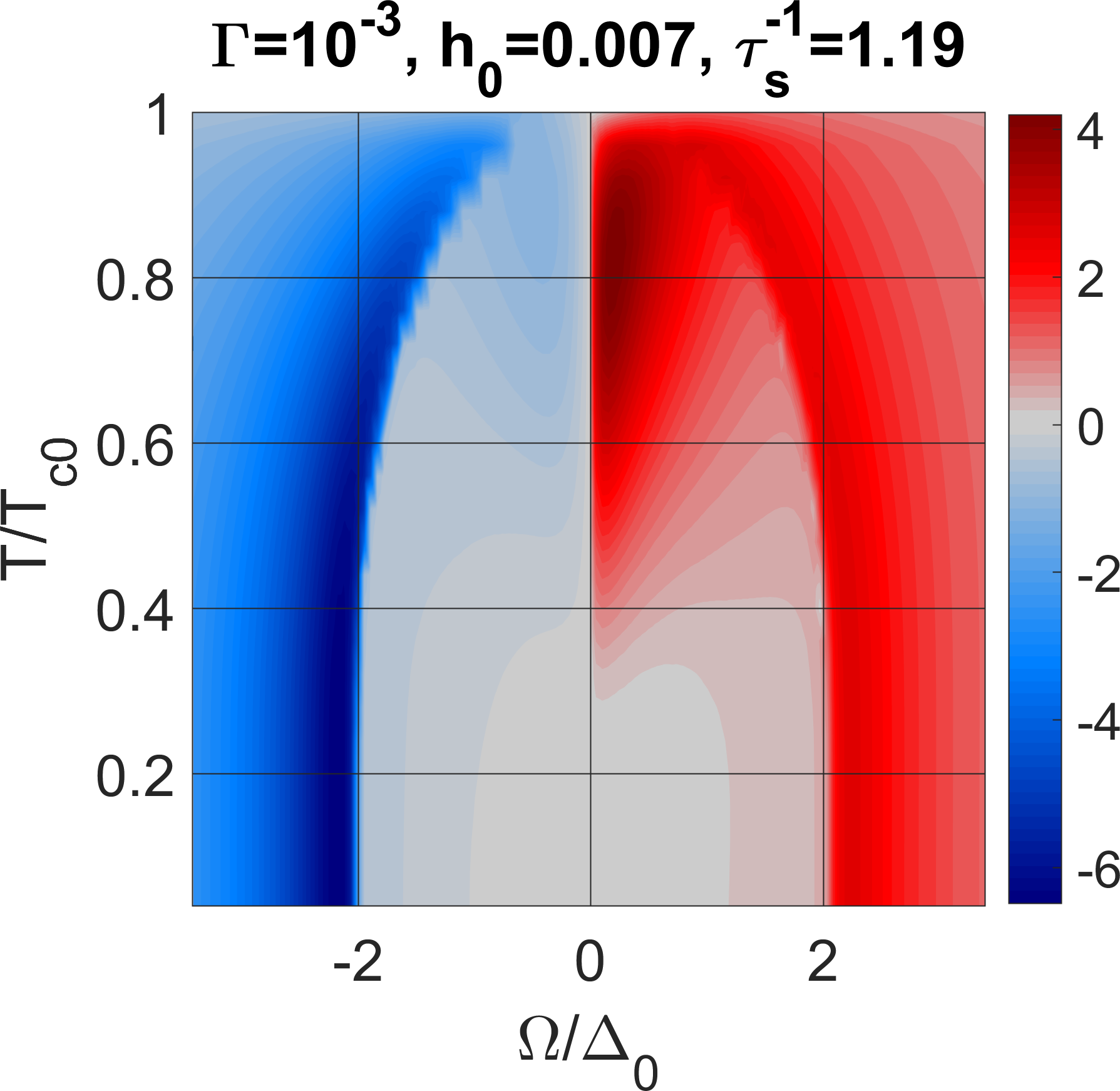} 
 \includegraphics[width=0.20\linewidth]
 {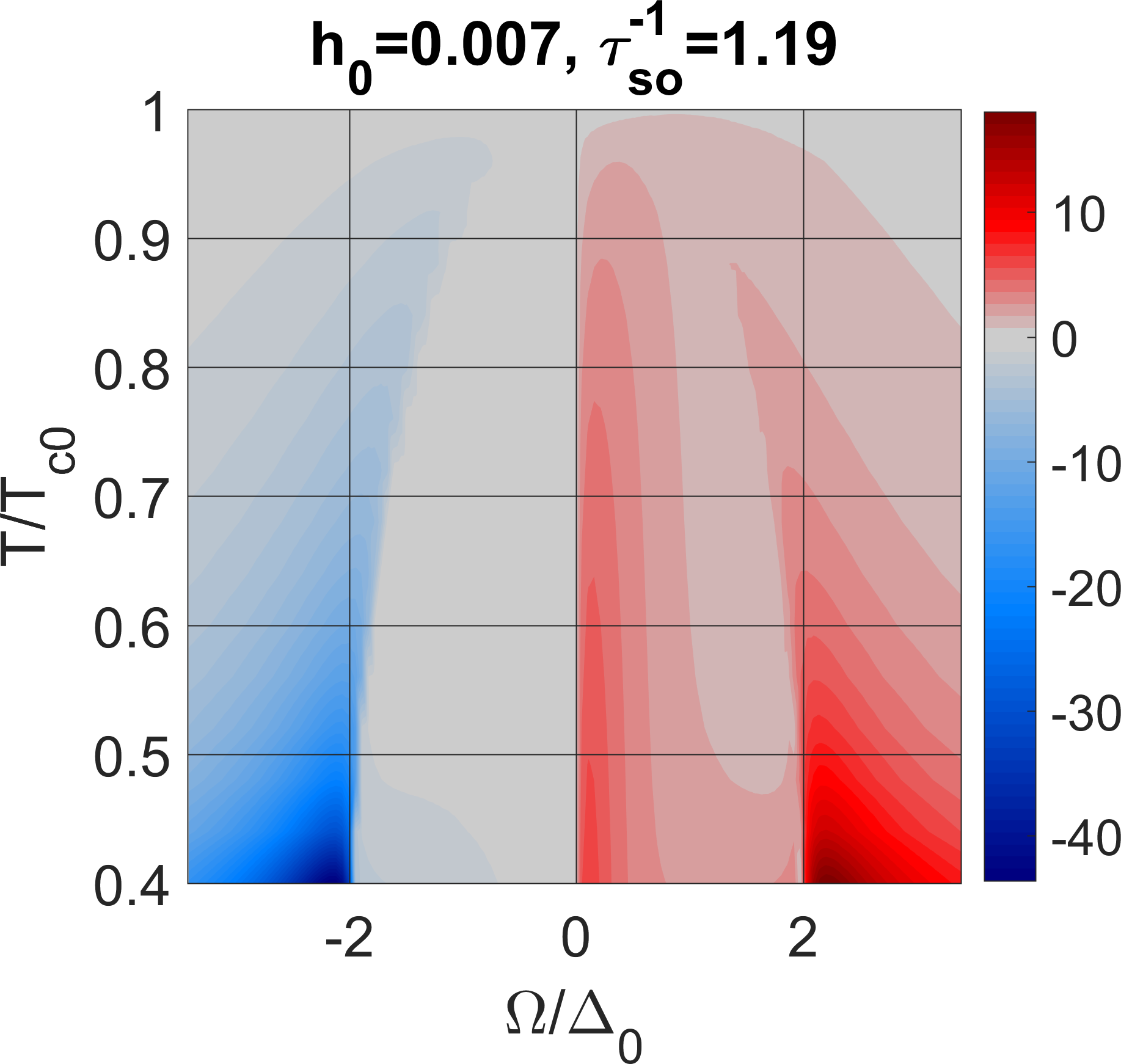}
 \includegraphics[width=0.18\linewidth]
 {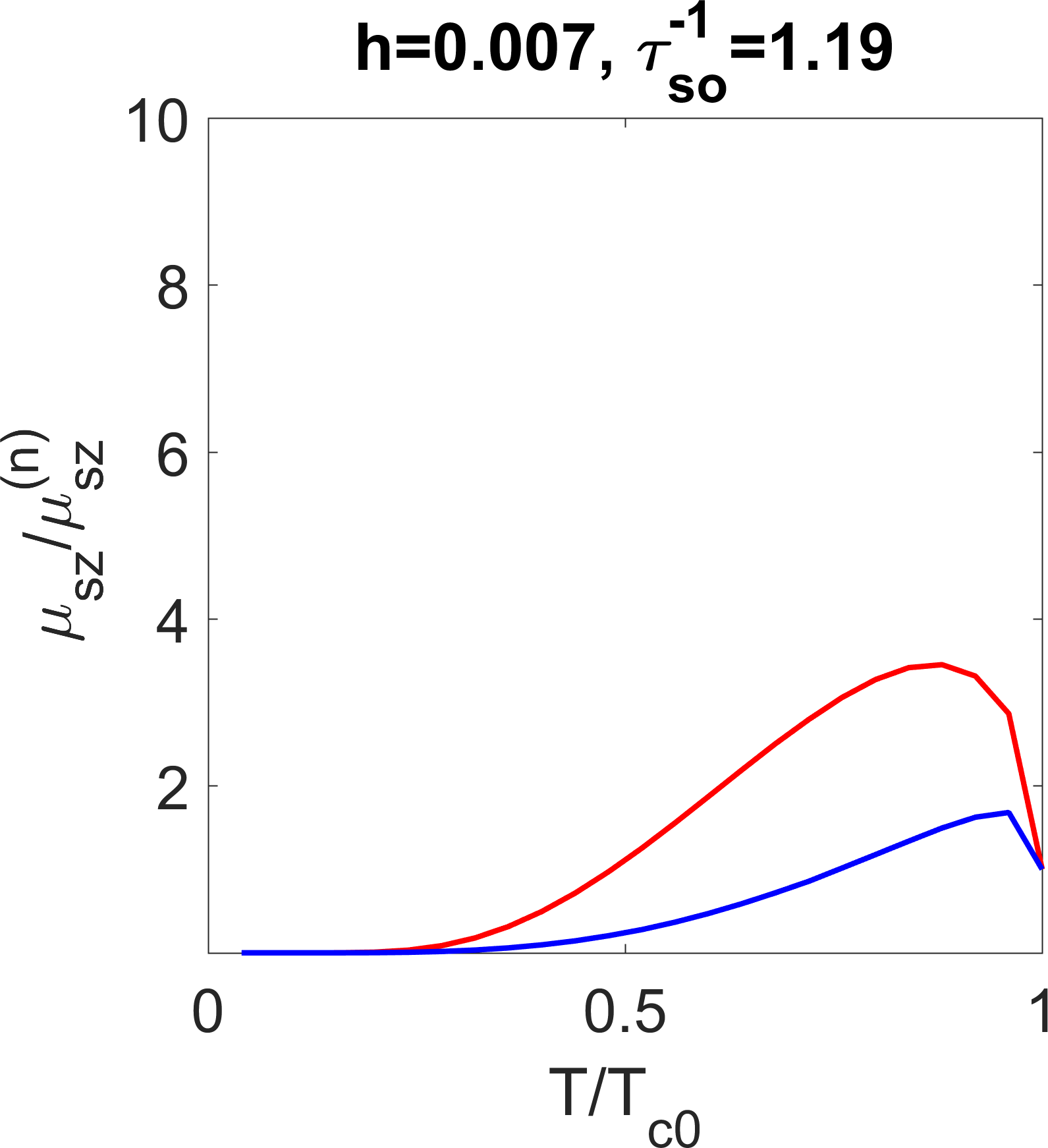} 
 \includegraphics[width=0.175\linewidth]
 {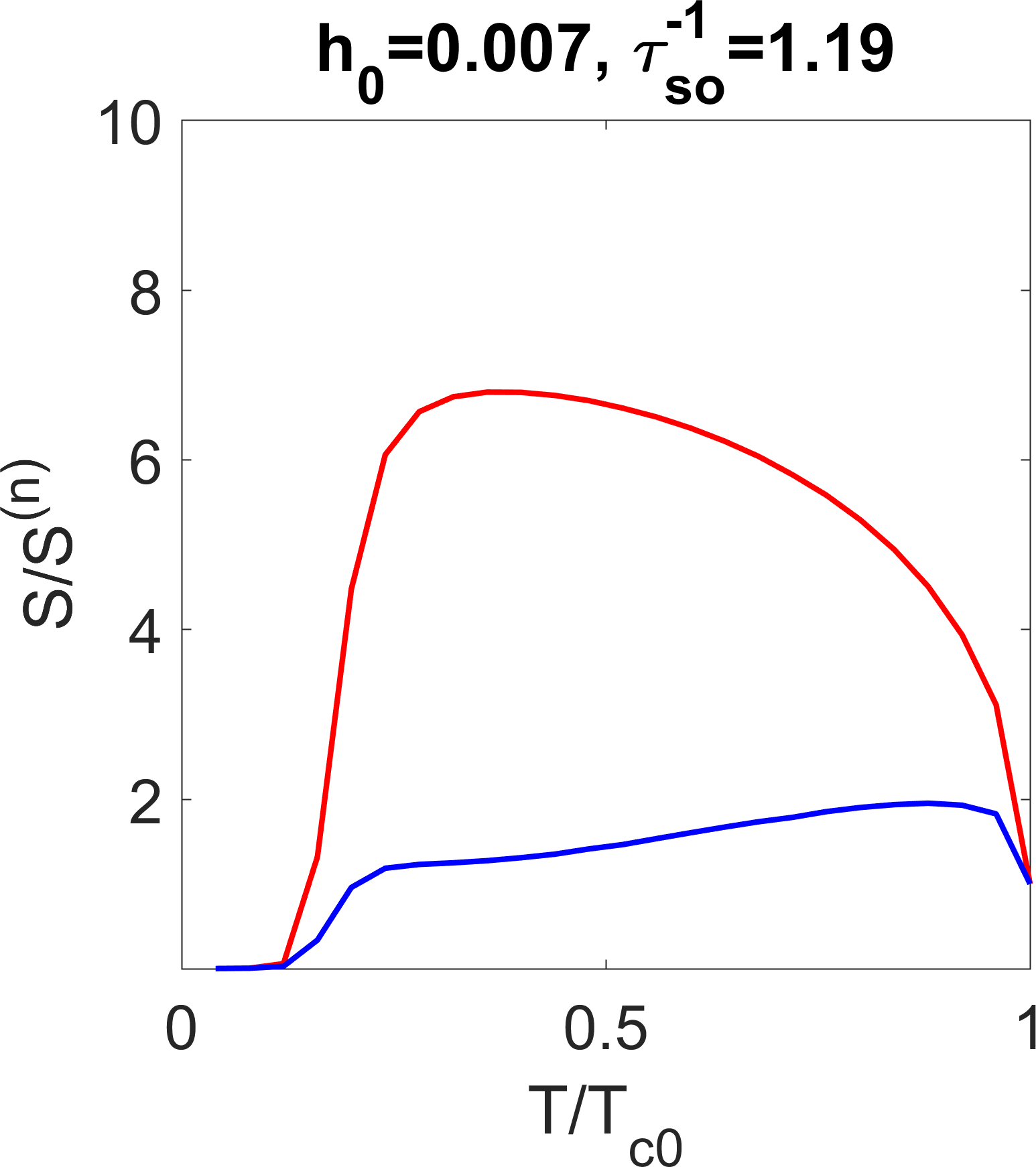} 
 \\
 \includegraphics[width=0.20\linewidth]
 {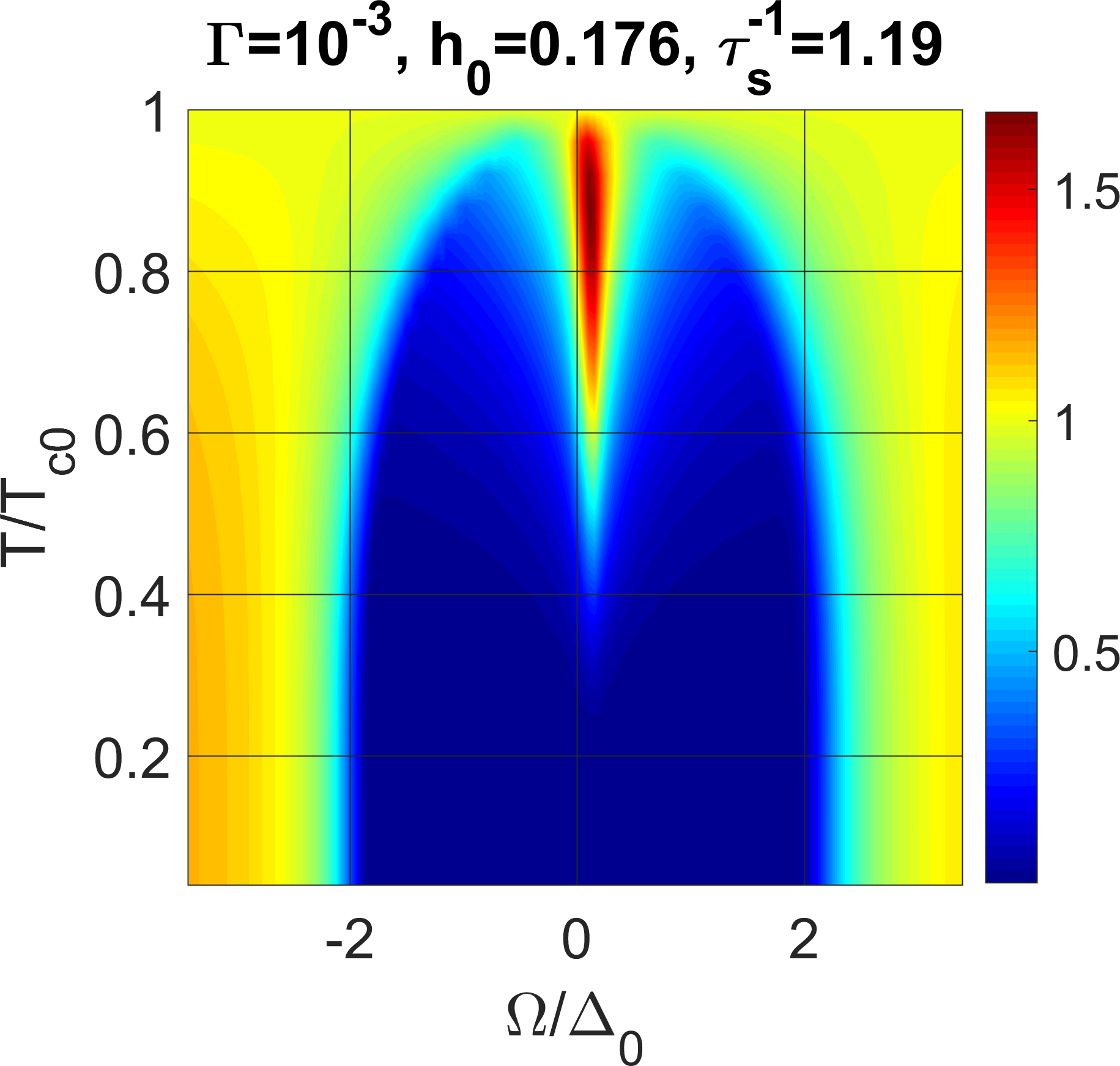} 
 \includegraphics[width=0.20\linewidth]
 {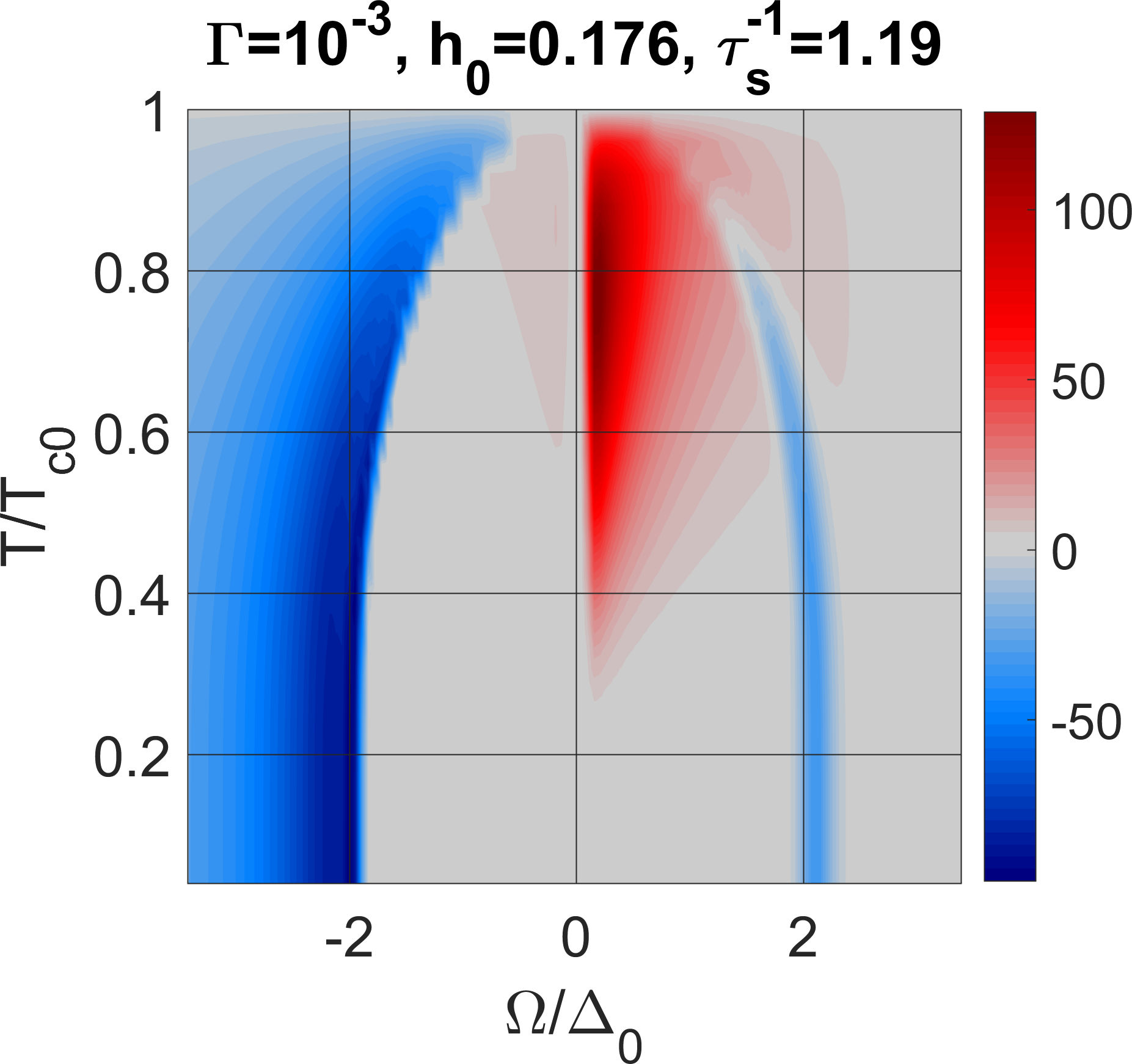} 
 \includegraphics[width=0.20\linewidth]
 {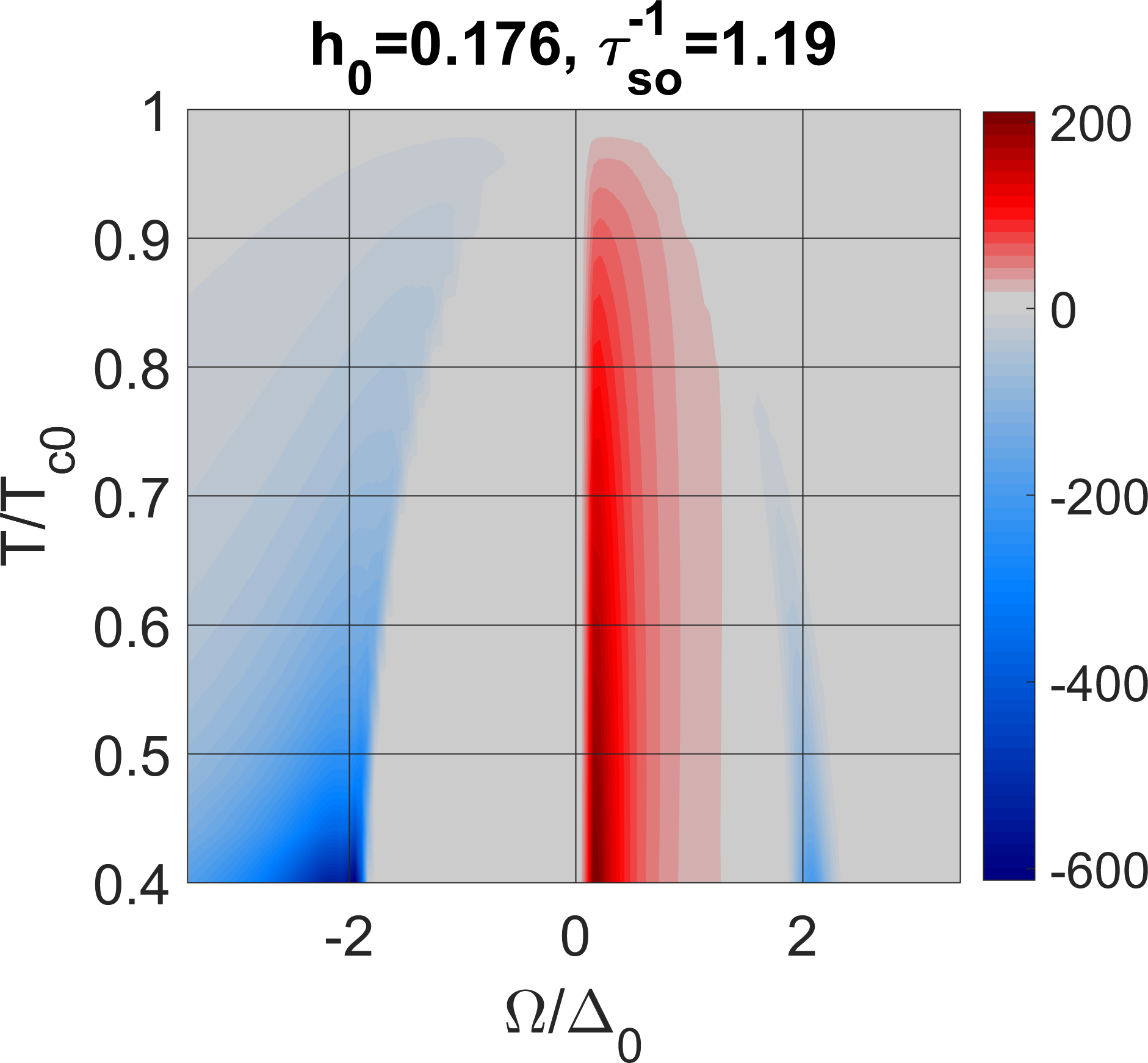}
 \includegraphics[width=0.18\linewidth]
 {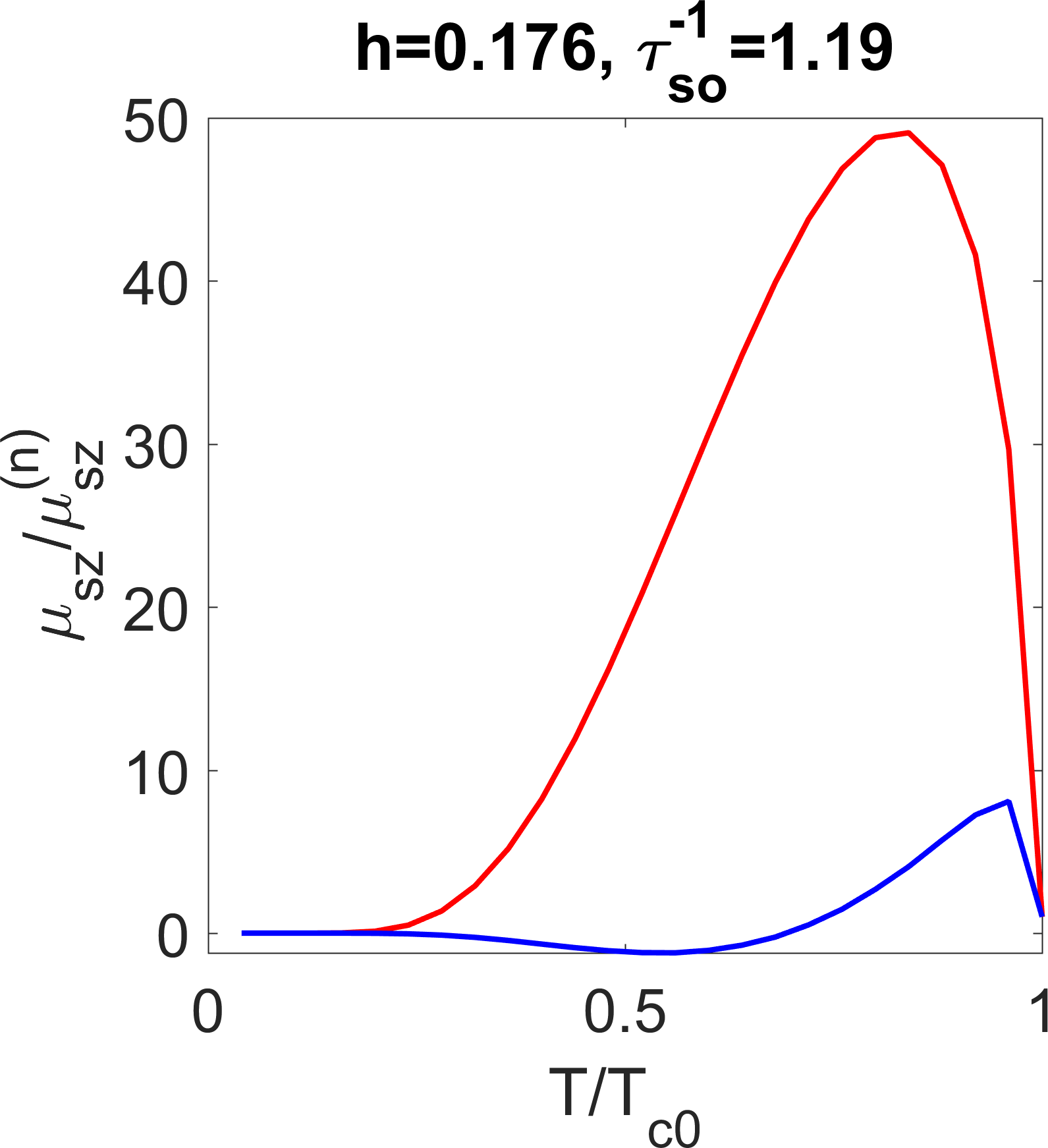} 
 \includegraphics[width=0.175\linewidth]
 {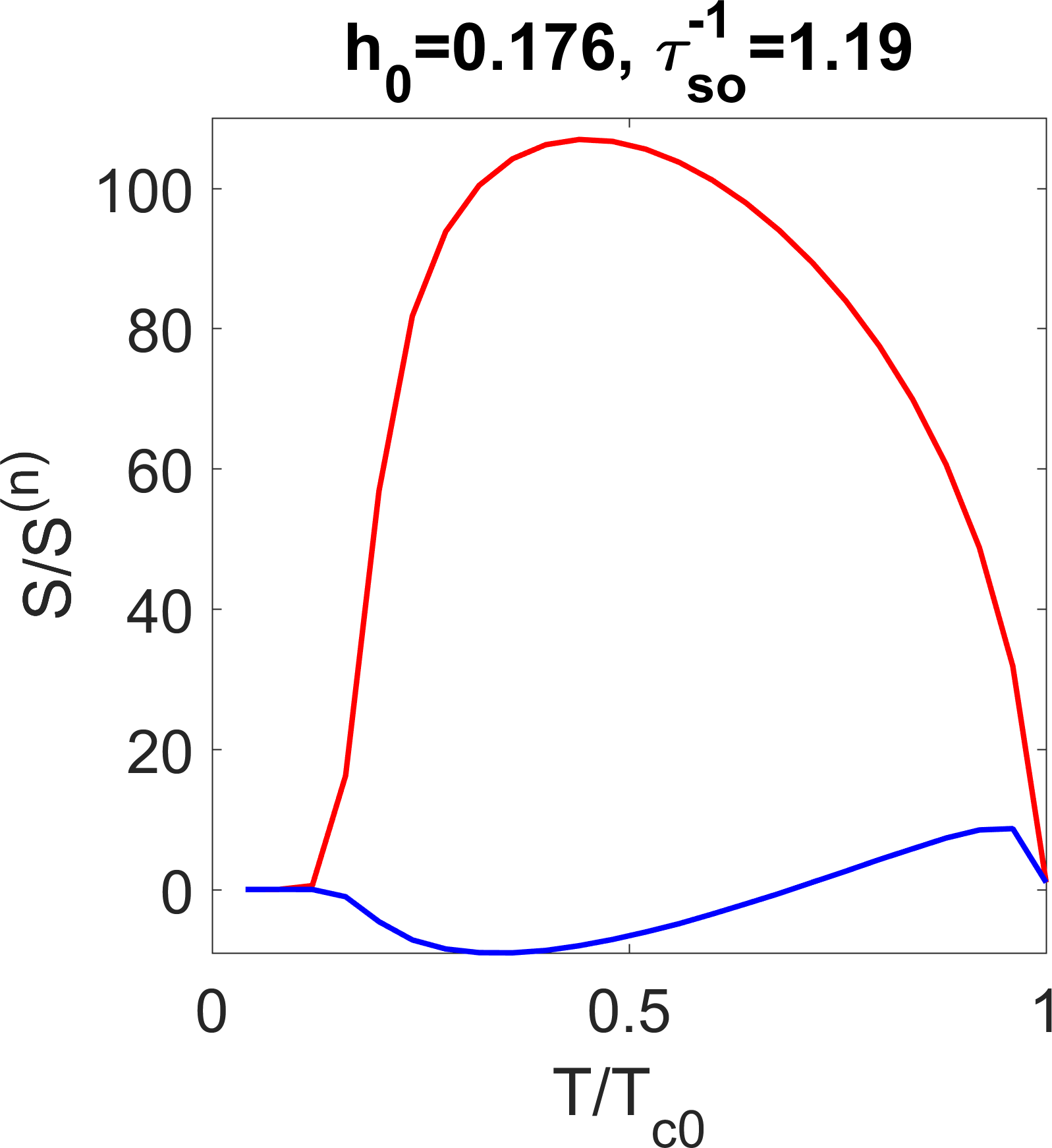} 
  \\
\includegraphics[width=0.20\linewidth]
 {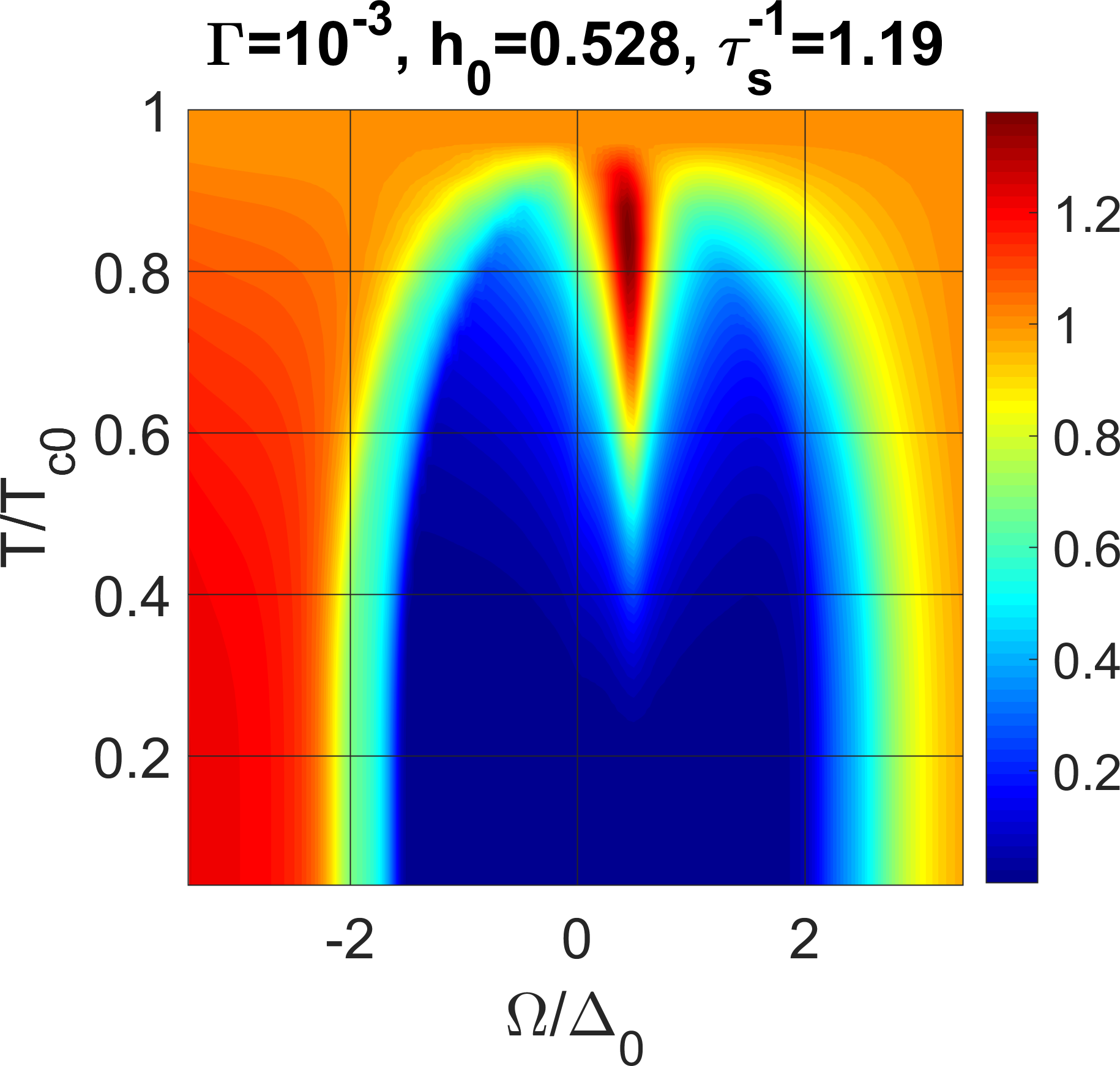} 
 \includegraphics[width=0.20\linewidth]
 {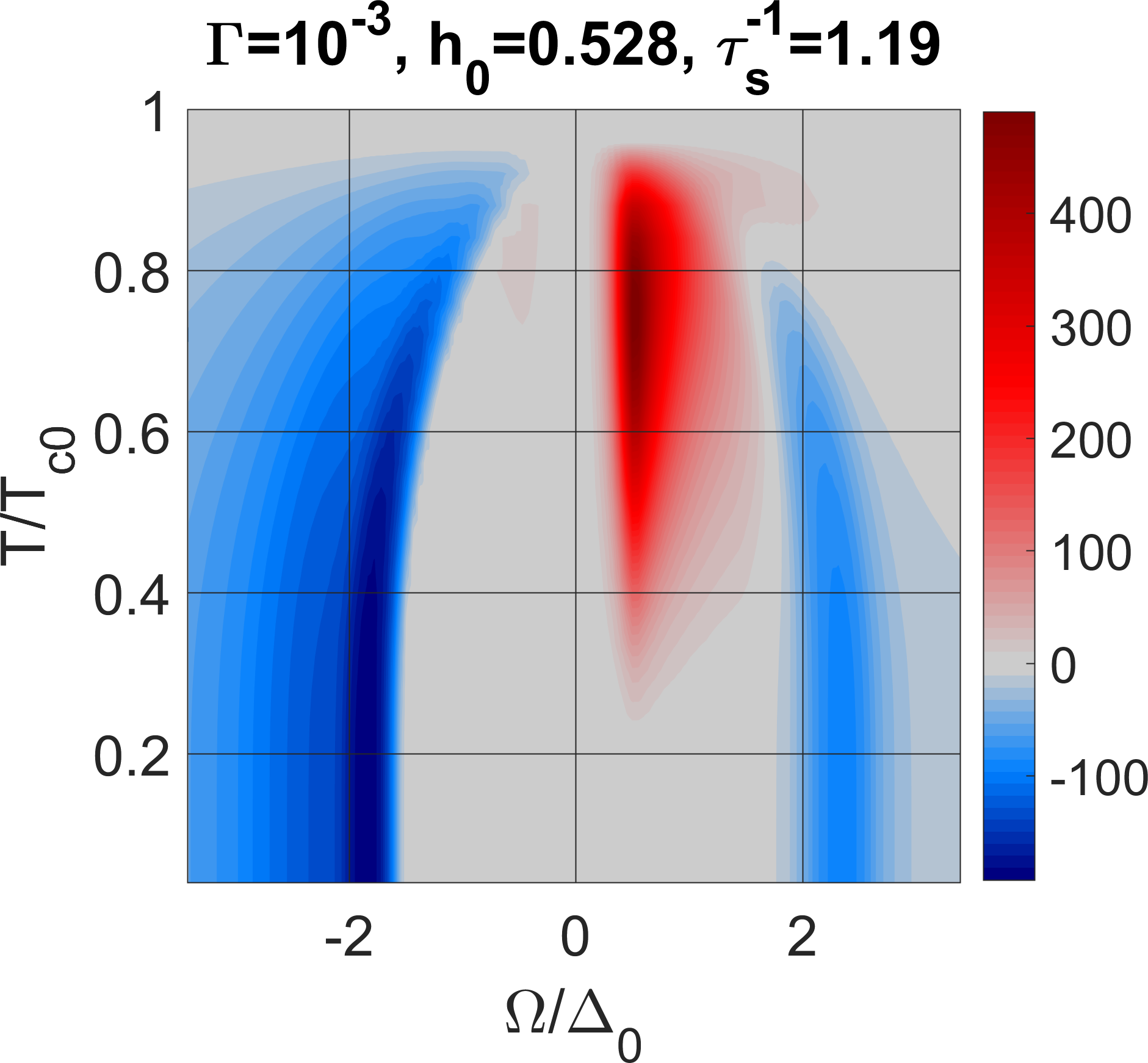} 
 \includegraphics[width=0.20\linewidth]
 {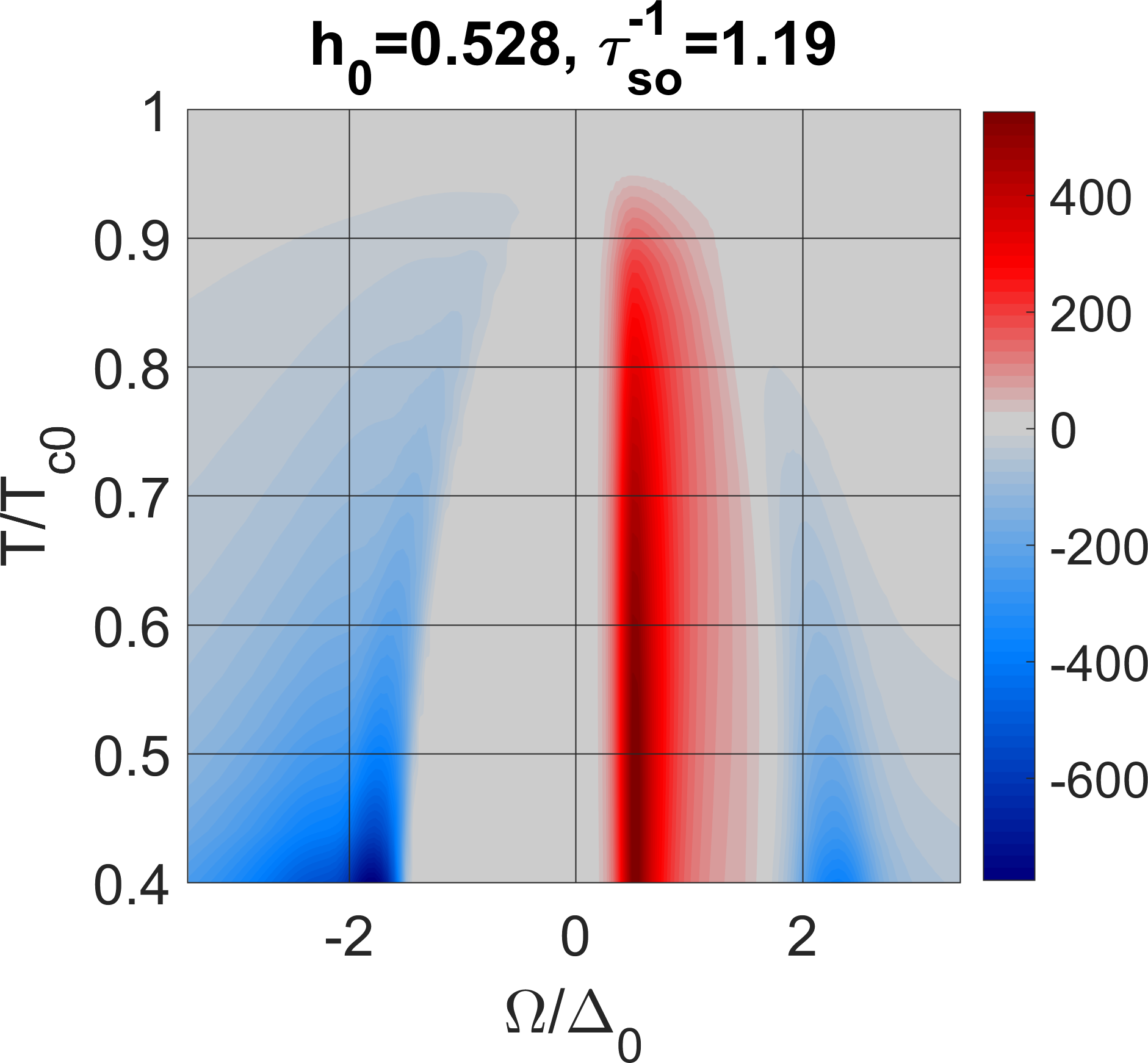}
 \includegraphics[width=0.18\linewidth]
 {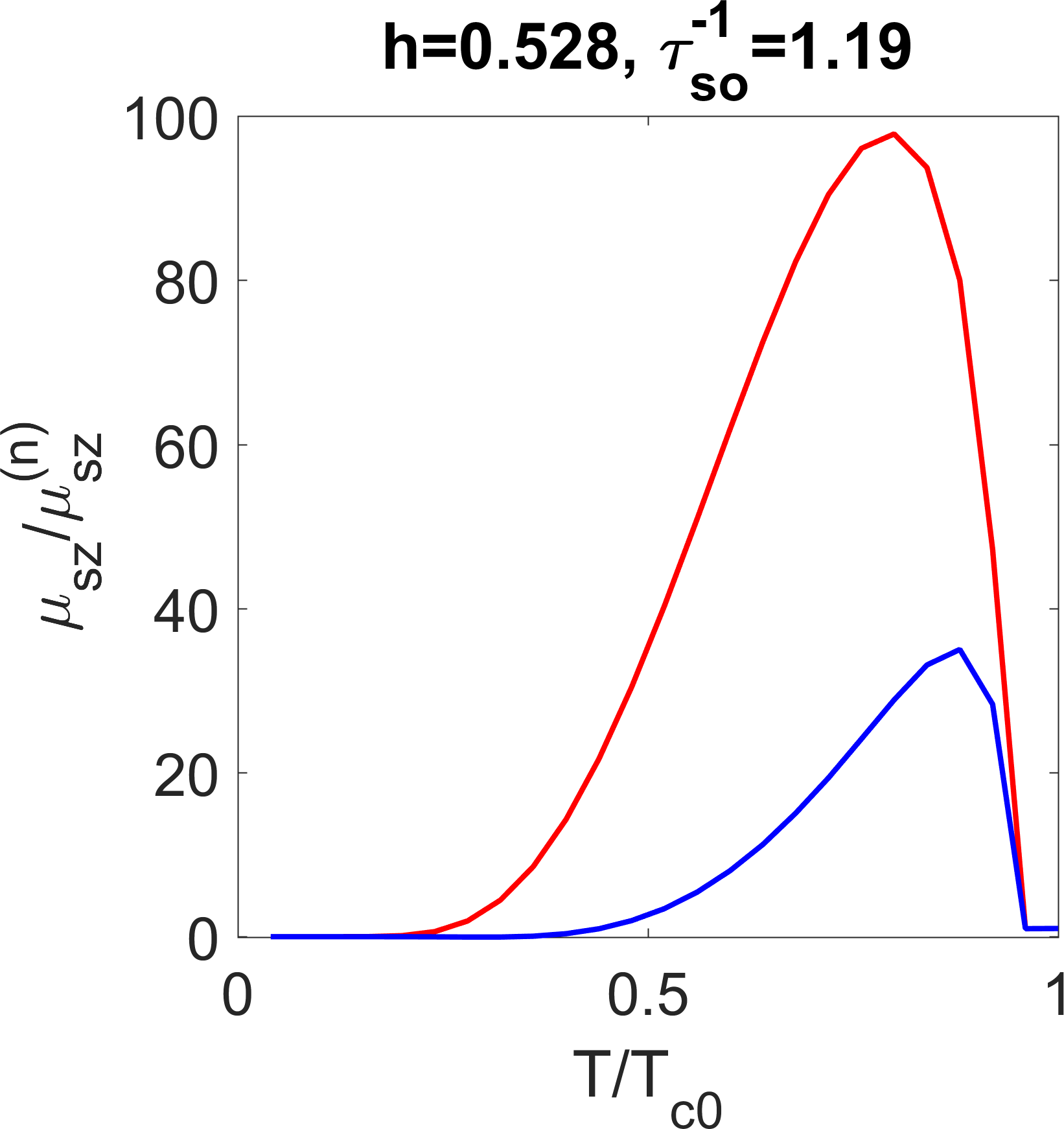} 
 \includegraphics[width=0.175\linewidth]
 {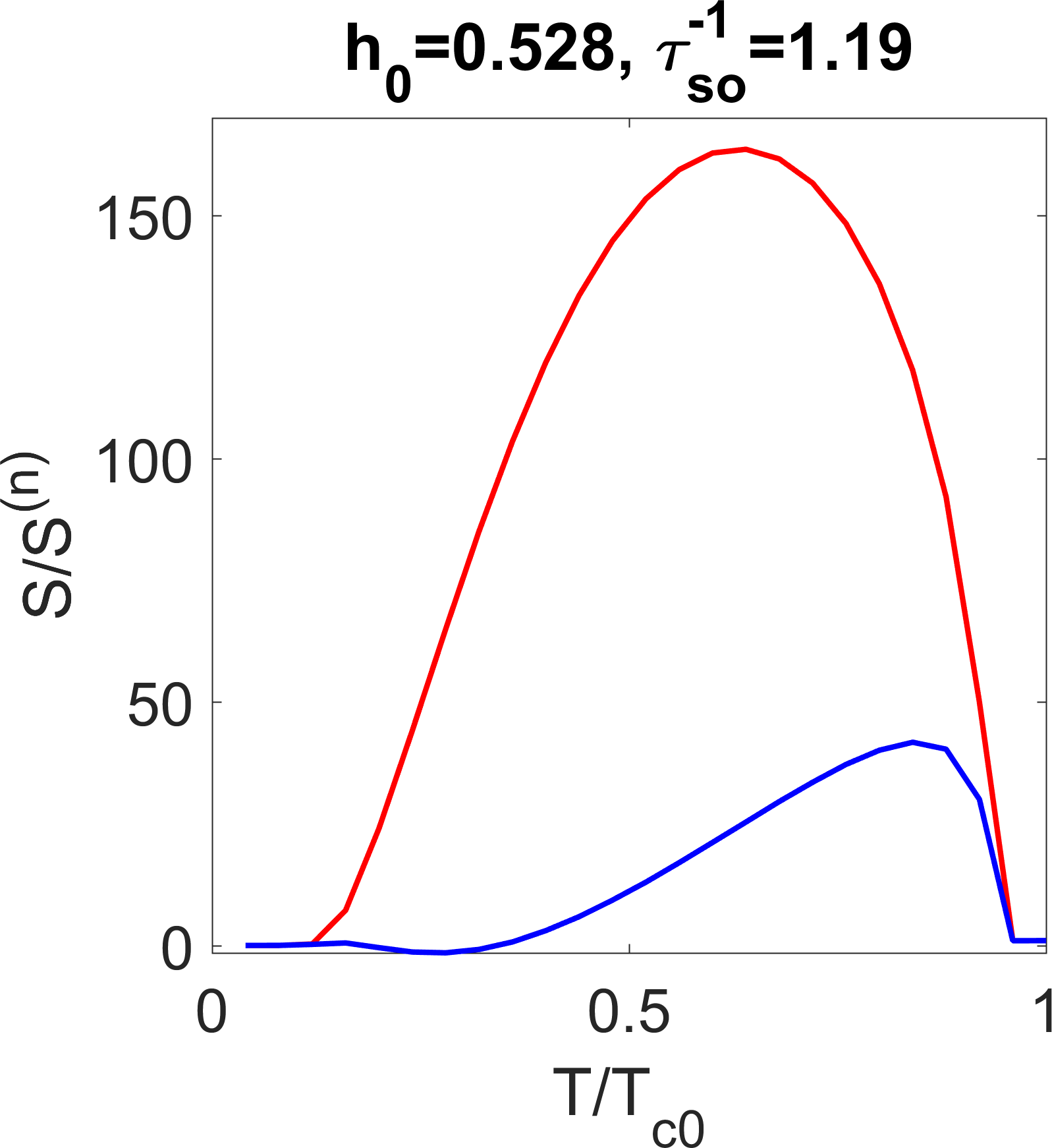} 
   \\
 \includegraphics[width=0.20\linewidth]
 {QhEq500omEq04bEq99Gamma0001AnomalousNormPolL2DT.png} 
 \includegraphics[width=0.20\linewidth]
 {SzhEq500omEq04bEq99Gamma0001AnomalousPolL2DT.png} 
 \includegraphics[width=0.20\linewidth]
 {VDhEq500omEq04bEq99Gamma0001AnomalousPolL2DT.png}
 \includegraphics[width=0.18\linewidth]
 {SzMaghEq500omEq04bEq99Gamma0001AnomalousPolL.png} 
 \includegraphics[width=0.175\linewidth]
 {VDMaghEq500omEq04bEq99Gamma0001AnomalousPolL.png} 
   \end{array}$}
  \caption{\label{Fig:EnergyPolarWT-h-scan}
(1st column): Pumped energy of electronic system $ W (T,\Omega)/ W (T_c,\Omega)$.
(2nd column): Pumped spin accumulation $( T_{c0}/h_\Omega^2)  \mu_z(T,\Omega)$. 
(3rd column): Non-local voltage generated by pumped spin accumulation $ (e T_{c0}/h_\Omega^2)V(\Omega,T)$. 
(4th column): Magnon-induced spin accumulation. 
(5th column):  nonlocal Seebeck coefficient in FI/SC/FM bilayer.
 Parameters are $(\tau_{so}T_{c0})^{-1} =1.19$, energy relaxation rate $\Gamma /T_{c0} =10^{-3}$.  We consider circular polarization  $h_{l,\Omega},h_{r,-\Omega} \neq 0$.  
Scan over $\Omega$, $T$, different values of Zeeman splitting. 
  }
 \end{figure*}

  Using the scheme described above we calculate 
  $\mu_{sz}(\Omega,T)$, $W(\Omega,T)$ and $V(\Omega,T)$ in the wide range of parameters. The series of calculation results for varying $h_0$ and $\tau_s$
  are shown in Figs.~\ref{Fig:EnergyPolarWT-so-scan}, \ref{Fig:EnergyPolarWT-h-scan}.  

\subsection{Numerical higher-order calculation in thin-film limit}
\label{SMSec:HigherOrder}

Numerical calculations can be also performed beyond low-order perturbation
theory. The Usadel equation in the thin-film limit can be written as
\begin{align}
  [\Omega(g), g] = 0
  \,,
  g^2 = 1
  \,,
\end{align}
where $\Omega=\partial_t\delta(t-t')+X$ where $X$ is the operator on the r.h.s. of Eq.~\eqref{SMEq:KeldyshUsadelT}. 
This is formally solved by
\begin{align}
  \label{eq:g-sgn-omega}
  g = \sgn(\Omega(g))
  \,,
\end{align}
where $\sgn$ is the sign function, defined as the analytic continuation $\sgn{}z=\sgn\Re z$ of $\sgn$ from real axis to complex plane, 
so it extends to an operator-valued function. 
For finite matrices, it can be
defined via the eigenvalue decomposition
$X=V\mathrm{diag}(\lambda_1,\ldots,\lambda_n)V^{-1}$ as
$\sgn(X) =
V\mathrm{diag}(\sgn\Re\lambda_1,\ldots,\sgn\Re\lambda_n)V^{-1}$.

To deal with the time convolutions, for periodic forces
$\vec{h}(t)=\vec{h}(t+2\pi\Omega^{-1})$, we can make a
Green function Floquet Ansatz,
\begin{align}
  g(t,t')
  &=
  \int_{-\infty}^\infty\frac{d\omega\,d\omega'}{4\pi^2}
  e^{-i\omega t + i\omega' t'} g(\omega,\omega')
  \,,
  \\
  g(\omega,\omega')
  &=
  \sum_k g_{0,k}(\omega) 2\pi\delta(\omega - \omega' + k\Omega)
  \,,
  \\
  g_{m,n}(\omega) &= g_{0,n-m}(\omega+m\Omega)
  \,.
\end{align}
One can now check that $(A\circ{}B)_{m,n}(\omega)=\sum_kA_{m,k}(\omega)B_{k,n}(\omega)$.
Moreover, $(\epsilon)_{m,n}(E)=(E + n\omega)\delta_{m,n}$,
and $(\vec{h})_{m,n}(E)=\int dt\,e^{-i(n-m)\Omega t}\vec{h}(t)$.
We take
\begin{align}
  \vec{h}
  &=
  h_0\hat{z}
  +
  \Re[h_{ac,x}e^{i\Omega t}]\hat{x}
  +
  \Re[h_{ac,y}e^{i\Omega t}]\hat{y}
  \,,
  \\
  \vec{h}_{m,n}(E)
  &=
  h_0\hat{z}\delta_{m,n}
  +
  \hat{x}\frac{1}{2}(h_{ac,x}\delta_{m,n+1} + h_{ac,x}^*\delta_{m,n-1})
  \\
  \notag
  &
  +
  \hat{y}\frac{1}{2}(h_{ac,y}\delta_{m,n+1} + h_{ac,y}^*\delta_{m,n-1})
  \,.
\end{align}
Hence, $\Omega\mapsto{}(\Omega)_{m,n}(E)$, and Eq.~\eqref{eq:g-sgn-omega}
becomes a matrix equation. The matrix size is infinite, but when the time-dependent
perturbations are not too large, when solving for $g_{0,0}(E)$ we can limit
the equations to $g_{m,n}(E)$, $|m|,|n|\le{}N$ for some cutoff $N$.
Second-order perturbation theory corresponds to $N=1$.
The iteration~\eqref{eq:g-sgn-omega} is reasonably
convergent, and can be solved numerically in a straightforward way
also for large $N$. However, if $\tau_{so}$ is small, Newton method
is preferable.

\begin{figure}[htb!]
  \centering
  \begin{minipage}[c]{0.5\linewidth}
  \includegraphics[width=\textwidth]
 {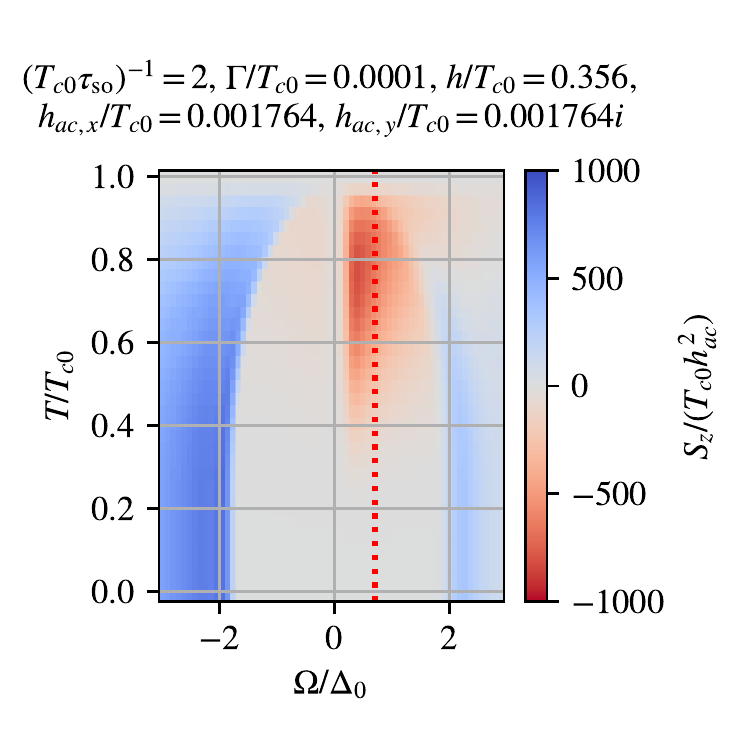} 
 \end{minipage}
 \begin{minipage}[c]{0.45\linewidth}
 \includegraphics[width=\textwidth]
 {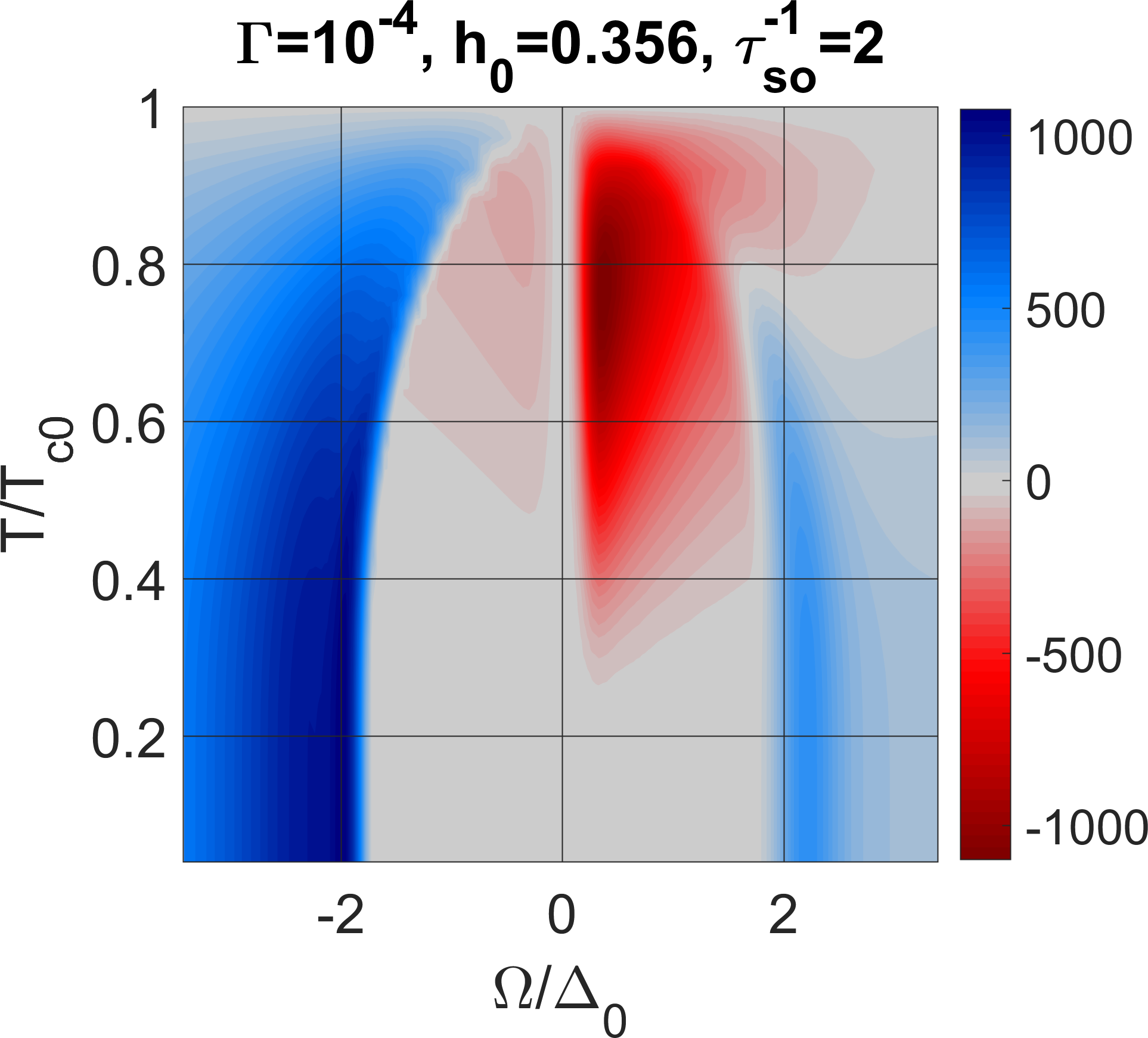} 
 \end{minipage}
 \newline
  \begin{minipage}[c]{0.5\linewidth}
  \includegraphics[width=\textwidth]
 {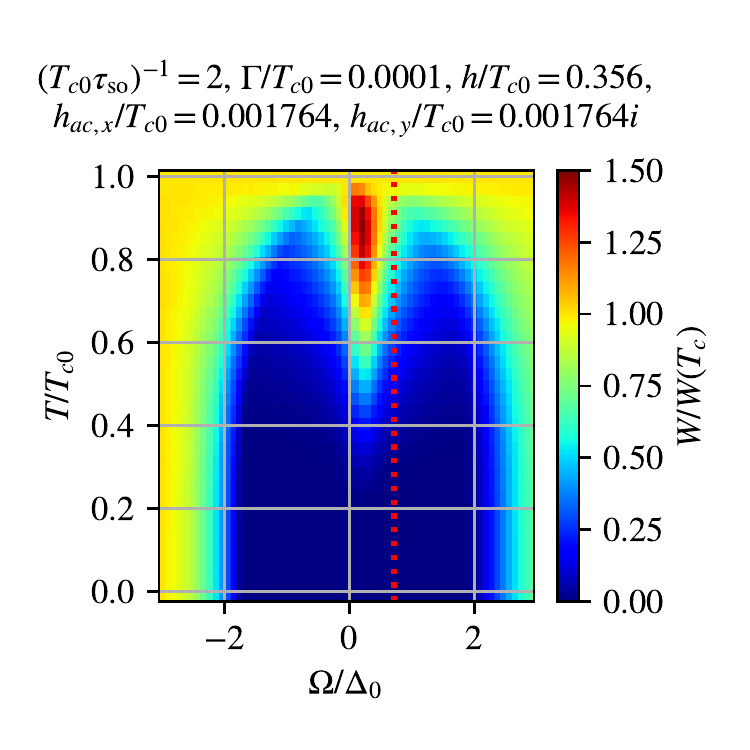} 
 \end{minipage}
  \begin{minipage}[c]{0.45\linewidth}
 \includegraphics[width=\textwidth]
 {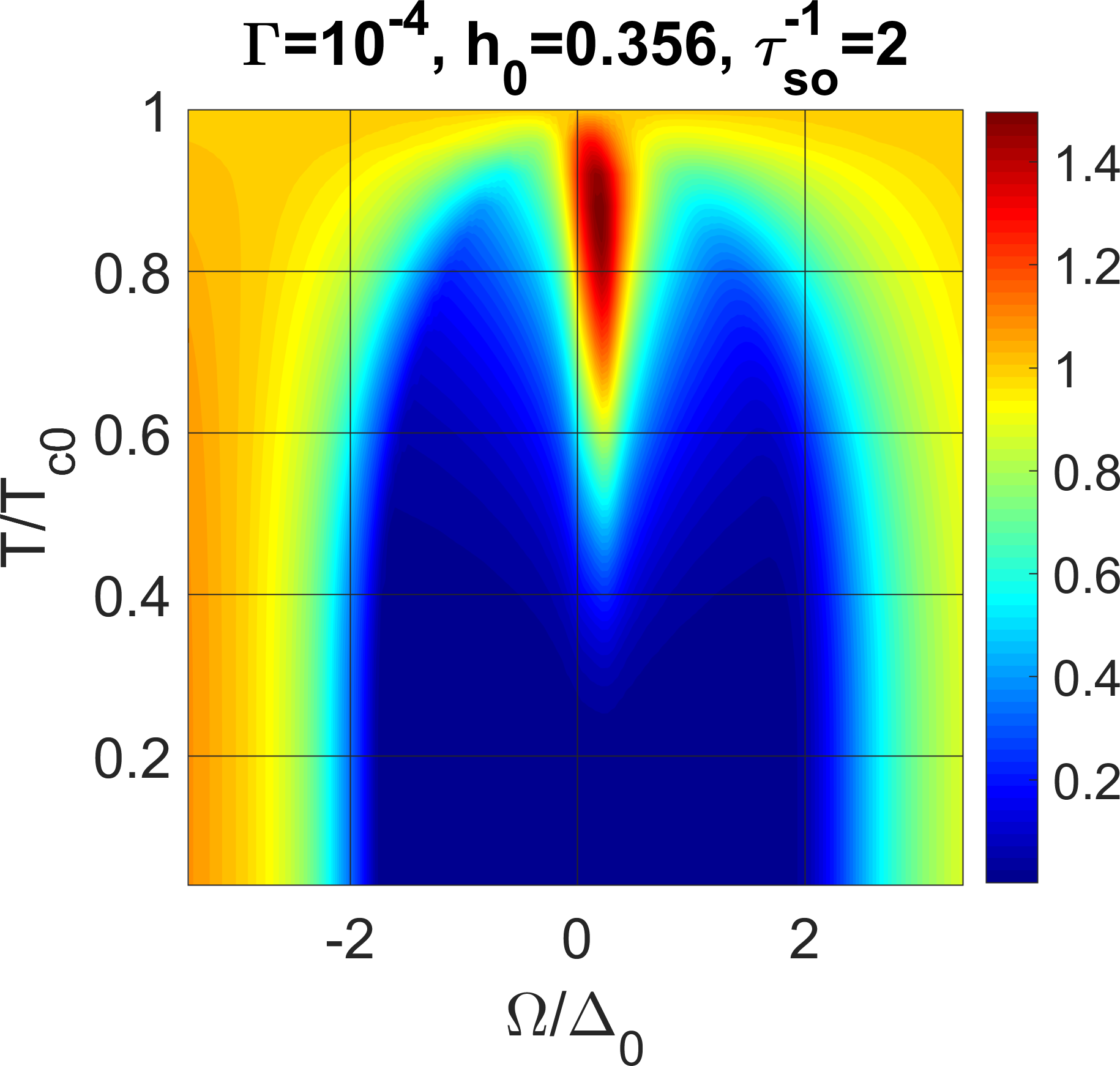} 
  \end{minipage}
 \caption{\label{Fig:jsQcomparison}
 Comparison of energy and spin accumulation calculated from the numerically exact solution (left panels) and second-order perturbation theory (right panels).
  }
 \end{figure}

The results are compared with the perturbation calculation of the previous section in Fig.~\ref{Fig:jsQcomparison}. For the small excitation amplitude chosen here, results coincide.

\subsection{Spin pumping by the stochastic field of magnons }
\label{SMSec:MagnonDerivation}

Here we show how to derive the electron-magnon collision integral in the case of a spin-split superconductor. The approach is relatively standard \cite{adachi2013theory,kamenev2011field}, but the details related with superconductivity are new. We present it here for the convenience of the reader.

    We can quantize time-dependent components of magnetization by introducing the bosonic creation and annihilation operators 
    \begin{align} \label{SMEQ:Hem}
    & H_{e-m} = \frac{1}{\sqrt{S_0}} \sum_{p,\bar p} 
     \Theta_{p,\bar p} \hat a_p^\dagger 
     \hat{\bm m} \hat a_{\bar p} 
     \\
   & \hat{\bm m} =  
     (\hat b \hat \sigma_+ + \hat b^\dagger \hat \sigma_-)\hat\tau_3,
    \end{align} 
 where $S_0$ is the value of the localized spin, the matrix $\Theta_{p,\bar p}$ describes spin-dependent scattering 
 at the surface and $\hat b$, $\hat b^\dagger$
are the magnon field operators at the interface. They are expressed through the  
operators of the magnon modes in the usual way $\hat b (\bm r) =  \sum_k \frac{e^{i\bm k\bm r}}{\sqrt{N_{FI}}} \hat b_k$, where $N_{FI}$ is the number of sites in the FI. 

   To calculate the matrix current driven by the electron-magnon interaction (\ref{SMEQ:Hem}) we 
   calculate the corresponding electron-magnon collision integral. 
   We work in the interaction representation with respect to $H_{em}$, so that the Heisenberg equations are 
     \begin{align} \label{SMEq:EqMotion1}
  & \partial_t \hat a_p = i [\hat H_{e-m}, \hat a_p] =
   - i \Theta_{p,{\bar p}} \hat{\bm m} \hat a_{\bar p} 
   \\ \label{SMEq:EqMotion2}
 & \partial_t \hat a^\dagger_p =
  i [\hat H_{e-m}, \hat a^\dagger_p] =
   i \Theta_{p,{\bar p}}  \hat a^\dagger_{\bar p} \hat{\bm m}
   \end{align}
 The contour-ordered GF is defined as  
  \begin{align} \label{SMEq:EvolutionOperator}
  & \hat G (t_1,t_2,p,p^\prime) = 
  \langle {\cal T}_c S a_p(t_1)a^\dagger_{p^\prime}(t_2) \rangle 
  \\
  &  S = {\cal T}_c \exp ( - i \int_c H_{e-m} dt )
  \end{align}
  The e-m collision integral $\hat J(t_1,t_2)$ is
  given by 
    \begin{align}
  & \hat J (t_1,t_2)  = 
  \partial_{t_1} \hat G (t_1,t_2) + 
  \partial_{t_2} \hat G (t_1,t_2)  
   \end{align}
  
  Using equations of motion 
  (\ref{SMEq:EqMotion1},\ref{SMEq:EqMotion2})
   and expanding the S-matrix $  S \approx 1- i \int_c  \hat H_{e-m} (t) dt $  we get
      \begin{align} 
  & - \partial_{t_1} \hat G (t_1,t_2) = 
  \\ \nonumber
  &   -  \langle {\cal T}_c  \int_c dt \hat H_{e-m} (t) 
    \hat{\bm m} (t_1) \hat\Theta_{\bar p\bar p^\prime} 
    \hat a_{\bar p^\prime}(t_1)
   \hat a^\dagger_{p}(t_2)
   \rangle
   =
   \\  \nonumber
 &  \langle {\cal T}_c  \int_c dt 
    \hat a_{p_1}^\dagger (t) 
      \hat{\bm m} (t) 
      \hat\Theta  
      \hat a_{\bar p_1} (t) 
     \hat{\bm m} (t_1) 
     \hat\Theta      
     \hat a_{\bar p^\prime}(t_1)
   a^\dagger_{p}(t_2)
   \rangle 
   = 
   \\ \nonumber
   & S_0^{-1}\langle {\cal T}_c  \int_c dt 
    \hat a_{p_1}^\dagger (t) \hat\sigma_+ 
    \hat\Theta  
    \hat a_{\bar p_1} (t) 
    \hat\sigma_- 
    \hat\Theta 
    \hat a_{\bar p^\prime}(t_1)
   a^\dagger_{p}(t_2) \hat b(t)\hat b^\dagger(t_1)
   \rangle + 
    \\ \nonumber
   & S_0^{-1} \langle {\cal T}_c  \int_c dt 
    \hat a_{p_1}^\dagger (t) \hat\sigma_- 
    \hat\Theta  
    \hat a_{\bar p_1} (t) 
    \hat\sigma_+ 
    \hat\Theta
    \hat a_{\bar p^\prime}(t_1)
   a^\dagger_{p}(t_2) \hat b^\dagger(t)\hat b(t_1)
   \rangle 
    = 
     \\ \nonumber
   &      \int_c dt
   \Sigma (t_1,t, \bar p^\prime, \bar p_1)
   G (t,t_2,\bar p_1, p).  
   \end{align}
  Here the self-energy is 
  \begin{align} 
  &S_0 \Sigma (t_1,t)= 
  \\ \nonumber
  &  \hat\sigma_- \hat\Theta G(t_1,t)
   \hat\Theta \hat\sigma_+  D^{(l)} (t,t_1)
   + 
    \hat\sigma_+ \hat\Theta 
   G(t_1,t)\hat\Theta
   \hat\sigma_-  D^{(r)} (t,t_1) 
  \end{align}
  where the left- and right-hand polarized magnon propagators are 
  \begin{align}
  & D^{(l)} (t,t_1) =\langle {\cal T}_c 
  \hat b(t) \hat b^\dagger (t_1) \rangle 
   \\
   & D^{(r)} (t,t_1) = D^{(l)} (t_1,t)
    \end{align}    
  
Differentiating with respect to the second time variable gives
  \begin{align} \nonumber
  & \partial_{t_2} \hat G (t_1,t_2) = 
    \langle {\cal T}_c  \int_c dt \hat H_{e-m} (t) 
   \hat a_{\bar p}(t_1) 
   a^\dagger_{p}(t_2) \hat\Theta\hat{\bm m} (t_2)
   \rangle
   =
   \\ \nonumber
 &     \langle {\cal T}_c  \int_c dt \hat a_{p_1}^\dagger (t)     \hat{\bm m} (t) \hat \Theta
      \hat a_{\bar p_1} (t) 
     \hat a_{\bar p}(t_1) 
   a^\dagger_{p}(t_2) \hat\Theta\hat{\bm m} (t_2)
   \rangle 
   = 
   \\ \nonumber
   &   S_0^{-1}  \langle {\cal T}_c  \int_c dt 
    \hat a_{\bar p} (t_1) 
    a^\dagger_{p_1}(t)
    \hat\sigma_+\hat \Theta
    \hat a_{\bar p_1} (t) 
     \hat a_{p}(t_2) \hat\sigma_-\hat \Theta
    \hat b(t)\hat b^\dagger(t_2)
   \rangle 
   +
   \\ \nonumber
  & S_0^{-1}  \langle {\cal T}_c  \int_c dt 
    \hat a_{\bar p} (t_1) 
    a^\dagger_{p_1}(t)
    \hat\sigma_-\hat \Theta
    \hat a_{\bar p_1} (t) 
     \hat a_{p}(t_2)\hat \Theta \hat\sigma_+
    \hat b(t)\hat b^\dagger(t_2)
   \rangle 
    = 
   \\ \nonumber
   & 
    \int_c dt
    G (t_1,t,\bar p, p_1)
   \Sigma (t,t_2, \bar p_1, p). 
   \end{align} 
  
  Thus the matrix double-time current which can be considered as the electron-magnon collision integral (CI) is given by  
 \begin{align} \label{SMEq:CImagnon}
 \hat J (t_1,t_2) = 
 \hat G \circ \Sigma - \Sigma \circ \hat G. 
 \end{align}
  The self-energy can be represented as the sum of 
  the parts associated with right- and left-hand polarized magnons $\Sigma= \Sigma^{(l)} + \Sigma^{(r)}$.  
  
 Switching to the Keldysh contour, we express the convolution product as matrix components coupling different segments of the contour, i.e.,
  \begin{align}
  &S_0 \Sigma^{(r)}_{11} = \hat\sigma_+\hat \Theta [ G_{11}D^{(r)}_{11} + G_{12}D^{(r)}_{21} ]\hat \Theta\hat\sigma_-
   \\
   &S_0 \Sigma^{(r)}_{22} = 
   \hat\sigma_+ \hat \Theta[G_{22}D^{(r)}_{22} + G_{21}D^{(r)}_{12}]\hat \Theta\hat\sigma_-
   \\
   & S_0 \Sigma^{(r)}_{12} = \hat\sigma_+ \hat \Theta[G_{12}D^{(r)}_{22} + G_{11}D^{(r)}_{12} ]\hat \Theta\hat\sigma_-
   \\
   &S_0 \Sigma^{(r)}_{21} = \hat\sigma_+ \hat \Theta
   [G_{21}D^{(r)}_{11} + G_{22}D^{(r)}_{21}]\hat \Theta\hat\sigma_-.
\end{align}   
 and 
  \begin{align}
  &S_0 \Sigma^{(l)}_{11} = \hat\sigma_-\hat \Theta [ G_{11}D^{(l)}_{11} + G_{12}D^{(l)}_{21} ]\hat \Theta\hat\sigma_+
   \\
   &S_0 \Sigma^{(l)}_{22} = 
   \hat\sigma_- \hat \Theta[G_{22}D^{(l)}_{22} + G_{21}D^{(l)}_{12}]\hat \Theta\hat\sigma_-+
   \\
   &S_0  \Sigma^{(l)}_{12} = \hat\sigma_- \hat \Theta[G_{12}D^{(l)}_{22} + G_{11}D^{(l)}_{12} ]\hat \Theta\hat\sigma_+
   \\
   &S_0 \Sigma^{(l)}_{21} = \hat\sigma_- \hat \Theta
   [G_{21}D^{(r)}_{11} + G_{22}D^{(r)}_{21}]\hat \Theta\hat\sigma_+.
\end{align}

 Next we use the RAK representation \cite{kamenev2011field} $G\to \hat H \hat G \hat H$ 
 which yields
 \begin{align}
 &S_0 \Sigma^R_{r} = 
 \hat\sigma_+ \hat \Theta[ D^R_rG^K + D^K_r G^R ]
 \hat \Theta\hat\sigma_- /2
 \\
  &S_0 \Sigma^A_{r} = 
  \hat\sigma_+ \hat \Theta [ D^A_rG^K + D^K_r G^A ]\hat \Theta \hat\sigma_-/2
  \\
  &S_0 \Sigma^K_{r} = 
  \hat\sigma_+ \hat \Theta [ D^K_rG^K + (D^R_r - D^A_r)(G^R-G^A) ] \hat \Theta
  \hat\sigma_-/2
 \end{align}
 
 \begin{align}
 &S_0 \Sigma^R_{l} = 
 \hat\sigma_- \hat \Theta[ D^R_l G^K + D^K_l G^R ]\hat \Theta\hat\sigma_+ /2
 \\
  & S_0\Sigma^A_{l} = 
  \hat\sigma_- \hat \Theta[ D^A_lG^K + D^K_l G^A ]\hat \Theta \hat\sigma_+/2
  \\
  & S_0\Sigma^K_{l} = 
  \hat\sigma_-\hat \Theta [ D^K_lG^K + (D^R_l - D^A_l)(G^R-G^A) ]\hat \Theta
  \hat\sigma_+/2
 \end{align}
  The left- and right- handed magnon propagators are 
  $D^{R/A/K}_{l,r} = \sum_k  D^{R/A/K}_{l,r} (\Omega, \omega_k)$
  where
  \begin{align}
 & D^{R/A/K}_l (\Omega, \omega_k)=  D^{R/A/K}_r (\Omega, -\omega_k) 
  \\
 & D^K_{r/l} = (D^R_{r/l} - D^A_{r/l}) n_B(\Omega/T_m),
  \end{align}
  where 
  \begin{align}
  D^{R}_r  = (D^{A}_r )^* = \frac{1}{(\Omega+i\alpha)-\omega_k}.
  \end{align}
 
 The dc part of the collision integral (\ref{SMEq:CImagnon}) is given by the sum of two terms corresponding to the left- and right-handed magnons $\hat J^K = \hat J_l + \hat J_r $ where e.g.
 \begin{align} \label{SMEq:JrFull}
 &S_0 \hat J_r = 
  \langle [\hat \Theta \hat\sigma_+ 
  \hat G \hat\Theta \hat \sigma_- ,
  \hat G] ^K (12) \rangle D^K_r  
  + 
 \\  \nonumber
 & \hat\sigma_+ \hat \Theta G^K(1) 
 \hat\sigma_- \hat \Theta G^K (2)D^R_r  
    +
    \\ \nonumber 
   & \hat\sigma_+ \hat \Theta G^{RA} (1)
   \hat\sigma_-\hat \Theta G^A (2)D^{RA}_r
  -
  \\ \nonumber
  & 
  G^K (1)
 \hat\sigma_+ \hat \Theta G^K (2)\hat\sigma_- 
 \hat \Theta D^R_r
  -
  \\ \nonumber
 &  G^R (1)
 \hat\sigma_+\hat \Theta G^{RA} (2)\hat\sigma_- 
 \hat \Theta D^{RA}_r  
 \end{align}
  where the angular brackets $\langle..\rangle$
  denote the average by momentum and disorder.

 Since only the first term in Eq.~(\ref{SMEq:JrFull}) depends on the distribution of magnons and the collision integral is zero in equilibrium we can write it as  
 $\hat J (\varepsilon,\Omega)=  \hat J_r +\hat J_l$ where
  \begin{align} \label{Eq:MagnonDrivenJr}
 &S_0 \hat J_r (\varepsilon, \Omega) =
  \hat\chi_{rl} (\varepsilon, \Omega)
  \delta D^{K}_r (\Omega)
   \\ \label{Eq:MagnonDrivenJl}
  &S_0 \hat J_l (\varepsilon, \Omega) = 
  \hat\chi_{lr} (\varepsilon, \Omega) 
  \delta D^{K}_l(\Omega).
    \end{align}  
    Here we denote the
 response functions 
 \begin{align}
 & \hat \chi_{rl} (\varepsilon, \Omega) = 
  \langle [\hat \Theta \hat\sigma_+ 
  \hat G \hat\Theta \hat \sigma_- ,
  \hat G] ^K (12) \rangle 
 \\
 & \hat \chi_{lr} (\varepsilon, \Omega) = 
 \langle [\hat \Theta \hat\sigma_- 
  \hat G \hat\Theta \hat \sigma_+ ,
  \hat G] ^K (12) \rangle ,
 \end{align}     
    where
    $G(1)=G(\varepsilon)$, $G(2)=G(\varepsilon+\Omega)$ 
  and $\delta D^{K}_{r/l}$ are the non-equilibrium 
  parts of magnon Keldysh functions. In the stationary case they can be parametrized by the magnon distribution function
  $\delta D^{K}_{r/l}(\Omega) = D^{RA}_{r/l}(\Omega) \delta f_m (\Omega)$. If the non-equilibrium is determined by the temperature difference between superconductor and magnon subsystem, the distribution function reads 
   $\delta f_m(\Omega) =  \coth (\Omega/2T_m)-\coth (\Omega/2T)$.
  For small damping $\alpha \ll \omega_k$ we can write $D^{RA}_{r} = 2i\delta (\omega_k-\Omega)$   
    We can sum by the magnon states 
  to replace $D^{RA}_{r/l}$ with the density of states for magnons as 
 \begin{align} \label{SMEq:MagnonDOS}
  & \frac{1}{N_{FI}}\sum_k  D^{RA}_{r} (\omega_k) \approx
  \\ \nonumber
  & i v_s  D_m(\Omega) (1 + {\rm Step}( \Omega))/2 
  \end{align}
where $m_M$ is the magnon mass and ${\rm Step}(x)$ is a step function, $v_s= V_{FI}/N_{FI}$ is the volume per spin,
$D_m(\Omega) = m_M^{3/2} |\Omega|^{1/2}$ is the magnon density of states. To avoid extra parameters in the model, we have set the magnon gap to vanish, but it can be easily added if needed.

Using the symmetry relation $D^K_l (\Omega)= D^K_r (- \Omega)$
we can write the total current 
 \begin{align} \label{Eq:TotalCurrent}
  \hat J (\varepsilon) = v_s \int_{-\infty}^{\infty} d\Omega 
   [ \hat\chi_{rl} (\varepsilon, \Omega) + \hat\chi_{lr} (\varepsilon, -\Omega) ] \delta D_r^K (\Omega), 
\end{align}   
   where  
   $v_s = V_{FI}/S_0 N_{FI}$ is the volume per spin in FI. 
   
   The correlators in Eqs.~(\ref{Eq:MagnonDrivenJr}--\ref {Eq:MagnonDrivenJl}) can be determined by calculating the matrix currents generated by the classical time-dependent exchange field 
    \begin{align}
 & \hat J_r^{cl} (\Omega,\varepsilon) = 
  [ \hat\chi_{rl}(\Omega,\varepsilon) +  \hat \chi_{lr}(-\Omega,\varepsilon) ]
  m_{r,\Omega} m_{l,-\Omega}
  \\
  & \hat J_l^{cl}  (\Omega,\varepsilon)= 
  [ \hat \chi_{lr} (\Omega,\varepsilon) +  \hat \chi_{rl} (-\Omega,\varepsilon) ]
  m_{l,\Omega} m_{r,-\Omega}
  \end{align}
  where $m_{r/l,\Omega} = (m_{x,\Omega} \pm im_{y,\Omega})$ are the right-hand and left-hand polarized components. The expressions for $\hat J_{r/l}^{cl}$
  can be found using the quasiclassical equations. Then, to get the magnon-driven current we 
 can replace the classical field amplitudes by the magnon propagators $m_{r,\Omega} m_{l,-\Omega} \to v_s \delta D^{K}_r/\Omega$ and 
 $m_{l,\Omega} m_{r,-\Omega} \to  v_s \delta D^{K}_l/\Omega$.

  Using this general matrix current we can calculate spin and energy currents
    as $ j_e=\nu \int_{-\infty}^{\infty} {\cal J}_e d\varepsilon$ and  $j_{sz} = \nu\int_{-\infty}^{\infty}  {\cal J}_{sz}d\varepsilon$
    where the spectral densities are  
    \begin{align} \label{SMbc:Jsmag}
 {\mathbfcal J}_s (\varepsilon) =  
  \frac{1}{8}  {\rm Tr} 
  [ {\bm \sigma}\hat\tau_3 \hat J (\varepsilon)] 
  \\ \label{SMbc:Jemag}
  {\cal J}_e (\varepsilon) =  \frac{\varepsilon}{4} 
    {\rm Tr} 
  [ \hat\tau_3 \hat J (\varepsilon)] 
  \end{align}
    
    As shown in the next section,  these currents can be in general expressed through the linear spin susceptibility 
    \begin{align}
  & j_{sz} =  \nu J_{sd} \int_{-\infty}^{\infty} 
  {\rm Im}(\chi_l)
  \delta D^{K}_l d\Omega
  \\
    &  j_e = \nu J_{sd}  \int_{-\infty}^{\infty}
    \Omega \;{\rm Im}(\chi_l)
  \delta D^{K}_l d\Omega
 \end{align}
    In the normal state and low-frequency regime we have seen in Sec.~\ref{SMSec:SpinDiffusionNormal} that $ \nu J_{sd}{\rm Im} \chi_l = \Omega {\rm Re} A_{\rm eff}^{\uparrow\downarrow} $. 
    Then taking into account (\ref{SMEq:MagnonDOS})
    \begin{align}
  & j_{sz} =  {\rm Re} A_{\rm eff}^{\uparrow\downarrow} \int_{0}^{\infty} 
  {\rm Im}(\chi_l)
  D_m(\Omega) \Omega  [ n_B(T_m)-  n_B(T)] d\Omega
  \\
    &  j_e = {\rm Re} A_{\rm eff}^{\uparrow\downarrow}  \int_{0}^{\infty}
     {\rm Im}(\chi_l)
  D_m(\Omega) \Omega^2  [ n_B(T_m)-  n_B(T)] d\Omega
 \end{align}
 coincides with that derived in \cite{bender2015interfacial,cornelissen2016magnon}. 
In the normal state the expression coincides with that derived in \cite{bender2015interfacial,cornelissen2016magnon}. 
 
 
  \subsection{Spin-energy pumping: general relations}
  \label{SMsec:SpinEnergyPumping}

For coherent magnetization precession at frequency $\Omega$,  boundary conditions for dc spin and energy currents are
 \begin{align} \label{SMbc:Js}
 {\mathbfcal J}_s (\varepsilon) =  \frac{i J_{sd}}{8}
 {\rm Tr} ( {\bm \sigma}  [ {\bm\sigma\bm m} \hat\tau_3\commacirc \hat g_h]^K )(\varepsilon)
 \\ \label{SMbc:Je}
  {\cal J}_e (\varepsilon) =  \frac{i J_{sd}}{4} 
  \varepsilon  
 {\rm Tr} ( [ {\bm\sigma\bm m} \hat\tau_3\commacirc \hat g_h]^K )(\varepsilon)
  \end{align}

  
   The currents can be written as 
  \begin{align} \label{SMEq:TrSigmaZ}
    {\mathcal J}_{sz} (\varepsilon)
    &= 
     (J_{sd}/2) m_{l,\Omega} m_{r,-\Omega} \times
    \\   \nonumber
  & \{ ( \chi^K_r
  (\varepsilon-\Omega,\varepsilon) 
   -  \chi^K_l (\varepsilon,\varepsilon-\Omega) 
    -
      \\ \nonumber
   &[\chi^K_l
  (\varepsilon+\Omega,\varepsilon) 
   -  \chi^K_r (\varepsilon,\varepsilon+\Omega) ] )
    \}
   ,\\
 \label{SMEq:TrSigma0}
  \mathcal J_{e}(\varepsilon) &= (\varepsilon J_{sd})
     m_{l,\Omega} m_{r,-\Omega} \times 
    \\   \nonumber
   & \{  ( \chi^K_r
  (\varepsilon-\Omega,\varepsilon) 
   -  \chi^K_l (\varepsilon,\varepsilon-\Omega) 
    +
      \\ \nonumber
   &[\chi^K_l
  (\varepsilon+\Omega,\varepsilon) 
   -  \chi^K_r (\varepsilon,\varepsilon+\Omega) ] ),
     \}
     \end{align} 
 where we introduce the linear response functions for the Keldysh GF
  \begin{align}
 & \frac{i}{4}{\rm Tr} [ \hat\sigma_- \hat\tau_3 g^K
  (\varepsilon-\Omega,\varepsilon) ]= m_{r,-\Omega}  \chi^K_r (\varepsilon-\Omega,\varepsilon)  
  \\
 &\frac{i}{4} {\rm Tr} [ \hat\sigma_+\hat\tau_3 g^K
  (\varepsilon+\Omega,\varepsilon)] = m_{l,\Omega}  \chi^K_l (\varepsilon+\Omega,\varepsilon),
\end{align}   
 and where $\hat\sigma_{\pm} = (\hat\sigma_x \pm i\hat\sigma_y )/2$. 
 
The total energy and spin currents are defined as $ j_e= \int {\cal J}_e d\varepsilon$ and  $j_{sz} = \int  {\cal J}_{sz}d\varepsilon$, respectively. They can be written as 
  \begin{align}
 & j_{sz} = \nu J_{sd}
  [\chi_r(-\Omega) - \chi_l(\Omega)] m_{l,\Omega} m_{r,-\Omega} ,
  \\
  & j_{e} = \nu J_{sd}
 \Omega  [\chi_r(-\Omega) - \chi_l(\Omega)] m_{l,\Omega} m_{r,-\Omega},
\end{align}
  where we introduce 
  \begin{align}
  & \chi_l(\Omega) = \int_{-\infty}^{\infty} 
  d\varepsilon \chi^K_l (\varepsilon,\varepsilon-\Omega) ,
   \\
 & \chi_r(\Omega) = \int_{-\infty}^{\infty}
 d\varepsilon 
 \chi^K_r (\varepsilon,\varepsilon-\Omega) .
  \end{align}
 The total spin and energy currents
 satisfy the relation 
 \begin{align}
  j_e = \Omega j_{sz} 
 \end{align}
 
   Using the relation $\chi_r(-\Omega) = \chi^*_l(\Omega)$
 we obtain for the currents 
   \begin{align} \label{SMEq:jsz}
 & j_{sz} = 2 J_{sd} \; {\rm Im }\chi_l(\Omega) 
 m_{l, \Omega} m_{r,-\Omega} ,
  \\ \label{SMEq:je}
 & j_{e} =
2 J_{sd}\; \Omega\; {\rm Im }\chi_l(\Omega) m_{l, \Omega} m_{r,-\Omega} . \end{align}   
    Let us establish the connection between spin current and the Gilbert damping coefficient.
 From the boundary condition \eqref{eq:spin_bc}, the linearized spin current can be written as
\begin{equation}
    j_{sl}(\Omega) = 2i J_{sd}[ \chi_{l}(\Omega) -\chi(0) + \Omega]m_{l,\Omega},
\end{equation}
where $\chi(0)$ is the {\it nonlinear} static spin susceptibility. Spin current can also be parametrized in terms of the damping-like and field-like components as
\begin{equation}
    \bm j(t) = -\alpha (\bm m\times \partial_t\bm m) - (\delta \Omega/\Omega) \partial_t\bm m,\label{eq:LLG_decomposition}
\end{equation}
where $\alpha$ is the Gilbert damping coefficient and $\delta \Omega$ is the FMR frequency shift. Comparing the two expressions for the current, the coefficients in \eqref{eq:LLG_decomposition} can expressed as 
 \begin{align}
 \alpha &=  \frac{{\rm Re}(j_{s l})}{\Omega m_{l, \Omega}}
 = 2 J_{sd} \frac{{\rm Im}(\chi_l) }{\Omega},\\
 \delta \Omega &= \frac{\Im(j_{sl})}{m_{l,\Omega}} = 2 J_{sd} \left\{\Re[\chi_l(\Omega)-\chi(0)] + \Omega\right\}.
 \end{align}
 The last term in the frequency shift drops out when comparing difference in the frequency shift between the superconducting state and the normal state.
 We find that the energy current can be in general written as 
  \begin{align}
 j_e =  \Omega^2 \alpha (\Omega)
  m_{l, \Omega} m_{r,-\Omega}.
 \end{align}   

As shown in Sec.~\ref{SMSec:MagnonDerivation} 
the magnon-driven currents are obtained by replacing the classical field amplitudes by magnon propagators $m_{r,\Omega} m_{l,-\Omega} \to  v_s \delta D^{K}_r$ and 
 $m_{l,\Omega} m_{r,-\Omega} \to v_s \delta D^{K}_l$
 \begin{align}
  & j_{sz} = 2\nu J_{sd} \int_{-\infty}^{\infty} 
  {\rm Im}(\chi_l)
  \delta D^{K}_l d\Omega
  \\
    &  j_e = 2\nu J_{sd}  \int_{-\infty}^{\infty}
    \Omega \;{\rm Im}(\chi_l)
  \delta D^{K}_l d\Omega
 \end{align}
 For thermal magnons with small temperature bias we can write $j_{sz/e} = G_{ms/me}\delta T$, with thermal spin and heat conductances defined as
  \begin{align}
  & G_{ms} = 2\mathcal{V}_S\nu h_0 v_s  m_M^{3/2} \int_0^\infty
     \Omega^{1/2}  {\rm Im} (\chi_{l}) \partial_T n_B d\Omega
  \\
    &  G_{me} = 2\mathcal{V}_S \nu h_0 v_s m_M^{3/2}  \int_0^\infty
       \Omega^{3/2}  {\rm Im} (\chi_{l}) \partial_T n_B    d\Omega
       \,,
 \end{align}
 where $\nu$ is the density of states and $\mathcal{V}_S$
 the volume.
 
 \subsection{Superconductor as an ultrasensitive magnon detector}
 \label{SMsec:detector}
 The giant magnon-induced voltage signal in the superconductor can be used to realize a bolometric or calorimetric magnon detector and an optimized device could reach single-magnon sensitivity down to tens of GHz of magnon frequencies. Assuming unit quantum efficiency, i.e., that the main damping mechanism of magnetization dynamics in the junction is provided by the coupling to the quasiparticles in the superconductor, we can then proceed analogously to the description of the thermoelectric read-out of the dissipated spin signal as in the case of thermoelectric detection of electromagnetic radiation, presented in Refs.~\onlinecite{heikkila2018thermoelectric,chakraborty2018thermoelectric}. In particular, the noise equivalent power NEP can be made of the order of the thermal fluctuation noise due to the heat contacts to the phonons and magnons. The optimum regime is one where the heat conductances $G_{\rm th}$ to both are of the same order of magnitude, in which case the thermal fluctuation noise is given by \cite{heikkila2018thermoelectric}
 \begin{equation}
     NEP_{\rm TFN}^2 = k_B T^2 G_{\rm th} (1+\sqrt{1+ZT_i})^2/ZT_i,
 \end{equation}
 where $ZT_i$ is the intrinsic thermoelectric figure of merit of the junction. For low temperatures and not too large spin polarization $P$ of the superconductor-ferromagnet contact, it is $ZT_i = P^2/(1-P^2)$ \cite{ozaeta2014predicted}. On the other hand, the energy resolution in a calorimetric detection is given by\cite{chakraborty2018thermoelectric}
 \begin{equation}
     \Delta E = NEP \sqrt{\tau_{\rm eff}},
 \end{equation}
 where $\tau_{\rm eff} \approx \tau_{\rm th}$, the thermal relaxation time in the superconductor. For an Al detector at $k_B T \sim 0.2 \Delta$, with a superconductor volume of $10^{-19}$ m$^{-3}$, we would then obtain $NEP \sim 10^{-19}$ W/$\sqrt{\rm Hz}$ and energy resolution enough for single-magnon detection accuracy for magnons with frequency above 200 GHz. On the other hand, with 100 times smaller detector sizes, still within reach of regular sample preparation techniques, the figures of merit could be 10 times smaller, and hence a single-magnon regime could be reached with magnon frequencies above 20 GHz.


%% file: main.bbl
\begin{thebibliography}{83}%
\makeatletter
\providecommand \@ifxundefined [1]{%
 \@ifx{#1\undefined}
}%
\providecommand \@ifnum [1]{%
 \ifnum #1\expandafter \@firstoftwo
 \else \expandafter \@secondoftwo
 \fi
}%
\providecommand \@ifx [1]{%
 \ifx #1\expandafter \@firstoftwo
 \else \expandafter \@secondoftwo
 \fi
}%
\providecommand \natexlab [1]{#1}%
\providecommand \enquote  [1]{``#1''}%
\providecommand \bibnamefont  [1]{#1}%
\providecommand \bibfnamefont [1]{#1}%
\providecommand \citenamefont [1]{#1}%
\providecommand \href@noop [0]{\@secondoftwo}%
\providecommand \href [0]{\begingroup \@sanitize@url \@href}%
\providecommand \@href[1]{\@@startlink{#1}\@@href}%
\providecommand \@@href[1]{\endgroup#1\@@endlink}%
\providecommand \@sanitize@url [0]{\catcode `\\12\catcode `\$12\catcode
  `\&12\catcode `\#12\catcode `\^12\catcode `\_12\catcode `\%12\relax}%
\providecommand \@@startlink[1]{}%
\providecommand \@@endlink[0]{}%
\providecommand \url  [0]{\begingroup\@sanitize@url \@url }%
\providecommand \@url [1]{\endgroup\@href {#1}{\urlprefix }}%
\providecommand \urlprefix  [0]{URL }%
\providecommand \Eprint [0]{\href }%
\providecommand \doibase [0]{https://doi.org/}%
\providecommand \selectlanguage [0]{\@gobble}%
\providecommand \bibinfo  [0]{\@secondoftwo}%
\providecommand \bibfield  [0]{\@secondoftwo}%
\providecommand \translation [1]{[#1]}%
\providecommand \BibitemOpen [0]{}%
\providecommand \bibitemStop [0]{}%
\providecommand \bibitemNoStop [0]{.\EOS\space}%
\providecommand \EOS [0]{\spacefactor3000\relax}%
\providecommand \BibitemShut  [1]{\csname bibitem#1\endcsname}%
\let\auto@bib@innerbib\@empty
\bibitem [{\citenamefont {Wolf}\ \emph {et~al.}(2001)\citenamefont {Wolf},
  \citenamefont {Awschalom}, \citenamefont {Buhrman}, \citenamefont {Daughton},
  \citenamefont {von Moln{\'a}r}, \citenamefont {Roukes}, \citenamefont
  {Chtchelkanova},\ and\ \citenamefont {Treger}}]{wolf2001spintronics}%
  \BibitemOpen
  \bibfield  {author} {\bibinfo {author} {\bibfnamefont {S.}~\bibnamefont
  {Wolf}}, \bibinfo {author} {\bibfnamefont {D.}~\bibnamefont {Awschalom}},
  \bibinfo {author} {\bibfnamefont {R.}~\bibnamefont {Buhrman}}, \bibinfo
  {author} {\bibfnamefont {J.}~\bibnamefont {Daughton}}, \bibinfo {author}
  {\bibfnamefont {v.~S.}\ \bibnamefont {von Moln{\'a}r}}, \bibinfo {author}
  {\bibfnamefont {M.}~\bibnamefont {Roukes}}, \bibinfo {author} {\bibfnamefont
  {A.~Y.}\ \bibnamefont {Chtchelkanova}},\ and\ \bibinfo {author}
  {\bibfnamefont {D.}~\bibnamefont {Treger}},\ }\href@noop {} {\bibfield
  {journal} {\bibinfo  {journal} {Science}\ }\textbf {\bibinfo {volume}
  {294}},\ \bibinfo {pages} {1488} (\bibinfo {year} {2001})}\BibitemShut
  {NoStop}%
\bibitem [{\citenamefont {{\v{Z}}uti{\'c}}\ \emph {et~al.}(2004)\citenamefont
  {{\v{Z}}uti{\'c}}, \citenamefont {Fabian},\ and\ \citenamefont
  {Sarma}}]{vzutic2004spintronics}%
  \BibitemOpen
  \bibfield  {author} {\bibinfo {author} {\bibfnamefont {I.}~\bibnamefont
  {{\v{Z}}uti{\'c}}}, \bibinfo {author} {\bibfnamefont {J.}~\bibnamefont
  {Fabian}},\ and\ \bibinfo {author} {\bibfnamefont {S.~D.}\ \bibnamefont
  {Sarma}},\ }\href@noop {} {\bibfield  {journal} {\bibinfo  {journal} {Rev.
  Mod. Phys.}\ }\textbf {\bibinfo {volume} {76}},\ \bibinfo {pages} {323}
  (\bibinfo {year} {2004})}\BibitemShut {NoStop}%
\bibitem [{\citenamefont {Bauer}\ \emph {et~al.}(2012)\citenamefont {Bauer},
  \citenamefont {Saitoh},\ and\ \citenamefont {van Wees}}]{Bauer2012a}%
  \BibitemOpen
  \bibfield  {author} {\bibinfo {author} {\bibfnamefont {G.~E.~W.}\
  \bibnamefont {Bauer}}, \bibinfo {author} {\bibfnamefont {E.}~\bibnamefont
  {Saitoh}},\ and\ \bibinfo {author} {\bibfnamefont {B.~J.}\ \bibnamefont {van
  Wees}},\ }\href {http://dx.doi.org/10.1038/nmat3301} {\bibfield  {journal}
  {\bibinfo  {journal} {Nat. Mater.}\ }\textbf {\bibinfo {volume} {11}},\
  \bibinfo {pages} {391} (\bibinfo {year} {2012})}\BibitemShut {NoStop}%
\bibitem [{\citenamefont {Tserkovnyak}\ \emph {et~al.}(2002)\citenamefont
  {Tserkovnyak}, \citenamefont {Brataas},\ and\ \citenamefont
  {Bauer}}]{tserkovnyak2002enhanced}%
  \BibitemOpen
  \bibfield  {author} {\bibinfo {author} {\bibfnamefont {Y.}~\bibnamefont
  {Tserkovnyak}}, \bibinfo {author} {\bibfnamefont {A.}~\bibnamefont
  {Brataas}},\ and\ \bibinfo {author} {\bibfnamefont {G.~E.}\ \bibnamefont
  {Bauer}},\ }\href@noop {} {\bibfield  {journal} {\bibinfo  {journal} {Phys.
  Rev. Lett.}\ }\textbf {\bibinfo {volume} {88}},\ \bibinfo {pages} {117601}
  (\bibinfo {year} {2002})}\BibitemShut {NoStop}%
\bibitem [{\citenamefont {Brataas}\ \emph {et~al.}(2002)\citenamefont
  {Brataas}, \citenamefont {Tserkovnyak}, \citenamefont {Bauer},\ and\
  \citenamefont {Halperin}}]{brataas2002spin}%
  \BibitemOpen
  \bibfield  {author} {\bibinfo {author} {\bibfnamefont {A.}~\bibnamefont
  {Brataas}}, \bibinfo {author} {\bibfnamefont {Y.}~\bibnamefont
  {Tserkovnyak}}, \bibinfo {author} {\bibfnamefont {G.~E.}\ \bibnamefont
  {Bauer}},\ and\ \bibinfo {author} {\bibfnamefont {B.~I.}\ \bibnamefont
  {Halperin}},\ }\href@noop {} {\bibfield  {journal} {\bibinfo  {journal}
  {Phys. Rev. B}\ }\textbf {\bibinfo {volume} {66}},\ \bibinfo {pages} {060404}
  (\bibinfo {year} {2002})}\BibitemShut {NoStop}%
\bibitem [{\citenamefont {Tserkovnyak}\ \emph {et~al.}(2005)\citenamefont
  {Tserkovnyak}, \citenamefont {Brataas}, \citenamefont {Bauer},\ and\
  \citenamefont {Halperin}}]{RevModPhys.77.1375}%
  \BibitemOpen
  \bibfield  {author} {\bibinfo {author} {\bibfnamefont {Y.}~\bibnamefont
  {Tserkovnyak}}, \bibinfo {author} {\bibfnamefont {A.}~\bibnamefont
  {Brataas}}, \bibinfo {author} {\bibfnamefont {G.~E.~W.}\ \bibnamefont
  {Bauer}},\ and\ \bibinfo {author} {\bibfnamefont {B.~I.}\ \bibnamefont
  {Halperin}},\ }\href {https://doi.org/10.1103/RevModPhys.77.1375} {\bibfield
  {journal} {\bibinfo  {journal} {Rev. Mod. Phys.}\ }\textbf {\bibinfo {volume}
  {77}},\ \bibinfo {pages} {1375} (\bibinfo {year} {2005})}\BibitemShut
  {NoStop}%
\bibitem [{\citenamefont {Nakayama}\ \emph {et~al.}(2013)\citenamefont
  {Nakayama}, \citenamefont {Althammer}, \citenamefont {Chen}, \citenamefont
  {Uchida}, \citenamefont {Kajiwara}, \citenamefont {Kikuchi}, \citenamefont
  {Ohtani}, \citenamefont {Gepr{\"a}gs}, \citenamefont {Opel}, \citenamefont
  {Takahashi} \emph {et~al.}}]{nakayama2013spin}%
  \BibitemOpen
  \bibfield  {author} {\bibinfo {author} {\bibfnamefont {H.}~\bibnamefont
  {Nakayama}}, \bibinfo {author} {\bibfnamefont {M.}~\bibnamefont {Althammer}},
  \bibinfo {author} {\bibfnamefont {Y.-T.}\ \bibnamefont {Chen}}, \bibinfo
  {author} {\bibfnamefont {K.-i.}\ \bibnamefont {Uchida}}, \bibinfo {author}
  {\bibfnamefont {Y.}~\bibnamefont {Kajiwara}}, \bibinfo {author}
  {\bibfnamefont {D.}~\bibnamefont {Kikuchi}}, \bibinfo {author} {\bibfnamefont
  {T.}~\bibnamefont {Ohtani}}, \bibinfo {author} {\bibfnamefont
  {S.}~\bibnamefont {Gepr{\"a}gs}}, \bibinfo {author} {\bibfnamefont
  {M.}~\bibnamefont {Opel}}, \bibinfo {author} {\bibfnamefont {S.}~\bibnamefont
  {Takahashi}}, \emph {et~al.},\ }\href@noop {} {\bibfield  {journal} {\bibinfo
   {journal} {Phys. Rev. Lett.}\ }\textbf {\bibinfo {volume} {110}},\ \bibinfo
  {pages} {206601} (\bibinfo {year} {2013})}\BibitemShut {NoStop}%
\bibitem [{\citenamefont {Weiler}\ \emph {et~al.}(2013)\citenamefont {Weiler},
  \citenamefont {Althammer}, \citenamefont {Schreier}, \citenamefont {Lotze},
  \citenamefont {Pernpeintner}, \citenamefont {Meyer}, \citenamefont {Huebl},
  \citenamefont {Gross}, \citenamefont {Kamra}, \citenamefont {Xiao} \emph
  {et~al.}}]{weiler2013experimental}%
  \BibitemOpen
  \bibfield  {author} {\bibinfo {author} {\bibfnamefont {M.}~\bibnamefont
  {Weiler}}, \bibinfo {author} {\bibfnamefont {M.}~\bibnamefont {Althammer}},
  \bibinfo {author} {\bibfnamefont {M.}~\bibnamefont {Schreier}}, \bibinfo
  {author} {\bibfnamefont {J.}~\bibnamefont {Lotze}}, \bibinfo {author}
  {\bibfnamefont {M.}~\bibnamefont {Pernpeintner}}, \bibinfo {author}
  {\bibfnamefont {S.}~\bibnamefont {Meyer}}, \bibinfo {author} {\bibfnamefont
  {H.}~\bibnamefont {Huebl}}, \bibinfo {author} {\bibfnamefont
  {R.}~\bibnamefont {Gross}}, \bibinfo {author} {\bibfnamefont
  {A.}~\bibnamefont {Kamra}}, \bibinfo {author} {\bibfnamefont
  {J.}~\bibnamefont {Xiao}}, \emph {et~al.},\ }\href@noop {} {\bibfield
  {journal} {\bibinfo  {journal} {Phys. Rev. Lett.}\ }\textbf {\bibinfo
  {volume} {111}},\ \bibinfo {pages} {176601} (\bibinfo {year}
  {2013})}\BibitemShut {NoStop}%
\bibitem [{\citenamefont {Uchida}\ \emph {et~al.}(2010)\citenamefont {Uchida},
  \citenamefont {Xiao}, \citenamefont {Adachi}, \citenamefont {Ohe},
  \citenamefont {Takahashi}, \citenamefont {Ieda}, \citenamefont {Ota},
  \citenamefont {Kajiwara}, \citenamefont {Umezawa}, \citenamefont {Kawai},
  \citenamefont {Bauer}, \citenamefont {Maekawa},\ and\ \citenamefont
  {Saitoh}}]{Uchida2010}%
  \BibitemOpen
  \bibfield  {author} {\bibinfo {author} {\bibfnamefont {K.}~\bibnamefont
  {Uchida}}, \bibinfo {author} {\bibfnamefont {J.}~\bibnamefont {Xiao}},
  \bibinfo {author} {\bibfnamefont {H.}~\bibnamefont {Adachi}}, \bibinfo
  {author} {\bibfnamefont {J.}~\bibnamefont {Ohe}}, \bibinfo {author}
  {\bibfnamefont {S.}~\bibnamefont {Takahashi}}, \bibinfo {author}
  {\bibfnamefont {J.}~\bibnamefont {Ieda}}, \bibinfo {author} {\bibfnamefont
  {T.}~\bibnamefont {Ota}}, \bibinfo {author} {\bibfnamefont {Y.}~\bibnamefont
  {Kajiwara}}, \bibinfo {author} {\bibfnamefont {H.}~\bibnamefont {Umezawa}},
  \bibinfo {author} {\bibfnamefont {H.}~\bibnamefont {Kawai}}, \bibinfo
  {author} {\bibfnamefont {G.~E.~W.}\ \bibnamefont {Bauer}}, \bibinfo {author}
  {\bibfnamefont {S.}~\bibnamefont {Maekawa}},\ and\ \bibinfo {author}
  {\bibfnamefont {E.}~\bibnamefont {Saitoh}},\ }\href
  {http://dx.doi.org/10.1038/nmat2856} {\bibfield  {journal} {\bibinfo
  {journal} {Nat. Mater.}\ }\textbf {\bibinfo {volume} {9}},\ \bibinfo {pages}
  {894} (\bibinfo {year} {2010})}\BibitemShut {NoStop}%
\bibitem [{\citenamefont {Chumak}\ \emph {et~al.}(2015)\citenamefont {Chumak},
  \citenamefont {Vasyuchka}, \citenamefont {Serga},\ and\ \citenamefont
  {Hillebrands}}]{chumak2015magnon}%
  \BibitemOpen
  \bibfield  {author} {\bibinfo {author} {\bibfnamefont {A.}~\bibnamefont
  {Chumak}}, \bibinfo {author} {\bibfnamefont {V.}~\bibnamefont {Vasyuchka}},
  \bibinfo {author} {\bibfnamefont {A.}~\bibnamefont {Serga}},\ and\ \bibinfo
  {author} {\bibfnamefont {B.}~\bibnamefont {Hillebrands}},\ }\href@noop {}
  {\bibfield  {journal} {\bibinfo  {journal} {Nat. Phys.}\ }\textbf {\bibinfo
  {volume} {11}},\ \bibinfo {pages} {453} (\bibinfo {year} {2015})}\BibitemShut
  {NoStop}%
\bibitem [{\citenamefont {Cornelissen}\ \emph {et~al.}(2015)\citenamefont
  {Cornelissen}, \citenamefont {Liu}, \citenamefont {Duine}, \citenamefont
  {Youssef},\ and\ \citenamefont {Van~Wees}}]{cornelissen2015long}%
  \BibitemOpen
  \bibfield  {author} {\bibinfo {author} {\bibfnamefont {L.}~\bibnamefont
  {Cornelissen}}, \bibinfo {author} {\bibfnamefont {J.}~\bibnamefont {Liu}},
  \bibinfo {author} {\bibfnamefont {R.}~\bibnamefont {Duine}}, \bibinfo
  {author} {\bibfnamefont {J.~B.}\ \bibnamefont {Youssef}},\ and\ \bibinfo
  {author} {\bibfnamefont {B.}~\bibnamefont {Van~Wees}},\ }\href@noop {}
  {\bibfield  {journal} {\bibinfo  {journal} {Nat. Phys.}\ }\textbf {\bibinfo
  {volume} {11}},\ \bibinfo {pages} {1022} (\bibinfo {year}
  {2015})}\BibitemShut {NoStop}%
\bibitem [{\citenamefont {Yang}\ \emph {et~al.}(2010)\citenamefont {Yang},
  \citenamefont {Yang}, \citenamefont {Takahashi}, \citenamefont {Maekawa},\
  and\ \citenamefont {Parkin}}]{yang2010extremely}%
  \BibitemOpen
  \bibfield  {author} {\bibinfo {author} {\bibfnamefont {H.}~\bibnamefont
  {Yang}}, \bibinfo {author} {\bibfnamefont {S.-H.}\ \bibnamefont {Yang}},
  \bibinfo {author} {\bibfnamefont {S.}~\bibnamefont {Takahashi}}, \bibinfo
  {author} {\bibfnamefont {S.}~\bibnamefont {Maekawa}},\ and\ \bibinfo {author}
  {\bibfnamefont {S.~S.}\ \bibnamefont {Parkin}},\ }\href@noop {} {\bibfield
  {journal} {\bibinfo  {journal} {Nat. Mater.}\ }\textbf {\bibinfo {volume}
  {9}},\ \bibinfo {pages} {586} (\bibinfo {year} {2010})}\BibitemShut {NoStop}%
\bibitem [{\citenamefont {Quay}\ \emph {et~al.}(2013)\citenamefont {Quay},
  \citenamefont {Chevallier}, \citenamefont {Bena},\ and\ \citenamefont
  {Aprili}}]{quay2013spin}%
  \BibitemOpen
  \bibfield  {author} {\bibinfo {author} {\bibfnamefont {C.}~\bibnamefont
  {Quay}}, \bibinfo {author} {\bibfnamefont {D.}~\bibnamefont {Chevallier}},
  \bibinfo {author} {\bibfnamefont {C.}~\bibnamefont {Bena}},\ and\ \bibinfo
  {author} {\bibfnamefont {M.}~\bibnamefont {Aprili}},\ }\href@noop {}
  {\bibfield  {journal} {\bibinfo  {journal} {Nat. Phys.}\ }\textbf {\bibinfo
  {volume} {9}},\ \bibinfo {pages} {84} (\bibinfo {year} {2013})}\BibitemShut
  {NoStop}%
\bibitem [{\citenamefont {Wolf}\ \emph {et~al.}(2013)\citenamefont {Wolf},
  \citenamefont {H\"ubler}, \citenamefont {Kolenda}, \citenamefont
  {v.~L\"ohneysen},\ and\ \citenamefont {Beckmann}}]{Wolf2013}%
  \BibitemOpen
  \bibfield  {author} {\bibinfo {author} {\bibfnamefont {M.~J.}\ \bibnamefont
  {Wolf}}, \bibinfo {author} {\bibfnamefont {F.}~\bibnamefont {H\"ubler}},
  \bibinfo {author} {\bibfnamefont {S.}~\bibnamefont {Kolenda}}, \bibinfo
  {author} {\bibfnamefont {H.}~\bibnamefont {v.~L\"ohneysen}},\ and\ \bibinfo
  {author} {\bibfnamefont {D.}~\bibnamefont {Beckmann}},\ }\href
  {https://doi.org/10.1103/PhysRevB.87.024517} {\bibfield  {journal} {\bibinfo
  {journal} {Phys. Rev. B}\ }\textbf {\bibinfo {volume} {87}},\ \bibinfo
  {pages} {024517} (\bibinfo {year} {2013})}\BibitemShut {NoStop}%
\bibitem [{\citenamefont {H{\"u}bler}\ \emph {et~al.}(2012)\citenamefont
  {H{\"u}bler}, \citenamefont {Wolf}, \citenamefont {Beckmann},\ and\
  \citenamefont {L{\"o}hneysen}}]{hubler2012long}%
  \BibitemOpen
  \bibfield  {author} {\bibinfo {author} {\bibfnamefont {F.}~\bibnamefont
  {H{\"u}bler}}, \bibinfo {author} {\bibfnamefont {M.}~\bibnamefont {Wolf}},
  \bibinfo {author} {\bibfnamefont {D.}~\bibnamefont {Beckmann}},\ and\
  \bibinfo {author} {\bibfnamefont {H.~v.}\ \bibnamefont {L{\"o}hneysen}},\
  }\href@noop {} {\bibfield  {journal} {\bibinfo  {journal} {Phys. Rev. Lett.}\
  }\textbf {\bibinfo {volume} {109}},\ \bibinfo {pages} {207001} (\bibinfo
  {year} {2012})}\BibitemShut {NoStop}%
\bibitem [{\citenamefont {Kolenda}\ \emph {et~al.}(2017)\citenamefont
  {Kolenda}, \citenamefont {S{\"u}rgers}, \citenamefont {Fischer},\ and\
  \citenamefont {Beckmann}}]{Kolenda2017}%
  \BibitemOpen
  \bibfield  {author} {\bibinfo {author} {\bibfnamefont {S.}~\bibnamefont
  {Kolenda}}, \bibinfo {author} {\bibfnamefont {C.}~\bibnamefont
  {S{\"u}rgers}}, \bibinfo {author} {\bibfnamefont {G.}~\bibnamefont
  {Fischer}},\ and\ \bibinfo {author} {\bibfnamefont {D.}~\bibnamefont
  {Beckmann}},\ }\href@noop {} {\bibfield  {journal} {\bibinfo  {journal}
  {Phys. Rev. B}\ }\textbf {\bibinfo {volume} {95}},\ \bibinfo {pages} {224505}
  (\bibinfo {year} {2017})}\BibitemShut {NoStop}%
\bibitem [{\citenamefont {Heidrich}\ and\ \citenamefont
  {Beckmann}(2019)}]{heidrich2019nonlocal}%
  \BibitemOpen
  \bibfield  {author} {\bibinfo {author} {\bibfnamefont {J.}~\bibnamefont
  {Heidrich}}\ and\ \bibinfo {author} {\bibfnamefont {D.}~\bibnamefont
  {Beckmann}},\ }\href@noop {} {\bibfield  {journal} {\bibinfo  {journal}
  {Phys. Rev. B}\ }\textbf {\bibinfo {volume} {100}},\ \bibinfo {pages}
  {134501} (\bibinfo {year} {2019})}\BibitemShut {NoStop}%
\bibitem [{\citenamefont {Linder}\ and\ \citenamefont
  {Robinson}(2015)}]{linder2015superconducting}%
  \BibitemOpen
  \bibfield  {author} {\bibinfo {author} {\bibfnamefont {J.}~\bibnamefont
  {Linder}}\ and\ \bibinfo {author} {\bibfnamefont {J.~W.}\ \bibnamefont
  {Robinson}},\ }\href@noop {} {\bibfield  {journal} {\bibinfo  {journal} {Nat.
  Phys.}\ }\textbf {\bibinfo {volume} {11}},\ \bibinfo {pages} {307} (\bibinfo
  {year} {2015})}\BibitemShut {NoStop}%
\bibitem [{\citenamefont {Han}\ \emph {et~al.}(2020)\citenamefont {Han},
  \citenamefont {Maekawa},\ and\ \citenamefont {Xie}}]{Han2019}%
  \BibitemOpen
  \bibfield  {author} {\bibinfo {author} {\bibfnamefont {W.}~\bibnamefont
  {Han}}, \bibinfo {author} {\bibfnamefont {S.}~\bibnamefont {Maekawa}},\ and\
  \bibinfo {author} {\bibfnamefont {X.-C.}\ \bibnamefont {Xie}},\ }\href@noop
  {} {\bibfield  {journal} {\bibinfo  {journal} {Nat. Mater.}\ }\textbf
  {\bibinfo {volume} {19}},\ \bibinfo {pages} {139} (\bibinfo {year}
  {2020})}\BibitemShut {NoStop}%
\bibitem [{\citenamefont {Bergeret}\ \emph {et~al.}(2018)\citenamefont
  {Bergeret}, \citenamefont {Silaev}, \citenamefont {Virtanen},\ and\
  \citenamefont {Heikkil\"a}}]{RevModPhys.90.041001}%
  \BibitemOpen
  \bibfield  {author} {\bibinfo {author} {\bibfnamefont {F.~S.}\ \bibnamefont
  {Bergeret}}, \bibinfo {author} {\bibfnamefont {M.}~\bibnamefont {Silaev}},
  \bibinfo {author} {\bibfnamefont {P.}~\bibnamefont {Virtanen}},\ and\
  \bibinfo {author} {\bibfnamefont {T.~T.}\ \bibnamefont {Heikkil\"a}},\ }\href
  {https://doi.org/10.1103/RevModPhys.90.041001} {\bibfield  {journal}
  {\bibinfo  {journal} {Rev. Mod. Phys.}\ }\textbf {\bibinfo {volume} {90}},\
  \bibinfo {pages} {041001} (\bibinfo {year} {2018})}\BibitemShut {NoStop}%
\bibitem [{\citenamefont {Ozaeta}\ \emph {et~al.}(2014)\citenamefont {Ozaeta},
  \citenamefont {Virtanen}, \citenamefont {Bergeret},\ and\ \citenamefont
  {Heikkil{\"a}}}]{ozaeta2014predicted}%
  \BibitemOpen
  \bibfield  {author} {\bibinfo {author} {\bibfnamefont {A.}~\bibnamefont
  {Ozaeta}}, \bibinfo {author} {\bibfnamefont {P.}~\bibnamefont {Virtanen}},
  \bibinfo {author} {\bibfnamefont {F.~S.}\ \bibnamefont {Bergeret}},\ and\
  \bibinfo {author} {\bibfnamefont {T.~T.}\ \bibnamefont {Heikkil{\"a}}},\
  }\href@noop {} {\bibfield  {journal} {\bibinfo  {journal} {Phys. Rev. Lett.}\
  }\textbf {\bibinfo {volume} {112}},\ \bibinfo {pages} {057001} (\bibinfo
  {year} {2014})}\BibitemShut {NoStop}%
\bibitem [{\citenamefont {Silaev}\ \emph {et~al.}(2015)\citenamefont {Silaev},
  \citenamefont {Virtanen}, \citenamefont {Bergeret},\ and\ \citenamefont
  {Heikkilä}}]{Silaev2015}%
  \BibitemOpen
  \bibfield  {author} {\bibinfo {author} {\bibfnamefont {M.}~\bibnamefont
  {Silaev}}, \bibinfo {author} {\bibfnamefont {P.}~\bibnamefont {Virtanen}},
  \bibinfo {author} {\bibfnamefont {F.~S.}\ \bibnamefont {Bergeret}},\ and\
  \bibinfo {author} {\bibfnamefont {T.~T.}\ \bibnamefont {Heikkilä}},\ }\href
  {https://link.aps.org/doi/10.1103/PhysRevLett.114.167002} {\bibfield
  {journal} {\bibinfo  {journal} {Phys. Rev. Lett.}\ }\textbf {\bibinfo
  {volume} {114}},\ \bibinfo {pages} {167002} (\bibinfo {year}
  {2015})}\BibitemShut {NoStop}%
\bibitem [{\citenamefont {Heikkil{\"a}}\ \emph {et~al.}(2019)\citenamefont
  {Heikkil{\"a}}, \citenamefont {Silaev}, \citenamefont {Virtanen},\ and\
  \citenamefont {Bergeret}}]{heikkila2019thermal}%
  \BibitemOpen
  \bibfield  {author} {\bibinfo {author} {\bibfnamefont {T.~T.}\ \bibnamefont
  {Heikkil{\"a}}}, \bibinfo {author} {\bibfnamefont {M.}~\bibnamefont
  {Silaev}}, \bibinfo {author} {\bibfnamefont {P.}~\bibnamefont {Virtanen}},\
  and\ \bibinfo {author} {\bibfnamefont {F.~S.}\ \bibnamefont {Bergeret}},\
  }\href@noop {} {\bibfield  {journal} {\bibinfo  {journal} {Prog. Surf. Sci.}\
  }\textbf {\bibinfo {volume} {94}},\ \bibinfo {pages} {100540} (\bibinfo
  {year} {2019})}\BibitemShut {NoStop}%
\bibitem [{\citenamefont {Krishtop}\ \emph {et~al.}(2015)\citenamefont
  {Krishtop}, \citenamefont {Houzet},\ and\ \citenamefont
  {Meyer}}]{krishtop2015nonequilibrium}%
  \BibitemOpen
  \bibfield  {author} {\bibinfo {author} {\bibfnamefont {T.}~\bibnamefont
  {Krishtop}}, \bibinfo {author} {\bibfnamefont {M.}~\bibnamefont {Houzet}},\
  and\ \bibinfo {author} {\bibfnamefont {J.~S.}\ \bibnamefont {Meyer}},\
  }\href@noop {} {\bibfield  {journal} {\bibinfo  {journal} {Phys. Rev. B}\
  }\textbf {\bibinfo {volume} {91}},\ \bibinfo {pages} {121407} (\bibinfo
  {year} {2015})}\BibitemShut {NoStop}%
\bibitem [{\citenamefont {Bobkova}\ and\ \citenamefont
  {Bobkov}(2015)}]{bobkova2015long}%
  \BibitemOpen
  \bibfield  {author} {\bibinfo {author} {\bibfnamefont {I.~V.}\ \bibnamefont
  {Bobkova}}\ and\ \bibinfo {author} {\bibfnamefont {A.}~\bibnamefont
  {Bobkov}},\ }\href@noop {} {\bibfield  {journal} {\bibinfo  {journal} {JETP
  Letters}\ }\textbf {\bibinfo {volume} {101}},\ \bibinfo {pages} {118}
  (\bibinfo {year} {2015})}\BibitemShut {NoStop}%
\bibitem [{\citenamefont {Bobkova}\ and\ \citenamefont
  {Bobkov}(2017)}]{bobkova2017thermospin}%
  \BibitemOpen
  \bibfield  {author} {\bibinfo {author} {\bibfnamefont {I.}~\bibnamefont
  {Bobkova}}\ and\ \bibinfo {author} {\bibfnamefont {A.}~\bibnamefont
  {Bobkov}},\ }\href@noop {} {\bibfield  {journal} {\bibinfo  {journal} {Phys.
  Rev. B}\ }\textbf {\bibinfo {volume} {96}},\ \bibinfo {pages} {104515}
  (\bibinfo {year} {2017})}\BibitemShut {NoStop}%
\bibitem [{\citenamefont {Virtanen}\ \emph {et~al.}(2016)\citenamefont
  {Virtanen}, \citenamefont {Heikkil{\"a}},\ and\ \citenamefont
  {Bergeret}}]{virtanen2016stimulated}%
  \BibitemOpen
  \bibfield  {author} {\bibinfo {author} {\bibfnamefont {P.}~\bibnamefont
  {Virtanen}}, \bibinfo {author} {\bibfnamefont {T.}~\bibnamefont
  {Heikkil{\"a}}},\ and\ \bibinfo {author} {\bibfnamefont {F.}~\bibnamefont
  {Bergeret}},\ }\href@noop {} {\bibfield  {journal} {\bibinfo  {journal}
  {Phys. Rev. B}\ }\textbf {\bibinfo {volume} {93}},\ \bibinfo {pages} {014512}
  (\bibinfo {year} {2016})}\BibitemShut {NoStop}%
\bibitem [{\citenamefont {Kuzmanovi{\'c}}\ \emph {et~al.}(2020)\citenamefont
  {Kuzmanovi{\'c}}, \citenamefont {Wu}, \citenamefont {Weideneder},
  \citenamefont {Quay},\ and\ \citenamefont {Aprili}}]{kuzmanovic2020evidence}%
  \BibitemOpen
  \bibfield  {author} {\bibinfo {author} {\bibfnamefont {M.}~\bibnamefont
  {Kuzmanovi{\'c}}}, \bibinfo {author} {\bibfnamefont {B.}~\bibnamefont {Wu}},
  \bibinfo {author} {\bibfnamefont {M.}~\bibnamefont {Weideneder}}, \bibinfo
  {author} {\bibfnamefont {C.}~\bibnamefont {Quay}},\ and\ \bibinfo {author}
  {\bibfnamefont {M.}~\bibnamefont {Aprili}},\ }\href@noop {} {\bibfield
  {journal} {\bibinfo  {journal} {Nat. Comm.}\ }\textbf {\bibinfo {volume}
  {11}},\ \bibinfo {pages} {1} (\bibinfo {year} {2020})}\BibitemShut {NoStop}%
\bibitem [{\citenamefont {Bell}\ \emph {et~al.}(2008)\citenamefont {Bell},
  \citenamefont {Milikisyants}, \citenamefont {Huber},\ and\ \citenamefont
  {Aarts}}]{bell2008spin}%
  \BibitemOpen
  \bibfield  {author} {\bibinfo {author} {\bibfnamefont {C.}~\bibnamefont
  {Bell}}, \bibinfo {author} {\bibfnamefont {S.}~\bibnamefont {Milikisyants}},
  \bibinfo {author} {\bibfnamefont {M.}~\bibnamefont {Huber}},\ and\ \bibinfo
  {author} {\bibfnamefont {J.}~\bibnamefont {Aarts}},\ }\href@noop {}
  {\bibfield  {journal} {\bibinfo  {journal} {Phys. Rev. Lett.}\ }\textbf
  {\bibinfo {volume} {100}},\ \bibinfo {pages} {047002} (\bibinfo {year}
  {2008})}\BibitemShut {NoStop}%
\bibitem [{\citenamefont {Wakamura}\ \emph {et~al.}(2015)\citenamefont
  {Wakamura}, \citenamefont {Akaike}, \citenamefont {Omori}, \citenamefont
  {Niimi}, \citenamefont {Takahashi}, \citenamefont {Fujimaki}, \citenamefont
  {Maekawa},\ and\ \citenamefont {Otani}}]{wakamura2015quasiparticle}%
  \BibitemOpen
  \bibfield  {author} {\bibinfo {author} {\bibfnamefont {T.}~\bibnamefont
  {Wakamura}}, \bibinfo {author} {\bibfnamefont {H.}~\bibnamefont {Akaike}},
  \bibinfo {author} {\bibfnamefont {Y.}~\bibnamefont {Omori}}, \bibinfo
  {author} {\bibfnamefont {Y.}~\bibnamefont {Niimi}}, \bibinfo {author}
  {\bibfnamefont {S.}~\bibnamefont {Takahashi}}, \bibinfo {author}
  {\bibfnamefont {A.}~\bibnamefont {Fujimaki}}, \bibinfo {author}
  {\bibfnamefont {S.}~\bibnamefont {Maekawa}},\ and\ \bibinfo {author}
  {\bibfnamefont {Y.}~\bibnamefont {Otani}},\ }\href@noop {} {\bibfield
  {journal} {\bibinfo  {journal} {Nat. Mater.}\ }\textbf {\bibinfo {volume}
  {14}},\ \bibinfo {pages} {675} (\bibinfo {year} {2015})}\BibitemShut
  {NoStop}%
\bibitem [{\citenamefont {Jeon}\ \emph
  {et~al.}(2018{\natexlab{a}})\citenamefont {Jeon}, \citenamefont {Ciccarelli},
  \citenamefont {Ferguson}, \citenamefont {Kurebayashi}, \citenamefont {Cohen},
  \citenamefont {Montiel}, \citenamefont {Eschrig}, \citenamefont {Robinson},\
  and\ \citenamefont {Blamire}}]{Jeon2018}%
  \BibitemOpen
  \bibfield  {author} {\bibinfo {author} {\bibfnamefont {K.-R.}\ \bibnamefont
  {Jeon}}, \bibinfo {author} {\bibfnamefont {C.}~\bibnamefont {Ciccarelli}},
  \bibinfo {author} {\bibfnamefont {A.~J.}\ \bibnamefont {Ferguson}}, \bibinfo
  {author} {\bibfnamefont {H.}~\bibnamefont {Kurebayashi}}, \bibinfo {author}
  {\bibfnamefont {L.~F.}\ \bibnamefont {Cohen}}, \bibinfo {author}
  {\bibfnamefont {X.}~\bibnamefont {Montiel}}, \bibinfo {author} {\bibfnamefont
  {M.}~\bibnamefont {Eschrig}}, \bibinfo {author} {\bibfnamefont {J.~W.~A.}\
  \bibnamefont {Robinson}},\ and\ \bibinfo {author} {\bibfnamefont {M.~G.}\
  \bibnamefont {Blamire}},\ }\href {https://doi.org/10.1038/s41563-018-0058-9}
  {\bibfield  {journal} {\bibinfo  {journal} {Nat. Mater.}\ }\textbf {\bibinfo
  {volume} {17}},\ \bibinfo {pages} {499} (\bibinfo {year}
  {2018}{\natexlab{a}})}\BibitemShut {NoStop}%
\bibitem [{\citenamefont {Jeon}\ \emph
  {et~al.}(2020{\natexlab{a}})\citenamefont {Jeon}, \citenamefont {Montiel},
  \citenamefont {Komori}, \citenamefont {Ciccarelli}, \citenamefont {Haigh},
  \citenamefont {Kurebayashi}, \citenamefont {Cohen}, \citenamefont {Chan},
  \citenamefont {Stenning}, \citenamefont {Lee} \emph
  {et~al.}}]{jeon2020tunable}%
  \BibitemOpen
  \bibfield  {author} {\bibinfo {author} {\bibfnamefont {K.-R.}\ \bibnamefont
  {Jeon}}, \bibinfo {author} {\bibfnamefont {X.}~\bibnamefont {Montiel}},
  \bibinfo {author} {\bibfnamefont {S.}~\bibnamefont {Komori}}, \bibinfo
  {author} {\bibfnamefont {C.}~\bibnamefont {Ciccarelli}}, \bibinfo {author}
  {\bibfnamefont {J.}~\bibnamefont {Haigh}}, \bibinfo {author} {\bibfnamefont
  {H.}~\bibnamefont {Kurebayashi}}, \bibinfo {author} {\bibfnamefont {L.~F.}\
  \bibnamefont {Cohen}}, \bibinfo {author} {\bibfnamefont {A.~K.}\ \bibnamefont
  {Chan}}, \bibinfo {author} {\bibfnamefont {K.~D.}\ \bibnamefont {Stenning}},
  \bibinfo {author} {\bibfnamefont {C.-M.}\ \bibnamefont {Lee}}, \emph
  {et~al.},\ }\href@noop {} {\bibfield  {journal} {\bibinfo  {journal} {Phys.
  Rev. X}\ }\textbf {\bibinfo {volume} {10}},\ \bibinfo {pages} {031020}
  (\bibinfo {year} {2020}{\natexlab{a}})}\BibitemShut {NoStop}%
\bibitem [{\citenamefont {Golovchanskiy}\ \emph {et~al.}(2020)\citenamefont
  {Golovchanskiy}, \citenamefont {Abramov}, \citenamefont {Stolyarov},
  \citenamefont {Chichkov}, \citenamefont {Silaev}, \citenamefont {Shchetinin},
  \citenamefont {Golubov}, \citenamefont {Ryazanov}, \citenamefont {Ustinov},\
  and\ \citenamefont {Kupriyanov}}]{PhysRevApplied.14.024086}%
  \BibitemOpen
  \bibfield  {author} {\bibinfo {author} {\bibfnamefont {I.}~\bibnamefont
  {Golovchanskiy}}, \bibinfo {author} {\bibfnamefont {N.}~\bibnamefont
  {Abramov}}, \bibinfo {author} {\bibfnamefont {V.}~\bibnamefont {Stolyarov}},
  \bibinfo {author} {\bibfnamefont {V.}~\bibnamefont {Chichkov}}, \bibinfo
  {author} {\bibfnamefont {M.}~\bibnamefont {Silaev}}, \bibinfo {author}
  {\bibfnamefont {I.}~\bibnamefont {Shchetinin}}, \bibinfo {author}
  {\bibfnamefont {A.}~\bibnamefont {Golubov}}, \bibinfo {author} {\bibfnamefont
  {V.}~\bibnamefont {Ryazanov}}, \bibinfo {author} {\bibfnamefont
  {A.}~\bibnamefont {Ustinov}},\ and\ \bibinfo {author} {\bibfnamefont
  {M.}~\bibnamefont {Kupriyanov}},\ }\href
  {https://doi.org/10.1103/PhysRevApplied.14.024086} {\bibfield  {journal}
  {\bibinfo  {journal} {Phys. Rev. Applied}\ }\textbf {\bibinfo {volume}
  {14}},\ \bibinfo {pages} {024086} (\bibinfo {year} {2020})}\BibitemShut
  {NoStop}%
\bibitem [{\citenamefont {Jeon}\ \emph
  {et~al.}(2020{\natexlab{b}})\citenamefont {Jeon}, \citenamefont {Jeon},
  \citenamefont {Zhou}, \citenamefont {Migliorini}, \citenamefont {Yoon},\ and\
  \citenamefont {Parkin}}]{Jeon2020giant}%
  \BibitemOpen
  \bibfield  {author} {\bibinfo {author} {\bibfnamefont {K.-R.}\ \bibnamefont
  {Jeon}}, \bibinfo {author} {\bibfnamefont {J.-C.}\ \bibnamefont {Jeon}},
  \bibinfo {author} {\bibfnamefont {X.}~\bibnamefont {Zhou}}, \bibinfo {author}
  {\bibfnamefont {A.}~\bibnamefont {Migliorini}}, \bibinfo {author}
  {\bibfnamefont {J.}~\bibnamefont {Yoon}},\ and\ \bibinfo {author}
  {\bibfnamefont {S.~S.~P.}\ \bibnamefont {Parkin}},\ }\href
  {https://doi.org/10.1021/acsnano.0c07187} {\bibfield  {journal} {\bibinfo
  {journal} {ACS Nano}\ }\textbf {\bibinfo {volume} {14}},\ \bibinfo {pages}
  {15874} (\bibinfo {year} {2020}{\natexlab{b}})}\BibitemShut {NoStop}%
\bibitem [{\citenamefont {Brataas}\ and\ \citenamefont
  {Tserkovnyak}(2004)}]{brataas2004spin}%
  \BibitemOpen
  \bibfield  {author} {\bibinfo {author} {\bibfnamefont {A.}~\bibnamefont
  {Brataas}}\ and\ \bibinfo {author} {\bibfnamefont {Y.}~\bibnamefont
  {Tserkovnyak}},\ }\href@noop {} {\bibfield  {journal} {\bibinfo  {journal}
  {Phys. Rev. Lett.}\ }\textbf {\bibinfo {volume} {93}},\ \bibinfo {pages}
  {087201} (\bibinfo {year} {2004})}\BibitemShut {NoStop}%
\bibitem [{\citenamefont {Morten}\ \emph {et~al.}(2008)\citenamefont {Morten},
  \citenamefont {Brataas}, \citenamefont {Bauer}, \citenamefont {Belzig},\ and\
  \citenamefont {Tserkovnyak}}]{morten2008proximity}%
  \BibitemOpen
  \bibfield  {author} {\bibinfo {author} {\bibfnamefont {J.~P.}\ \bibnamefont
  {Morten}}, \bibinfo {author} {\bibfnamefont {A.}~\bibnamefont {Brataas}},
  \bibinfo {author} {\bibfnamefont {G.~E.}\ \bibnamefont {Bauer}}, \bibinfo
  {author} {\bibfnamefont {W.}~\bibnamefont {Belzig}},\ and\ \bibinfo {author}
  {\bibfnamefont {Y.}~\bibnamefont {Tserkovnyak}},\ }\href@noop {} {\bibfield
  {journal} {\bibinfo  {journal} {EPL (Europhysics Letters)}\ }\textbf
  {\bibinfo {volume} {84}},\ \bibinfo {pages} {57008} (\bibinfo {year}
  {2008})}\BibitemShut {NoStop}%
\bibitem [{\citenamefont {Inoue}\ \emph {et~al.}(2017)\citenamefont {Inoue},
  \citenamefont {Ichioka},\ and\ \citenamefont {Adachi}}]{inoue2017spin}%
  \BibitemOpen
  \bibfield  {author} {\bibinfo {author} {\bibfnamefont {M.}~\bibnamefont
  {Inoue}}, \bibinfo {author} {\bibfnamefont {M.}~\bibnamefont {Ichioka}},\
  and\ \bibinfo {author} {\bibfnamefont {H.}~\bibnamefont {Adachi}},\
  }\href@noop {} {\bibfield  {journal} {\bibinfo  {journal} {Phys. Rev. B}\
  }\textbf {\bibinfo {volume} {96}},\ \bibinfo {pages} {024414} (\bibinfo
  {year} {2017})}\BibitemShut {NoStop}%
\bibitem [{\citenamefont {Kato}\ \emph
  {et~al.}(2019{\natexlab{a}})\citenamefont {Kato}, \citenamefont {Ohnuma},
  \citenamefont {Matsuo}, \citenamefont {Rech}, \citenamefont {Jonckheere},\
  and\ \citenamefont {Martin}}]{kato2019microscopic}%
  \BibitemOpen
  \bibfield  {author} {\bibinfo {author} {\bibfnamefont {T.}~\bibnamefont
  {Kato}}, \bibinfo {author} {\bibfnamefont {Y.}~\bibnamefont {Ohnuma}},
  \bibinfo {author} {\bibfnamefont {M.}~\bibnamefont {Matsuo}}, \bibinfo
  {author} {\bibfnamefont {J.}~\bibnamefont {Rech}}, \bibinfo {author}
  {\bibfnamefont {T.}~\bibnamefont {Jonckheere}},\ and\ \bibinfo {author}
  {\bibfnamefont {T.}~\bibnamefont {Martin}},\ }\href@noop {} {\bibfield
  {journal} {\bibinfo  {journal} {Phys. Rev. B}\ }\textbf {\bibinfo {volume}
  {99}},\ \bibinfo {pages} {144411} (\bibinfo {year}
  {2019}{\natexlab{a}})}\BibitemShut {NoStop}%
\bibitem [{\citenamefont {Silaev}(2020{\natexlab{a}})}]{silaev2020large}%
  \BibitemOpen
  \bibfield  {author} {\bibinfo {author} {\bibfnamefont {M.}~\bibnamefont
  {Silaev}},\ }\href@noop {} {\bibfield  {journal} {\bibinfo  {journal} {Phys.
  Rev. B}\ }\textbf {\bibinfo {volume} {102}},\ \bibinfo {pages} {180502}
  (\bibinfo {year} {2020}{\natexlab{a}})}\BibitemShut {NoStop}%
\bibitem [{\citenamefont {Silaev}(2020{\natexlab{b}})}]{silaev2020finite}%
  \BibitemOpen
  \bibfield  {author} {\bibinfo {author} {\bibfnamefont {M.}~\bibnamefont
  {Silaev}},\ }\href@noop {} {\bibfield  {journal} {\bibinfo  {journal} {Phys.
  Rev. B}\ }\textbf {\bibinfo {volume} {102}},\ \bibinfo {pages} {144521}
  (\bibinfo {year} {2020}{\natexlab{b}})}\BibitemShut {NoStop}%
\bibitem [{\citenamefont {Tanhayi~Ahari}\ and\ \citenamefont
  {Tserkovnyak}(2020)}]{tanhayi2020superconductivity}%
  \BibitemOpen
  \bibfield  {author} {\bibinfo {author} {\bibfnamefont {M.}~\bibnamefont
  {Tanhayi~Ahari}}\ and\ \bibinfo {author} {\bibfnamefont {Y.}~\bibnamefont
  {Tserkovnyak}},\ }\href@noop {} {\bibfield  {journal} {\bibinfo  {journal}
  {arXiv e-prints}\ ,\ \bibinfo {pages} {arXiv}} (\bibinfo {year}
  {2020})}\BibitemShut {NoStop}%
\bibitem [{\citenamefont {Trif}\ and\ \citenamefont
  {Tserkovnyak}(2013)}]{trif2013dynamic}%
  \BibitemOpen
  \bibfield  {author} {\bibinfo {author} {\bibfnamefont {M.}~\bibnamefont
  {Trif}}\ and\ \bibinfo {author} {\bibfnamefont {Y.}~\bibnamefont
  {Tserkovnyak}},\ }\href@noop {} {\bibfield  {journal} {\bibinfo  {journal}
  {Phys. Rev. Lett.}\ }\textbf {\bibinfo {volume} {111}},\ \bibinfo {pages}
  {087602} (\bibinfo {year} {2013})}\BibitemShut {NoStop}%
\bibitem [{\citenamefont {Ojaj{\"a}rvi}\ \emph {et~al.}(2020)\citenamefont
  {Ojaj{\"a}rvi}, \citenamefont {Manninen}, \citenamefont {Heikkil{\"a}},\ and\
  \citenamefont {Virtanen}}]{ojajarvi2020nonlinear}%
  \BibitemOpen
  \bibfield  {author} {\bibinfo {author} {\bibfnamefont {R.}~\bibnamefont
  {Ojaj{\"a}rvi}}, \bibinfo {author} {\bibfnamefont {J.}~\bibnamefont
  {Manninen}}, \bibinfo {author} {\bibfnamefont {T.~T.}\ \bibnamefont
  {Heikkil{\"a}}},\ and\ \bibinfo {author} {\bibfnamefont {P.}~\bibnamefont
  {Virtanen}},\ }\href@noop {} {\bibfield  {journal} {\bibinfo  {journal}
  {Phys. Rev. B}\ }\textbf {\bibinfo {volume} {101}},\ \bibinfo {pages}
  {115406} (\bibinfo {year} {2020})}\BibitemShut {NoStop}%
\bibitem [{\citenamefont {Jeon}\ \emph
  {et~al.}(2018{\natexlab{b}})\citenamefont {Jeon}, \citenamefont {Ciccarelli},
  \citenamefont {Kurebayashi}, \citenamefont {Wunderlich}, \citenamefont
  {Cohen}, \citenamefont {Komori}, \citenamefont {Robinson},\ and\
  \citenamefont {Blamire}}]{jeon2018spin}%
  \BibitemOpen
  \bibfield  {author} {\bibinfo {author} {\bibfnamefont {K.-R.}\ \bibnamefont
  {Jeon}}, \bibinfo {author} {\bibfnamefont {C.}~\bibnamefont {Ciccarelli}},
  \bibinfo {author} {\bibfnamefont {H.}~\bibnamefont {Kurebayashi}}, \bibinfo
  {author} {\bibfnamefont {J.}~\bibnamefont {Wunderlich}}, \bibinfo {author}
  {\bibfnamefont {L.~F.}\ \bibnamefont {Cohen}}, \bibinfo {author}
  {\bibfnamefont {S.}~\bibnamefont {Komori}}, \bibinfo {author} {\bibfnamefont
  {J.~W.}\ \bibnamefont {Robinson}},\ and\ \bibinfo {author} {\bibfnamefont
  {M.~G.}\ \bibnamefont {Blamire}},\ }\href@noop {} {\bibfield  {journal}
  {\bibinfo  {journal} {Phys. Rev. Applied}\ }\textbf {\bibinfo {volume}
  {10}},\ \bibinfo {pages} {014029} (\bibinfo {year}
  {2018}{\natexlab{b}})}\BibitemShut {NoStop}%
\bibitem [{\citenamefont {Gershenzon}\ \emph {et~al.}(1990)\citenamefont
  {Gershenzon}, \citenamefont {Gershenzon}, \citenamefont {Gol’tsman},
  \citenamefont {Lyul’kin}, \citenamefont {Semenov},\ and\ \citenamefont
  {Sergeev}}]{gershenzon1990electron}%
  \BibitemOpen
  \bibfield  {author} {\bibinfo {author} {\bibfnamefont {E.}~\bibnamefont
  {Gershenzon}}, \bibinfo {author} {\bibfnamefont {M.}~\bibnamefont
  {Gershenzon}}, \bibinfo {author} {\bibfnamefont {G.}~\bibnamefont
  {Gol’tsman}}, \bibinfo {author} {\bibfnamefont {A.}~\bibnamefont
  {Lyul’kin}}, \bibinfo {author} {\bibfnamefont {A.}~\bibnamefont
  {Semenov}},\ and\ \bibinfo {author} {\bibfnamefont {A.}~\bibnamefont
  {Sergeev}},\ }\href@noop {} {\bibfield  {journal} {\bibinfo  {journal} {Sov.
  Phys. JETP}\ }\textbf {\bibinfo {volume} {70}},\ \bibinfo {pages} {505}
  (\bibinfo {year} {1990})}\BibitemShut {NoStop}%
\bibitem [{\citenamefont {Klapwijk}\ \emph {et~al.}(1986)\citenamefont
  {Klapwijk}, \citenamefont {van~der Plas},\ and\ \citenamefont
  {Mooij}}]{klapwijk1986electron}%
  \BibitemOpen
  \bibfield  {author} {\bibinfo {author} {\bibfnamefont {T.}~\bibnamefont
  {Klapwijk}}, \bibinfo {author} {\bibfnamefont {P.}~\bibnamefont {van~der
  Plas}},\ and\ \bibinfo {author} {\bibfnamefont {J.}~\bibnamefont {Mooij}},\
  }\href@noop {} {\bibfield  {journal} {\bibinfo  {journal} {Phys. Rev. B}\
  }\textbf {\bibinfo {volume} {33}},\ \bibinfo {pages} {1474} (\bibinfo {year}
  {1986})}\BibitemShut {NoStop}%
\bibitem [{\citenamefont {Brataas}\ \emph {et~al.}(2008)\citenamefont
  {Brataas}, \citenamefont {Tserkovnyak},\ and\ \citenamefont
  {Bauer}}]{brataas2008scattering}%
  \BibitemOpen
  \bibfield  {author} {\bibinfo {author} {\bibfnamefont {A.}~\bibnamefont
  {Brataas}}, \bibinfo {author} {\bibfnamefont {Y.}~\bibnamefont
  {Tserkovnyak}},\ and\ \bibinfo {author} {\bibfnamefont {G.~E.}\ \bibnamefont
  {Bauer}},\ }\href@noop {} {\bibfield  {journal} {\bibinfo  {journal} {Phys.
  Rev. Lett.}\ }\textbf {\bibinfo {volume} {101}},\ \bibinfo {pages} {037207}
  (\bibinfo {year} {2008})}\BibitemShut {NoStop}%
\bibitem [{\citenamefont {Meservey}\ and\ \citenamefont
  {Tedrow}(1994)}]{meservey1994spin}%
  \BibitemOpen
  \bibfield  {author} {\bibinfo {author} {\bibfnamefont {R.}~\bibnamefont
  {Meservey}}\ and\ \bibinfo {author} {\bibfnamefont {P.}~\bibnamefont
  {Tedrow}},\ }\href@noop {} {\bibfield  {journal} {\bibinfo  {journal} {Phys.
  Rep.}\ }\textbf {\bibinfo {volume} {238}},\ \bibinfo {pages} {173} (\bibinfo
  {year} {1994})}\BibitemShut {NoStop}%
\bibitem [{\citenamefont {Hijano}\ \emph {et~al.}(2020)\citenamefont {Hijano},
  \citenamefont {Ili{\'c}}, \citenamefont {Rouco}, \citenamefont {Orellana},
  \citenamefont {Ilyn}, \citenamefont {Rogero}, \citenamefont {Virtanen},
  \citenamefont {Heikkil{\"a}}, \citenamefont {Khorshidian}, \citenamefont
  {Spies} \emph {et~al.}}]{hijano2020coexistence}%
  \BibitemOpen
  \bibfield  {author} {\bibinfo {author} {\bibfnamefont {A.}~\bibnamefont
  {Hijano}}, \bibinfo {author} {\bibfnamefont {S.}~\bibnamefont {Ili{\'c}}},
  \bibinfo {author} {\bibfnamefont {M.}~\bibnamefont {Rouco}}, \bibinfo
  {author} {\bibfnamefont {C.~G.}\ \bibnamefont {Orellana}}, \bibinfo {author}
  {\bibfnamefont {M.}~\bibnamefont {Ilyn}}, \bibinfo {author} {\bibfnamefont
  {C.}~\bibnamefont {Rogero}}, \bibinfo {author} {\bibfnamefont
  {P.}~\bibnamefont {Virtanen}}, \bibinfo {author} {\bibfnamefont
  {T.}~\bibnamefont {Heikkil{\"a}}}, \bibinfo {author} {\bibfnamefont
  {S.}~\bibnamefont {Khorshidian}}, \bibinfo {author} {\bibfnamefont
  {M.}~\bibnamefont {Spies}}, \emph {et~al.},\ }\href@noop {} {\bibfield
  {journal} {\bibinfo  {journal} {arXiv:2012.15549}\ } (\bibinfo {year}
  {2020})}\BibitemShut {NoStop}%
\bibitem [{spi()}]{spinrelnote}%
  \BibitemOpen
  \href@noop {} {}\bibinfo {note} {This picture is valid for weak spin
  relaxation with $\tau_{s}^{-1} \ll \Delta$. For larger $\tau_{s}^{-1}$ the
  spectrum becomes more complicated as the spin ceases to be a good quantum
  number. Our quasiclassical approach takes into account this spin
  mixing.}\BibitemShut {Stop}%
\bibitem [{Sup()}]{SupplMat}%
  \BibitemOpen
  \href@noop {} {}\bibinfo {note} {Supplementary material file includes
  derivation of the general relation between spin and energy currents generated
  by magnetization dynamics; derivation of kinetic equations with spin-energy
  coupling and calculation of the anomalous parts of interfacial spin and
  energy currents; derivation of the boundary conditions for Green's functions
  in S/FI system with stochastic magnetization field, e.g. thermal magnons;
  derivation of the second-order perturbation theory equations, results of such
  calculations for various parameters and the comparison with the
  non-perturbative numerical solution of the Keldysh-Usadel equation with a
  time-dependent Zeeman field}\BibitemShut {NoStop}%
\bibitem [{\citenamefont {Silaev}(2020{\natexlab{c}})}]{silaev2020ff}%
  \BibitemOpen
  \bibfield  {author} {\bibinfo {author} {\bibfnamefont {M.~A.}\ \bibnamefont
  {Silaev}},\ }\href {https://doi.org/10.1103/PhysRevB.102.144521} {\bibfield
  {journal} {\bibinfo  {journal} {Phys. Rev. B}\ }\textbf {\bibinfo {volume}
  {102}},\ \bibinfo {pages} {144521} (\bibinfo {year}
  {2020}{\natexlab{c}})}\BibitemShut {NoStop}%
\bibitem [{\citenamefont {Simensen}\ \emph {et~al.}(2021)\citenamefont
  {Simensen}, \citenamefont {Johnsen}, \citenamefont {Linder},\ and\
  \citenamefont {Brataas}}]{PhysRevB.103.024524}%
  \BibitemOpen
  \bibfield  {author} {\bibinfo {author} {\bibfnamefont {H.~T.}\ \bibnamefont
  {Simensen}}, \bibinfo {author} {\bibfnamefont {L.~G.}\ \bibnamefont
  {Johnsen}}, \bibinfo {author} {\bibfnamefont {J.}~\bibnamefont {Linder}},\
  and\ \bibinfo {author} {\bibfnamefont {A.}~\bibnamefont {Brataas}},\ }\href
  {https://doi.org/10.1103/PhysRevB.103.024524} {\bibfield  {journal} {\bibinfo
   {journal} {Phys. Rev. B}\ }\textbf {\bibinfo {volume} {103}},\ \bibinfo
  {pages} {024524} (\bibinfo {year} {2021})}\BibitemShut {NoStop}%
\bibitem [{\citenamefont {Tokuyasu}\ \emph {et~al.}(1988)\citenamefont
  {Tokuyasu}, \citenamefont {Sauls},\ and\ \citenamefont
  {Rainer}}]{Tokuyasu1988}%
  \BibitemOpen
  \bibfield  {author} {\bibinfo {author} {\bibfnamefont {T.}~\bibnamefont
  {Tokuyasu}}, \bibinfo {author} {\bibfnamefont {J.~A.}\ \bibnamefont
  {Sauls}},\ and\ \bibinfo {author} {\bibfnamefont {D.}~\bibnamefont
  {Rainer}},\ }\href {https://link.aps.org/doi/10.1103/PhysRevB.38.8823}
  {\bibfield  {journal} {\bibinfo  {journal} {Phys. Rev. B}\ }\textbf {\bibinfo
  {volume} {38}},\ \bibinfo {pages} {8823} (\bibinfo {year}
  {1988})}\BibitemShut {NoStop}%
\bibitem [{\citenamefont {Ohnuma}\ \emph {et~al.}(2014)\citenamefont {Ohnuma},
  \citenamefont {Adachi}, \citenamefont {Saitoh},\ and\ \citenamefont
  {Maekawa}}]{ohnuma2014enhanced}%
  \BibitemOpen
  \bibfield  {author} {\bibinfo {author} {\bibfnamefont {Y.}~\bibnamefont
  {Ohnuma}}, \bibinfo {author} {\bibfnamefont {H.}~\bibnamefont {Adachi}},
  \bibinfo {author} {\bibfnamefont {E.}~\bibnamefont {Saitoh}},\ and\ \bibinfo
  {author} {\bibfnamefont {S.}~\bibnamefont {Maekawa}},\ }\href@noop {}
  {\bibfield  {journal} {\bibinfo  {journal} {Phys. Rev. B}\ }\textbf {\bibinfo
  {volume} {89}},\ \bibinfo {pages} {174417} (\bibinfo {year}
  {2014})}\BibitemShut {NoStop}%
\bibitem [{\citenamefont {Millis}\ \emph {et~al.}(1988)\citenamefont {Millis},
  \citenamefont {Rainer},\ and\ \citenamefont
  {Sauls}}]{millis1988quasiclassical}%
  \BibitemOpen
  \bibfield  {author} {\bibinfo {author} {\bibfnamefont {A.}~\bibnamefont
  {Millis}}, \bibinfo {author} {\bibfnamefont {D.}~\bibnamefont {Rainer}},\
  and\ \bibinfo {author} {\bibfnamefont {J.}~\bibnamefont {Sauls}},\
  }\href@noop {} {\bibfield  {journal} {\bibinfo  {journal} {Phys. Rev. B}\
  }\textbf {\bibinfo {volume} {38}},\ \bibinfo {pages} {4504} (\bibinfo {year}
  {1988})}\BibitemShut {NoStop}%
\bibitem [{\citenamefont {Semenov}\ \emph {et~al.}(2016)\citenamefont
  {Semenov}, \citenamefont {Devyatov}, \citenamefont {De~Visser},\ and\
  \citenamefont {Klapwijk}}]{semenov2016coherent}%
  \BibitemOpen
  \bibfield  {author} {\bibinfo {author} {\bibfnamefont {A.}~\bibnamefont
  {Semenov}}, \bibinfo {author} {\bibfnamefont {I.}~\bibnamefont {Devyatov}},
  \bibinfo {author} {\bibfnamefont {P.}~\bibnamefont {De~Visser}},\ and\
  \bibinfo {author} {\bibfnamefont {T.}~\bibnamefont {Klapwijk}},\ }\href@noop
  {} {\bibfield  {journal} {\bibinfo  {journal} {Phys. Rev. Lett.}\ }\textbf
  {\bibinfo {volume} {117}},\ \bibinfo {pages} {047002} (\bibinfo {year}
  {2016})}\BibitemShut {NoStop}%
\bibitem [{\citenamefont {Linder}\ \emph {et~al.}(2016)\citenamefont {Linder},
  \citenamefont {Amundsen},\ and\ \citenamefont
  {Ouassou}}]{linder2016microwave}%
  \BibitemOpen
  \bibfield  {author} {\bibinfo {author} {\bibfnamefont {J.}~\bibnamefont
  {Linder}}, \bibinfo {author} {\bibfnamefont {M.}~\bibnamefont {Amundsen}},\
  and\ \bibinfo {author} {\bibfnamefont {J.~A.}\ \bibnamefont {Ouassou}},\
  }\href@noop {} {\bibfield  {journal} {\bibinfo  {journal} {Sci. Rep.}\
  }\textbf {\bibinfo {volume} {6}},\ \bibinfo {pages} {1} (\bibinfo {year}
  {2016})}\BibitemShut {NoStop}%
\bibitem [{\citenamefont {Eliashberg}(1971)}]{eliashberg1971inelastic}%
  \BibitemOpen
  \bibfield  {author} {\bibinfo {author} {\bibfnamefont {G.}~\bibnamefont
  {Eliashberg}},\ }\href@noop {} {\bibfield  {journal} {\bibinfo  {journal}
  {Zh. Eksp. Teor. Fiz.;(USSR)}\ }\textbf {\bibinfo {volume} {61}} (\bibinfo
  {year} {1971})}\BibitemShut {NoStop}%
\bibitem [{\citenamefont {Gor'kov}\ and\ \citenamefont
  {Kopnin}(1975)}]{gor1975vortex}%
  \BibitemOpen
  \bibfield  {author} {\bibinfo {author} {\bibfnamefont {L.~P.}\ \bibnamefont
  {Gor'kov}}\ and\ \bibinfo {author} {\bibfnamefont {N.}~\bibnamefont
  {Kopnin}},\ }\href@noop {} {\bibfield  {journal} {\bibinfo  {journal} {Soviet
  Physics Uspekhi}\ }\textbf {\bibinfo {volume} {18}},\ \bibinfo {pages} {496}
  (\bibinfo {year} {1975})}\BibitemShut {NoStop}%
\bibitem [{\citenamefont {Larkin}\ and\ \citenamefont
  {Ovchinnikov}(1977)}]{larkin1977non}%
  \BibitemOpen
  \bibfield  {author} {\bibinfo {author} {\bibfnamefont {A.}~\bibnamefont
  {Larkin}}\ and\ \bibinfo {author} {\bibfnamefont {Y.}~\bibnamefont
  {Ovchinnikov}},\ }\href@noop {} {\bibfield  {journal} {\bibinfo  {journal}
  {Zh. Eksp. Teor. Fiz.}\ }\textbf {\bibinfo {volume} {73}},\ \bibinfo {pages}
  {7} (\bibinfo {year} {1977})}\BibitemShut {NoStop}%
\bibitem [{\citenamefont {Artemenko}\ and\ \citenamefont
  {Volkov}(1979)}]{artemenko1979electric}%
  \BibitemOpen
  \bibfield  {author} {\bibinfo {author} {\bibfnamefont {S.~N.}\ \bibnamefont
  {Artemenko}}\ and\ \bibinfo {author} {\bibfnamefont {A.}~\bibnamefont
  {Volkov}},\ }\href@noop {} {\bibfield  {journal} {\bibinfo  {journal} {Soviet
  Physics Uspekhi}\ }\textbf {\bibinfo {volume} {22}},\ \bibinfo {pages} {295}
  (\bibinfo {year} {1979})}\BibitemShut {NoStop}%
\bibitem [{\citenamefont {Adachi}\ \emph {et~al.}(2011)\citenamefont {Adachi},
  \citenamefont {Ohe}, \citenamefont {Takahashi},\ and\ \citenamefont
  {Maekawa}}]{PhysRevB.83.094410}%
  \BibitemOpen
  \bibfield  {author} {\bibinfo {author} {\bibfnamefont {H.}~\bibnamefont
  {Adachi}}, \bibinfo {author} {\bibfnamefont {J.-i.}\ \bibnamefont {Ohe}},
  \bibinfo {author} {\bibfnamefont {S.}~\bibnamefont {Takahashi}},\ and\
  \bibinfo {author} {\bibfnamefont {S.}~\bibnamefont {Maekawa}},\ }\href
  {https://doi.org/10.1103/PhysRevB.83.094410} {\bibfield  {journal} {\bibinfo
  {journal} {Phys. Rev. B}\ }\textbf {\bibinfo {volume} {83}},\ \bibinfo
  {pages} {094410} (\bibinfo {year} {2011})}\BibitemShut {NoStop}%
\bibitem [{\citenamefont {Adachi}\ \emph {et~al.}(2013)\citenamefont {Adachi},
  \citenamefont {Uchida}, \citenamefont {Saitoh},\ and\ \citenamefont
  {Maekawa}}]{adachi2013theory}%
  \BibitemOpen
  \bibfield  {author} {\bibinfo {author} {\bibfnamefont {H.}~\bibnamefont
  {Adachi}}, \bibinfo {author} {\bibfnamefont {K.-i.}\ \bibnamefont {Uchida}},
  \bibinfo {author} {\bibfnamefont {E.}~\bibnamefont {Saitoh}},\ and\ \bibinfo
  {author} {\bibfnamefont {S.}~\bibnamefont {Maekawa}},\ }\href@noop {}
  {\bibfield  {journal} {\bibinfo  {journal} {Rep. Prog. Phys.}\ }\textbf
  {\bibinfo {volume} {76}},\ \bibinfo {pages} {036501} (\bibinfo {year}
  {2013})}\BibitemShut {NoStop}%
\bibitem [{\citenamefont {Kato}\ \emph
  {et~al.}(2019{\natexlab{b}})\citenamefont {Kato}, \citenamefont {Ohnuma},
  \citenamefont {Matsuo}, \citenamefont {Rech}, \citenamefont {Jonckheere},\
  and\ \citenamefont {Martin}}]{Kato2019}%
  \BibitemOpen
  \bibfield  {author} {\bibinfo {author} {\bibfnamefont {T.}~\bibnamefont
  {Kato}}, \bibinfo {author} {\bibfnamefont {Y.}~\bibnamefont {Ohnuma}},
  \bibinfo {author} {\bibfnamefont {M.}~\bibnamefont {Matsuo}}, \bibinfo
  {author} {\bibfnamefont {J.}~\bibnamefont {Rech}}, \bibinfo {author}
  {\bibfnamefont {T.}~\bibnamefont {Jonckheere}},\ and\ \bibinfo {author}
  {\bibfnamefont {T.}~\bibnamefont {Martin}},\ }\href@noop {} {\bibfield
  {journal} {\bibinfo  {journal} {Phys. Rev. B}\ }\textbf {\bibinfo {volume}
  {99}},\ \bibinfo {pages} {144411} (\bibinfo {year}
  {2019}{\natexlab{b}})}\BibitemShut {NoStop}%
\bibitem [{\citenamefont {Srivastava}\ and\ \citenamefont
  {Aiyar}(1987)}]{srivastava1987}%
  \BibitemOpen
  \bibfield  {author} {\bibinfo {author} {\bibfnamefont {C.~M.}\ \bibnamefont
  {Srivastava}}\ and\ \bibinfo {author} {\bibfnamefont {R.}~\bibnamefont
  {Aiyar}},\ }\href {https://doi.org/10.1088/0022-3719/20/8/013} {\bibfield
  {journal} {\bibinfo  {journal} {J. Phys. C}\ }\textbf {\bibinfo {volume}
  {20}},\ \bibinfo {pages} {1119} (\bibinfo {year} {1987})}\BibitemShut
  {NoStop}%
\bibitem [{\citenamefont {Cherepanov}\ \emph {et~al.}(1993)\citenamefont
  {Cherepanov}, \citenamefont {Kolokolov},\ and\ \citenamefont
  {L'vov}}]{cherepanov1993saga}%
  \BibitemOpen
  \bibfield  {author} {\bibinfo {author} {\bibfnamefont {V.}~\bibnamefont
  {Cherepanov}}, \bibinfo {author} {\bibfnamefont {I.}~\bibnamefont
  {Kolokolov}},\ and\ \bibinfo {author} {\bibfnamefont {V.}~\bibnamefont
  {L'vov}},\ }\href {https://doi.org/10.1016/0370-1573(93)90107-O} {\bibfield
  {journal} {\bibinfo  {journal} {Phys. Rep.}\ }\textbf {\bibinfo {volume}
  {229}},\ \bibinfo {pages} {81} (\bibinfo {year} {1993})}\BibitemShut
  {NoStop}%
\bibitem [{\citenamefont {Maki}(1973)}]{maki1973}%
  \BibitemOpen
  \bibfield  {author} {\bibinfo {author} {\bibfnamefont {K.}~\bibnamefont
  {Maki}},\ }\href {https://doi.org/10.1103/PhysRevB.8.191} {\bibfield
  {journal} {\bibinfo  {journal} {Phys. Rev. B}\ }\textbf {\bibinfo {volume}
  {8}},\ \bibinfo {pages} {191} (\bibinfo {year} {1973})}\BibitemShut {NoStop}%
\bibitem [{\citenamefont {Bender}\ and\ \citenamefont
  {Tserkovnyak}(2015)}]{bender2015interfacial}%
  \BibitemOpen
  \bibfield  {author} {\bibinfo {author} {\bibfnamefont {S.~A.}\ \bibnamefont
  {Bender}}\ and\ \bibinfo {author} {\bibfnamefont {Y.}~\bibnamefont
  {Tserkovnyak}},\ }\href {https://doi.org/10.1103/PhysRevB.91.140402}
  {\bibfield  {journal} {\bibinfo  {journal} {Phys. Rev. B}\ }\textbf {\bibinfo
  {volume} {91}},\ \bibinfo {pages} {140402} (\bibinfo {year}
  {2015})}\BibitemShut {NoStop}%
\bibitem [{\citenamefont {Cornelissen}\ \emph {et~al.}(2016)\citenamefont
  {Cornelissen}, \citenamefont {Peters}, \citenamefont {Bauer}, \citenamefont
  {Duine},\ and\ \citenamefont {van Wees}}]{cornelissen2016magnon}%
  \BibitemOpen
  \bibfield  {author} {\bibinfo {author} {\bibfnamefont {L.~J.}\ \bibnamefont
  {Cornelissen}}, \bibinfo {author} {\bibfnamefont {K.~J.~H.}\ \bibnamefont
  {Peters}}, \bibinfo {author} {\bibfnamefont {G.~E.~W.}\ \bibnamefont
  {Bauer}}, \bibinfo {author} {\bibfnamefont {R.~A.}\ \bibnamefont {Duine}},\
  and\ \bibinfo {author} {\bibfnamefont {B.~J.}\ \bibnamefont {van Wees}},\
  }\href {https://doi.org/10.1103/PhysRevB.94.014412} {\bibfield  {journal}
  {\bibinfo  {journal} {Phys. Rev. B}\ }\textbf {\bibinfo {volume} {94}},\
  \bibinfo {pages} {014412} (\bibinfo {year} {2016})}\BibitemShut {NoStop}%
\bibitem [{Note1()}]{Note1}%
  \BibitemOpen
  \bibinfo {note} {Fig. 4 assumes for simplicity that the electron-phonon
  coupling in Nb is the same as in Al.}\BibitemShut {Stop}%
\bibitem [{\citenamefont {Heikkil{\"a}}\ \emph {et~al.}(2018)\citenamefont
  {Heikkil{\"a}}, \citenamefont {Ojaj{\"a}rvi}, \citenamefont {Maasilta},
  \citenamefont {Strambini}, \citenamefont {Giazotto},\ and\ \citenamefont
  {Bergeret}}]{heikkila2018thermoelectric}%
  \BibitemOpen
  \bibfield  {author} {\bibinfo {author} {\bibfnamefont {T.}~\bibnamefont
  {Heikkil{\"a}}}, \bibinfo {author} {\bibfnamefont {R.}~\bibnamefont
  {Ojaj{\"a}rvi}}, \bibinfo {author} {\bibfnamefont {I.}~\bibnamefont
  {Maasilta}}, \bibinfo {author} {\bibfnamefont {E.}~\bibnamefont {Strambini}},
  \bibinfo {author} {\bibfnamefont {F.}~\bibnamefont {Giazotto}},\ and\
  \bibinfo {author} {\bibfnamefont {F.}~\bibnamefont {Bergeret}},\ }\href@noop
  {} {\bibfield  {journal} {\bibinfo  {journal} {Phys. Rev. Appl.}\ }\textbf
  {\bibinfo {volume} {10}},\ \bibinfo {pages} {034053} (\bibinfo {year}
  {2018})}\BibitemShut {NoStop}%
\bibitem [{\citenamefont {Chakraborty}\ and\ \citenamefont
  {Heikkil{\"a}}(2018)}]{chakraborty2018thermoelectric}%
  \BibitemOpen
  \bibfield  {author} {\bibinfo {author} {\bibfnamefont {S.}~\bibnamefont
  {Chakraborty}}\ and\ \bibinfo {author} {\bibfnamefont {T.~T.}\ \bibnamefont
  {Heikkil{\"a}}},\ }\href@noop {} {\bibfield  {journal} {\bibinfo  {journal}
  {J. Appl. Phys.}\ }\textbf {\bibinfo {volume} {124}},\ \bibinfo {pages}
  {123902} (\bibinfo {year} {2018})}\BibitemShut {NoStop}%
\bibitem [{\citenamefont {Govenius}\ \emph {et~al.}(2014)\citenamefont
  {Govenius}, \citenamefont {Lake}, \citenamefont {Tan}, \citenamefont
  {Pietil{\"a}}, \citenamefont {Julin}, \citenamefont {Maasilta}, \citenamefont
  {Virtanen},\ and\ \citenamefont {M{\"o}tt{\"o}nen}}]{govenius2014microwave}%
  \BibitemOpen
  \bibfield  {author} {\bibinfo {author} {\bibfnamefont {J.}~\bibnamefont
  {Govenius}}, \bibinfo {author} {\bibfnamefont {R.}~\bibnamefont {Lake}},
  \bibinfo {author} {\bibfnamefont {K.}~\bibnamefont {Tan}}, \bibinfo {author}
  {\bibfnamefont {V.}~\bibnamefont {Pietil{\"a}}}, \bibinfo {author}
  {\bibfnamefont {J.}~\bibnamefont {Julin}}, \bibinfo {author} {\bibfnamefont
  {I.}~\bibnamefont {Maasilta}}, \bibinfo {author} {\bibfnamefont
  {P.}~\bibnamefont {Virtanen}},\ and\ \bibinfo {author} {\bibfnamefont
  {M.}~\bibnamefont {M{\"o}tt{\"o}nen}},\ }\href@noop {} {\bibfield  {journal}
  {\bibinfo  {journal} {Phys. Rev. B}\ }\textbf {\bibinfo {volume} {90}},\
  \bibinfo {pages} {064505} (\bibinfo {year} {2014})}\BibitemShut {NoStop}%
\bibitem [{\citenamefont {Govenius}\ \emph {et~al.}(2016)\citenamefont
  {Govenius}, \citenamefont {Lake}, \citenamefont {Tan},\ and\ \citenamefont
  {M{\"o}tt{\"o}nen}}]{govenius2016detection}%
  \BibitemOpen
  \bibfield  {author} {\bibinfo {author} {\bibfnamefont {J.}~\bibnamefont
  {Govenius}}, \bibinfo {author} {\bibfnamefont {R.}~\bibnamefont {Lake}},
  \bibinfo {author} {\bibfnamefont {K.}~\bibnamefont {Tan}},\ and\ \bibinfo
  {author} {\bibfnamefont {M.}~\bibnamefont {M{\"o}tt{\"o}nen}},\ }\href@noop
  {} {\bibfield  {journal} {\bibinfo  {journal} {Phys. Rev. Lett.}\ }\textbf
  {\bibinfo {volume} {117}},\ \bibinfo {pages} {030802} (\bibinfo {year}
  {2016})}\BibitemShut {NoStop}%
\bibitem [{\citenamefont {Kokkoniemi}\ \emph {et~al.}(2019)\citenamefont
  {Kokkoniemi}, \citenamefont {Govenius}, \citenamefont {Vesterinen},
  \citenamefont {Lake}, \citenamefont {Gunyh{\'o}}, \citenamefont {Tan},
  \citenamefont {Simbierowicz}, \citenamefont {Gr{\"o}nberg}, \citenamefont
  {Lehtinen}, \citenamefont {Prunnila} \emph
  {et~al.}}]{kokkoniemi2019nanobolometer}%
  \BibitemOpen
  \bibfield  {author} {\bibinfo {author} {\bibfnamefont {R.}~\bibnamefont
  {Kokkoniemi}}, \bibinfo {author} {\bibfnamefont {J.}~\bibnamefont
  {Govenius}}, \bibinfo {author} {\bibfnamefont {V.}~\bibnamefont
  {Vesterinen}}, \bibinfo {author} {\bibfnamefont {R.~E.}\ \bibnamefont
  {Lake}}, \bibinfo {author} {\bibfnamefont {A.~M.}\ \bibnamefont
  {Gunyh{\'o}}}, \bibinfo {author} {\bibfnamefont {K.~Y.}\ \bibnamefont {Tan}},
  \bibinfo {author} {\bibfnamefont {S.}~\bibnamefont {Simbierowicz}}, \bibinfo
  {author} {\bibfnamefont {L.}~\bibnamefont {Gr{\"o}nberg}}, \bibinfo {author}
  {\bibfnamefont {J.}~\bibnamefont {Lehtinen}}, \bibinfo {author}
  {\bibfnamefont {M.}~\bibnamefont {Prunnila}}, \emph {et~al.},\ }\href@noop {}
  {\bibfield  {journal} {\bibinfo  {journal} {Communications Physics}\ }\textbf
  {\bibinfo {volume} {2}},\ \bibinfo {pages} {1} (\bibinfo {year}
  {2019})}\BibitemShut {NoStop}%
\bibitem [{\citenamefont {Bergeret}\ and\ \citenamefont
  {Tokatly}(2016)}]{bergeret2016manifestation}%
  \BibitemOpen
  \bibfield  {author} {\bibinfo {author} {\bibfnamefont {F.~S.}\ \bibnamefont
  {Bergeret}}\ and\ \bibinfo {author} {\bibfnamefont {I.~V.}\ \bibnamefont
  {Tokatly}},\ }\href@noop {} {\bibfield  {journal} {\bibinfo  {journal} {Phys.
  Rev. B}\ }\textbf {\bibinfo {volume} {94}},\ \bibinfo {pages} {180502}
  (\bibinfo {year} {2016})}\BibitemShut {NoStop}%
\bibitem [{\citenamefont {Tokatly}(2017)}]{Tokatly2017}%
  \BibitemOpen
  \bibfield  {author} {\bibinfo {author} {\bibfnamefont {I.~V.}\ \bibnamefont
  {Tokatly}},\ }\href {https://link.aps.org/doi/10.1103/PhysRevB.96.060502}
  {\bibfield  {journal} {\bibinfo  {journal} {Phys. Rev. B}\ }\textbf {\bibinfo
  {volume} {96}},\ \bibinfo {pages} {060502} (\bibinfo {year}
  {2017})}\BibitemShut {NoStop}%
\bibitem [{\citenamefont {Huang}\ \emph {et~al.}(2018)\citenamefont {Huang},
  \citenamefont {Tokatly},\ and\ \citenamefont
  {Bergeret}}]{huang2018extrinsic}%
  \BibitemOpen
  \bibfield  {author} {\bibinfo {author} {\bibfnamefont {C.}~\bibnamefont
  {Huang}}, \bibinfo {author} {\bibfnamefont {I.~V.}\ \bibnamefont {Tokatly}},\
  and\ \bibinfo {author} {\bibfnamefont {F.~S.}\ \bibnamefont {Bergeret}},\
  }\href@noop {} {\bibfield  {journal} {\bibinfo  {journal} {Phys. Rev. B}\
  }\textbf {\bibinfo {volume} {98}},\ \bibinfo {pages} {144515} (\bibinfo
  {year} {2018})}\BibitemShut {NoStop}%
\bibitem [{\citenamefont {Otrokov}\ \emph {et~al.}(2019)\citenamefont
  {Otrokov}, \citenamefont {Klimovskikh}, \citenamefont {Bentmann},
  \citenamefont {Estyunin}, \citenamefont {Zeugner}, \citenamefont {Aliev},
  \citenamefont {Ga{\ss}}, \citenamefont {Wolter}, \citenamefont {Koroleva},
  \citenamefont {Shikin} \emph {et~al.}}]{otrokov2019prediction}%
  \BibitemOpen
  \bibfield  {author} {\bibinfo {author} {\bibfnamefont {M.~M.}\ \bibnamefont
  {Otrokov}}, \bibinfo {author} {\bibfnamefont {I.~I.}\ \bibnamefont
  {Klimovskikh}}, \bibinfo {author} {\bibfnamefont {H.}~\bibnamefont
  {Bentmann}}, \bibinfo {author} {\bibfnamefont {D.}~\bibnamefont {Estyunin}},
  \bibinfo {author} {\bibfnamefont {A.}~\bibnamefont {Zeugner}}, \bibinfo
  {author} {\bibfnamefont {Z.~S.}\ \bibnamefont {Aliev}}, \bibinfo {author}
  {\bibfnamefont {S.}~\bibnamefont {Ga{\ss}}}, \bibinfo {author} {\bibfnamefont
  {A.}~\bibnamefont {Wolter}}, \bibinfo {author} {\bibfnamefont
  {A.}~\bibnamefont {Koroleva}}, \bibinfo {author} {\bibfnamefont {A.~M.}\
  \bibnamefont {Shikin}}, \emph {et~al.},\ }\href@noop {} {\bibfield  {journal}
  {\bibinfo  {journal} {Nature}\ }\textbf {\bibinfo {volume} {576}},\ \bibinfo
  {pages} {416} (\bibinfo {year} {2019})}\BibitemShut {NoStop}%
\bibitem [{\citenamefont {Dynes}\ \emph {et~al.}(1984)\citenamefont {Dynes},
  \citenamefont {Garno}, \citenamefont {Hertel},\ and\ \citenamefont
  {Orlando}}]{Dynes1984}%
  \BibitemOpen
  \bibfield  {author} {\bibinfo {author} {\bibfnamefont {R.~C.}\ \bibnamefont
  {Dynes}}, \bibinfo {author} {\bibfnamefont {J.~P.}\ \bibnamefont {Garno}},
  \bibinfo {author} {\bibfnamefont {G.~B.}\ \bibnamefont {Hertel}},\ and\
  \bibinfo {author} {\bibfnamefont {T.~P.}\ \bibnamefont {Orlando}},\ }\href
  {https://doi.org/10.1103/PhysRevLett.53.2437} {\bibfield  {journal} {\bibinfo
   {journal} {Phys. Rev. Lett.}\ }\textbf {\bibinfo {volume} {53}},\ \bibinfo
  {pages} {2437} (\bibinfo {year} {1984})}\BibitemShut {NoStop}%
\bibitem [{\citenamefont {Abrikosov}\ and\ \citenamefont
  {Gor’kov}(1962)}]{abrikosov1962spin}%
  \BibitemOpen
  \bibfield  {author} {\bibinfo {author} {\bibfnamefont {A.}~\bibnamefont
  {Abrikosov}}\ and\ \bibinfo {author} {\bibfnamefont {L.}~\bibnamefont
  {Gor’kov}},\ }\href@noop {} {\bibfield  {journal} {\bibinfo  {journal}
  {Sov. Phys. JETP}\ }\textbf {\bibinfo {volume} {15}},\ \bibinfo {pages} {752}
  (\bibinfo {year} {1962})}\BibitemShut {NoStop}%
\bibitem [{\citenamefont {Kamenev}(2011)}]{kamenev2011field}%
  \BibitemOpen
  \bibfield  {author} {\bibinfo {author} {\bibfnamefont {A.}~\bibnamefont
  {Kamenev}},\ }\href@noop {} {\emph {\bibinfo {title} {Field theory of
  non-equilibrium systems}}}\ (\bibinfo  {publisher} {Cambridge University
  Press},\ \bibinfo {year} {2011})\BibitemShut {NoStop}%
\end{thebibliography}%
